\documentclass{aa}
\usepackage{graphicx}
\usepackage[flushleft]{threeparttable}
\usepackage{amsmath}

\def\la{\lower.5ex\hbox{$\; \buildrel < \over \sim \;$}}
\def\ga{\lower.5ex\hbox{$\; \buildrel > \over \sim \;$}}

\begin{document}      

   \title{Predicting HCN, HCO$^{+}$, multi-transition CO, and dust emission of star-forming galaxies}
   \subtitle{Extension to luminous infrared galaxies and the role of cosmic ray ionization}

   \author{B.~Vollmer\inst{1}, J.~Freundlich\inst{1}, P.~Gratier\inst{2}, Th.~Liz\'ee\inst{1}, M.~Lendrin\inst{1}, J.~Braine\inst{2}, M.~Soida\inst{3}}

   \offprints{B.~Vollmer, e-mail: Bernd.Vollmer@astro.unistra.fr}

   \institute{Universit\'e de Strasbourg, CNRS, Observatoire Astronomique de Strasbourg, UMR 7550, 67000 Strasbourg, France \and
              Laboratoire d'astrophysique de Bordeaux, Univ. Bordeaux, CNRS, B18N, all\'e Geoffroy Saint-Hilaire, 33615 Pessac, France \and
              Astronomical Observatory, Jagiellonian University, ul. Orla 171, 30-244 Krak\'ow, Poland
              }

   \date{Received / Accepted}

   \authorrunning{Vollmer et al.}

   \abstract{
     The specific star-formation rate of
     star-forming `main sequence' galaxies significantly decreased since $z \sim 1.5$, due to the decreasing molecular gas fraction and star formation efficiency.
     The gas velocity dispersion decreased within the same redshift range, apparently correlated with the star formation efficiency (inverse of the molecular gas depletion time).
     However, the radio--infrared (IR) correlation has not changed significantly since $z \sim 1.5$.
     The theory of turbulent clumpy starforming gas disks together with the scaling relations of the interstellar medium describes the large and small-scale properties of galactic
     gas disks. Here we extend our previous work on infrared, multi-transition molecular line, and radio continuum emission of local and high-z starforming and starburst
     galaxies to local and $z \sim 0.5$ luminous infrared galaxies.
     The model reproduces the IR luminosities, CO, HCN, and HCO$^+$ line luminosities,
     and the CO spectral line energy distributions of these galaxies. We derive  CO(1-0) and HCN(1-0) conversion factors for all galaxy samples.
     The relation between the star formation rate per unit area and H$_2$ surface density cannot be fit simply for all redshifts. 
     There is a tight correlation between the star formation efficiency and the product of the gas turbulent velocity dispersion and the angular velocity of the galaxies.
     Galaxies of lower stellar masses can in principle compensate their gas consumption via star formation by radial viscous gas accretion.
     The limiting stellar mass increases with redshift.
     Whereas the radio continuum emission is directly proportional to the density of cosmic ray (CR) electrons, the molecular line emission depends on the CR ionization rate via the
     gas chemistry. The normalization of the CR ionization rate found for the different galaxy samples is about a factor of three to five higher than the normalization for the Solar
     neighborhood. This means that the mean yield of low energy CR particles for a given star formation rate per unit area is
     about three to ten times higher in external galaxies than observed by Voyager~I.
   }

\maketitle
\nolinenumbers

\section{Introduction \label{sec:intro}}

The star formation process has changed with time and a theory of star formation must account for these changes.  The cosmic star-formation
rate (SFR) density declined by a factor of approximately ten since the peak at $z\sim 2$, often referred to as 'cosmic noon' (Madau et al. 1998, Hopkins \& Beacom 2006). The mean SFR ($\dot{M}_*$) per stellar mass (specific SFR, hereafter sSFR) decreases simultaneously such that star-forming galaxies form a `main sequence' in the star-formation/stellar-mass 
space (e.g., Brinchmann et al. 2004; Daddi et al. 2007; Elbaz et al. 2007; Noeske et al. 2007a,b; Salim et al. 2007; Whitaker et al. 2012). 
Galactic starbursts, such as ultraluminous infrared galaxies (ULIRGs) or submillimeter-galaxies, represent outliers from the main sequence.
The slope and offset of the $\dot{M}_*$--$M_*$ main sequence relation change with redshift (Speagle et al. 2014). The most prominent change is 
that the sSFR ($\dot{M}_*/M_*$) increases significantly with redshift.
Pannella et al. (2015) estimated that the sSFR of a $M_* \sim 3 \times 10^{10}$~M$_{\odot}$ galaxy was 6 times higher at $z=1$ than today.
The slope and scatter of this correlation and the evolution of its normalization with cosmic time contain crucial and still poorly known information on galaxy evolution
(e.g., Karim et al. 2011; Rodighiero et al. 2011; Wuyts et al. 2011; Sargent et al. 2012).

The gas content of a galaxy largely determines its SFR but the redshift appears to play a role.
In local starforming galaxies the molecular gas depletion time is correlated with the sSFR (Saintonge et al. 2017).
Since $z \sim 1.5$, the mean molecular gas fraction ($M_{\rm mol}/M_*$) decreased by about a factor of four and the molecular gas depletion time, defined as M(H$_2$)/SFR, the inverse of the star formation efficiency (SFE = SFR/M(H$_2$)), increased from less than a Gyr to about 2~Gyr today
(Tacconi et al. 2020). The high gas velocity dispersion of $z \sim 2$ disks is well established (F\"{o}rster-Schreiber \& Wuyts 2020).
Typical velocity dispersions of the ionized gas are $\sim 45$~km\,s$^{-1}$ at $z\sim 2$, compared with $\sim 25$~km\,s$^{-1}$ at $z \sim 0$, varying as
$v_{\rm turb} \sim 23 + 10z$~km\,s$^{-1}$ (F\"{o}rster-Schreiber \& Wuyts 2020).
Molecular gas measurements at $z > 0$ are rare but follow a similar evolution albeit with $\sim 10$-$15$~km\,s$^{-1}$ lower velocity dispersions (\"{U}bler et al. 2019).
Weak or no trends are found between the gas velocity dispersion and global galaxy parameters such as $M_*$, SFR, or SFR surface density
(Genzel et al. 2011, Johnson et al. 2018, \"{U}bler et al. 2019).

In Vollmer et al. (2017, hereafter V17) we took a closer look at the gas content and  associated SFR in main sequence and starburst galaxies
at $z=0$ and $z \sim 1$--$2$. The time-dependent gas chemistry of the turbulent gas clouds was taken into account to 
produce model IR, CO, HCN, and HCO$^+$ fluxes for comparison with observations. 
Our relatively simple analytical model appeared to capture the essential physics of clumpy galactic gaseous disks.
Most importantly, the model yields radial profiles of the H$_2$ surface density and the gas velocity dispersion.

There are only few observations of the molecular gas in galaxies during the long period between cosmic noon and now.  The PHIBSS2 sample
of luminous infrared galaxies (LIRGs) at $z \sim 0.5$ (Freundlich et al. 2019, hereafter F19), which lie within the scatter of the main sequence,
provides observations of a large sample during this critical period in the
history of the universe. F19 found a mean molecular gas fraction 
$M_{\rm mol}/M_* =0.28 \pm 0.04$ and depletion time $t_{\rm dep} =$M(H$_2$)/SFR $ 0.84 \pm 0.07$~Gyr compared to
the local value of $t_{\rm dep} = 2.2$~Gyr with a scatter of $0.3$~dex (Leroy et al. 2013).
These authors argue that the star formation rate is mainly driven by the molecular gas fraction (see also Tacconi et al. 2020).
The large molecular gas reservoirs fueling star formation are thought to be maintained by a continuous supply of fresh gas from
the cosmic web and minor mergers (e.g., Keres et al. (2005, 2009), Ocvirk et al. (2008), Sancisi et al. 2008, Dekel et al. 2009).
However, as shown by the lower $t_{\rm dep}$ in the past, the SFR also varies with redshift.  
Our model tries to explain (i) the link between the molecular gas depletion time and the gas velocity dispersion and (ii)
the evolution of these quantities with redshift.

One of the tightest correlations in astronomy is the relation between the 
radio continuum (synchrotron) and
the far-infrared (FIR) dust emission (e.g., Mirabel \& Sanders 1984; Yun et al. 2001; Bell 2003; Basu et al. 2015; Molnar et al. 2021).
Given the multiple energy-loss mechanisms of cosmic-ray (CR) electrons in galaxies, the tightness of the infrared (IR)--radio continuum correlation is surprising.
In Vollmer et al. (2022, hereafter V22) we extended the analytical model of V17 by including a simplified prescription for the synchrotron emissivity.
The magnetic field strength is determined by the equipartition between the turbulent kinetic and the magnetic energy densities.
The observed IR-radio correlations for local and high-z SF and starburst galaxies were
reproduced by the model within $2\sigma$ of the joint uncertainty of model and data for all datasets. 

In this work we extend the models of V17 and V22 to local and $z \sim 0.5$ LIRGs to see if our model reproduces the radio continuum and
molecular line emission of these strongly starforming galaxies, which fill the gap of model galaxies
between $5 \times 10^{10}$ (local starforming galaxies) and $3 \times 10^{11}$~L$_{\odot}$ (ULIRGs and high-z starforming galaxies) in Vollmer et al. (2017; their Fig.4).
The V17 model for the molecular line emission and the V22 model for the radio continuum emission are applied to the extended galaxy sample.
The new modelling aspect is the calculation of the CR ionization rate based on the model radio continuum emission, which is calibrated by observations.
These ionization rates are used for the calculation of the molecular abundances (Sect.~2.1.4 in V17). In contrast to V17, the molecular line emission
is calculated by RADEX (van der Tak 2007).

The integrated infrared, molecular line, and radio continuum emission of the local and $z \sim 0.5$ LIRGs are calculated to confirm our model and study differences
(e.g., dust temperature, radio spectral index, and spectral line energy distribution) for different galaxy parameters.
The model results for all galaxy samples are used to study the gas supply for star formation by radial accretion within the galactic disks. 
In addition, the molecular line emission is used to constrain the CR ionization rate.
On the one hand, ionization by CR heats the gas, so molecular line emission depends on the density of the CR protons, and affects the gas chemistry.
On the other hand, the radio continuum emission is proportional to the density of the cosmic ray (CR) electrons. 
The combination of the molecular line emission and radio continuum models thus (i) verifies the validity of our model and (ii) yields the CR ionization rate in external galaxies,
which can be compared to the Galactic and local CR ionization rates. Our estimate of the extragalactic CR ionization rates is
complementary to those obtained from measurements of HCO$^+$/CO, OH/CO (Gaches et al. 2019; Luo et al. 2023), DCO$^+$/HCO$^+$ (Caselli et al. 1998),
HCO$^+$/N2H$^+$ (Ceccarelli et al. 2014), or H$_3^+$ (Indriolo et al. 2015; Neufeld \& Wolfire 2017).

This article is structured as follows: the galaxy samples are presented in Sect.~\ref{sec:samples}. An important input for the
analytic model is the galactic rotation curve. The extraction of these curves for the PHIBSS2 sample is described in Sect.~\ref{sec:rotcurv}.
The analytical model of turbulent clumpy star forming gas disks and their IR, molecular line and radio continuum emission is described
in Sect.~\ref{sec:model}. The model results are presented in Sect.~\ref{sec:modelresults}. The molecular gas depletion times, gas viscosity, and accretion rates are discussed in Sects.~\ref{sec:sfes} and \ref{sec:viscosity}.
Sect.~\ref{sec:crir} is devoted to the CR ionization fraction. 
The meaning of the variables used for the model of the gas disk (Sect.~\ref{sec:gasdisk} and Appendix~\ref{sec:dgasdisk}), the CR ionization rate
and molecular abundances (Sect.~\ref{sec:cosmicrays}), radio continuum and molecular line emission 
(Sect.~\ref{sec:radcont} and Appendix~\ref{sec:radconta}) are given in Tables~\ref{tab:parameters}, \ref{tab:parameters1}, and \ref{tab:parameters2}, respectively.

\section{Galaxy samples \label{sec:samples}}

Most galaxies form stars at a rate proportional to their stellar mass. The tight relation between star formation and stellar
mass is called the main sequence of star forming galaxies, in place from redshift $\sim 0$ up to $\sim 4$ (e.g., Speagle et al. 2014).
Galaxies with much higher SFRs than predicted by the main sequence are called starburst galaxies.

In addition to the four galaxy samples presented in V17 (local and high-z starforming and starburst galaxies) 
we include the PHIBSS2 intermediate-redshift ($z\sim0.5 - 0.8$) main sequence galaxies, which are luminous infrared galaxies (LIRGs, L$_{IR} > 10^{11}L_\odot$), 
and local LIRGs which resemble high-z main sequence galaxies in terms of $f_{\rm mol}$ and $\sigma_v$ (DYNAMO; Green et al. 2014; Fisher et al. 2019).
Three of our six galaxy samples consist of main sequence galaxies (local spirals, intermediate-z and high-z star-forming galaxies)
and three are starburst samples (low-z LIRGS and starbursts/ULIRGs, high-z starbursts).
High-z and dusty starburst galaxies with SFRs higher than
$200$~M$_{\odot}$yr$^{-1}$ are usually called submillimeter (submm) galaxies (e.g., Bothwell et al. 2013).
The physical properties of the galaxies are given in Tables~B.1 to B.4 of V17.

F19 measured mean $f_{\rm mol}$ and $t_{\rm dep}$ across the main sequence at z$\sim$0.5,
to study the connection between star formation and molecular gas and its evolution with redshift.
The PHIBSS2 LIRGs naturally fill in the luminosity gap between the local main sequence galaxies and the high-z main sequence and starburst galaxies.
PHIBSS2 represents the first systematic census of molecular gas in this redshift range. 
Our local LIRG sample is a subset of the DYNAMO survey galaxies (Green et al. 2014), which were identified as disk-galaxies
and for which CO measurements are available (Fisher et al. 2019). In fact, $10$ out of $14$ galaxies of this sample are LIRGs, the other galaxies
have IR luminosities  $L_{\rm IR} < 10^{11}$~L$_{\odot}$.
We added the galaxy IRAS~08339+6517 (Fisher et al. 2022; $L_{\rm IR} > 10^{11}$~L$_{\odot}$) to our local LIRG sample.
The LIRGs are described in Table~\ref{tab:gphibss2} (F19) and, for the local LIRGs, Table \ref{tab:gdynamo}. 

\section{PHIBSS2 rotation curves \label{sec:rotcurv}}

The rotation curve is a crucial model input because it defines the depth of the gravitational potential. For the local starforming galaxies we used the parametrized rotation curves
of Leroy et al. (2008). For all other galaxy samples except PHIBSS2 we assumed constant rotation curves.  

Since the signal-to-noise ratio of the PHIBSS~2 CO observations is quite low, the derived CO linewidths are quite uncertain.
However, PHIBSS~2 has the advantage of providing HST I-band images, which can be converted to a radial mass profile using a constant mass-to-light ratio.
In this way a realistic rotation curve can be obtained. As a cross-check, we produced mock CO spectra based on our model calculations.
The linewidths of these mock data are consistent with the observed CO linewidths.
F19 performed single Sersic fits with {\rm galfit} (Peng et al. 2002, 2010)
on publicly available high-resolution ($0.03$~arcsec per pixel) HST Advanced Camera for Survey (ACS)
images in the F814W I-band. This band is optimal as it is available for all the galaxies of the sample and probes the blue stellar population at
z=0.5--0.8 while avoiding the rest-frame UV light from very young stars. Since we have no additional photometric imaging data, we do not take dust
reddening into account. At the PHIBSS2 redshifts the restframe g-r color becomes approximately i-z. The integrated i-z color of the PHIBBS2
galaxies is about g-r$\sim 0.5$. According to Fig.~6 of Bell et al. (2003), we expect an uncertainty of about a factor of two on the mass-to-light ratio
and thus on the estimate of the mass surface density. Since the inner part of the galaxies, where a bulge might be present, are supposed to be redder
than the disk, the mass surface density can be underestimated up to a factor of two there. 

\begin{figure*}
  \centering
  \resizebox{14cm}{!}{\includegraphics{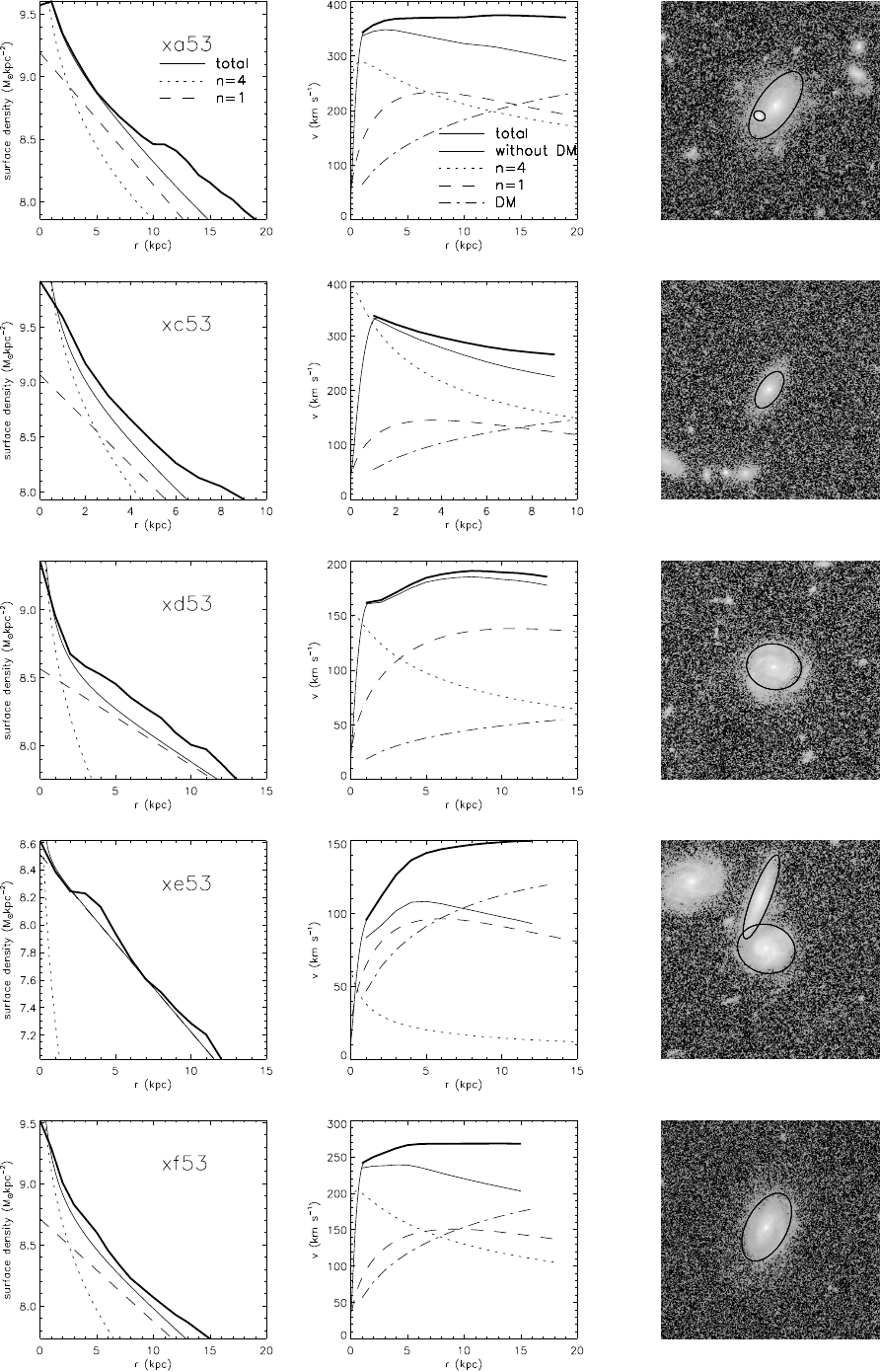}}
  \caption{PHIBSS2 stellar surface density profiles and rotation curves.
    Left panels: decomposition of the stellar surface density profiles (bulge with Sersic index $n=4$ and disk with $n=1$).
    Middle panel: rotation curve decomposition (bulge, disk, dark matter).
    Right panels: HST i-band images (F19) with the galaxy delimited by an ellipse.
  \label{fig:profiles_phibss2_1}}
\end{figure*}

Based on the position of the galaxy center, the inclination, and the position angle obtained by {\rm galfit} we produced radial profiles of the
I-band surface brightness for each galaxy, after masking any sources not belonging to the galaxy (Fig.~\ref{fig:profiles_phibss2_1}).
The mass surface density profile was then obtained by normalizing the integral of the Sersic fit by 
the total stellar mass obtained by SED fitting
(Table~\ref{tab:gphibss2}). Each profile was further decomposed into two components: a disk component with a Sersic index of $n=1$ and a spherical bulge component with $n=4$ (see Section 3.4 of F19). These components were optimized to fit the inner high-surface-density part of the profiles.
Finally, we added a dark matter component to obtain flat rotation curves.
To do so, a mass ratio between the dark matter halo and the stellar mass content of $M_{\rm DM}/M_*=20$ was assumed in agreement with the
findings of, e.g., Moster et al. (2013). The halo mass profile was determined using the standard description (e.g., van den Bosch 2001).
The rotation curve was determined based on the enclosed mass at each distance from the galaxy center.
Most of the rotation curves steeply rise in the inner kpc and become flat at radii larger than half of the maximum radius.

\begin{figure}
  \centering
  \resizebox{\hsize}{!}{\includegraphics{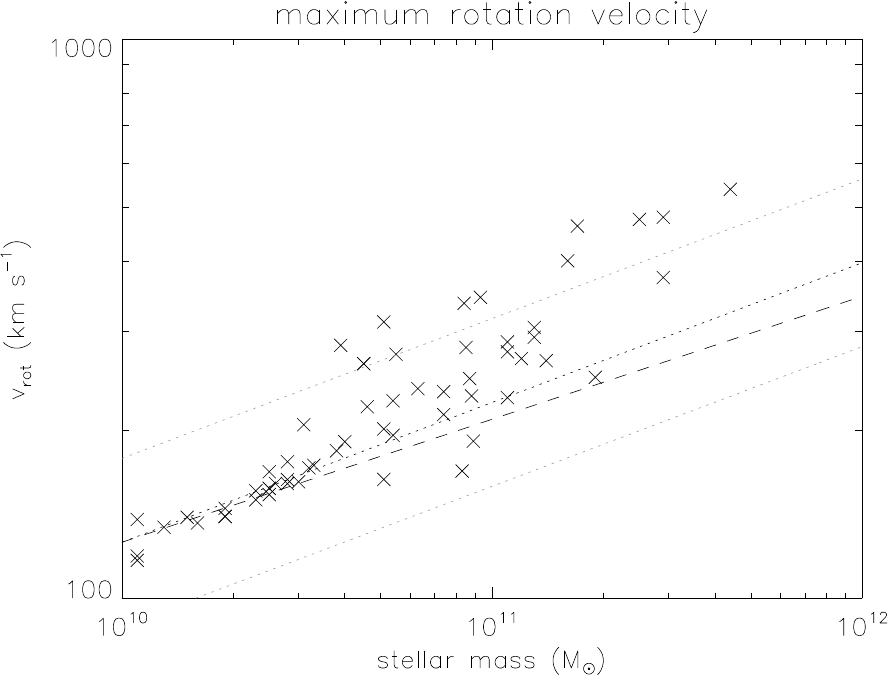}}
  \caption{PHIBSS2 galaxies (F19). Maximum rotation velocity as a function of the stellar mass.
    Black dashed and dotted lines: $v_{\rm rot} \propto M_*^{\beta}$ with $\beta=0.22$ and $0.25$, respectively.
    Gray dotted lines: scatter of $0.15$~dex (di Teodoro et al. 2021).
  \label{fig:mstarvmax}}
\end{figure}

As a check, we compare the maximum (model) rotation velocity of the PHIBSS2 galaxies (Fig.~\ref{fig:mstarvmax}) with the $z=0$ Tully-Fisher relation
(e.g., Bell \& de Jong 2001, di Teodoro et al. 2021).
For stellar masses smaller than $M_* \sim 5 \times 10^{10}$~M$_{\odot}$ the relation has
the expected slope $v_{\rm max} \propto M_*^{\frac{1}{4.3}}$ (Bell \& de Jong 2001). 
At higher masses the scatter significantly increases.

As can be seen in Fig.~\ref{fig:mstarvmax}, $9$ of the $59$ model PHIBBS2 galaxies have rotation velocities slightly in excess of the Tully-Fisher relation,
adopting a dispersion of $0.15$~dex (di Teodoro et al. 2021). These maximum rotation velocities come from the bright compact stellar disks rather than the dark matter
(dash-dotted line in Fig.~\ref{fig:profiles_phibss2_1}).
Since the model reasonably reproduces the observed CO fluxes, these high rotation velocities may well be real.

\section{The analytical model \label{sec:model}}

Galactic gas disks are modeled as turbulent clumpy star forming accretion disks. The turbulent nature of the ISM is taken into account by
applying scaling relations for the gas density and velocity dispersion to gas clouds of given sizes. The scaling relations change once the clouds become self-gravitating.
The cloud temperature is calculated via the balance of heating and cooling. In this way the gas density, temperature, and
velocity dispersion are determined for each cloud. The molecular abundances depend on the density, temperature,
and cosmic ray ionization rate, which is calculated by using a steady state model calibrated by the cosmic ray ionization in the
solar neighborhood. The molecular abundances, gas density, velocity dispersion, and temperature serve as input for the calculation of the molecular line emission.
The radio continuum emission is calculated via a steady state model calibrated by the observed infrared-radio correlations.
In this work we make simulations to predict the emission at many different wavelengths of observed galaxies.

\subsection{The gas disk \label{sec:gasdisk}}

We use the analytical model described in V17 and Liz\'ee et al. (2022) with a minor modification concerning the self-gravitating gas clouds (see Fig.~2 of V17).
This model was also used in V22 to calculate the integrated radio continuum emission of the galaxies.
The model considers the warm, cool neutral, and molecular phases of the interstellar medium as a single, turbulent gas
in vertical hydrostatic equilibrium. 
The gas is assumed to be clumpy, so that the local density is enhanced relative to the average density of the disk. 
Using the local density, the free fall time $t_{\rm l,ff}$ of an individual cloud controls the timescale for star formation.

Compared to the version presented in V17, we improved the code by making it modular and permitting the use of RADEX (van der Tak 2007).
Since the code now calculates the IR and line emission at a given radius, we set the mass fraction of the self-gravitating clouds with respect
to the diffuse clouds (Eq.~7 of V17) $y=0.5$ for all galaxies, which is the mean value of all galaxies modelled in V17. We verified that this modification does not change the results by
more than $\sim 30$\,\%. In addition, we divided the Toomre parameter $Q$ adopted for the local starburst galaxies in V17
by a factor of two (except for Arp 220E and Arp 220W). This modification led to CO spectral line energy distributions (SLED), which better reproduce the
available observations (see Sect.~\ref{sec:toomre}).

The model simultaneously calculates radial profiles of $\Sigma_{\rm gas}, \sigma_v$, 
the SFR, and the volume filling factor, all of which are large-scale properties, and at small scales $f_{\rm mol}$ and the IR SED and molecular line emission. 
The molecular line emission is calculated via RADEX.
For the details of our analytical model, we refer to Appendix~A of Liz\'ee et al. (2022).
The stellar mass profile is given by the Sersic fit to the observations so the calculations are of the gas distribution and kinematics.  
The dust is an assumed constant mass fraction (1\% of the hydrogen mass). 

The rotation curve and the surface density profile of the stellar disk serve as model inputs. The model contains three main free parameters:
(i) the Toomre parameter $Q$ of the gas, where a $Q=1$ disk is a disk with a maximal gas content, (ii) the external accretion rate
$\dot{M}$, where an increasing $\dot{M}$ at constant $Q$ raises $v_{\rm turb}$, and (iii) $\delta = l_{\rm driv}/l_{\rm cl}$ (see Appendix~\ref{sec:variables}
for the meaning of the variables). 
For each radius $R$, the
model yields the SFR per unit area $\dot{\Sigma}_{*}$, the H{\sc i} emission, and the infrared continuum and molecular line emission for a given set of
$Q$, $\dot{M}$, and $\delta$. Liz\'ee et al. (2022) showed that these radial profiles do not significantly depend on $\delta$
for $2 \le \delta \le 10$. We thus set $\delta=5$. 
The constant that links the rate of energy injection by supernovae to the star formation rate is set at $\xi = 9.2 \times 10^{-8}$~(pc\,yr$^{-1}$)$^2$
(Eq.~\ref{eq:energyflux}).

The details of the gas disk model are described in Appendix~\ref{sec:dgasdisk}.

\subsection{Cosmic ray ionization rate \label{sec:cosmicrays}}

V17 and Liz\'ee et al (2022) used a constant CR ionization rate. Here we calculate the CR ionization rate which is then injected into 
the time-dependent gas-grain code Nautilus, which uses the total H$_2$ CR ionization rate $\zeta_{{\rm H}_2}$ as input (Wakelam et al. 2012).
Cosmic rays with energies above $1$~MeV, interact with atoms and molecules of the interstellar medium.
These high-energy particles can directly ionize a species. This direct process is dominant for H$_2$ , H, O, N, He,
and CO. Electrons produced in the direct process can cause secondary ionization before being thermalized.
In addition to secondary ionization, the electrons produced in direct cosmic-ray ionization can electronically excite
both molecular and atomic hydrogen. The radiative relaxation of H$_2$ (and H) produces UV photons that ionize and
dissociate molecules. Whereas direct ionization is dominant for CO, secondary ionization and dissociation are dominant for HCN.
The CR ionization rate thus has an influence on the chemical abundances.

The spectral particle density of cosmic rays is related to the mean flux or spectrum through
\begin{equation}
\label{eq:neo}
n(E)=4\,\pi \frac{J}{v(E)}\ ,
\end{equation}
where $v(E)$ is the velocity of a particle with energy $E$.
The ionization rate of hydrogen by cosmic rays is
\begin{equation}
\zeta_{\rm H}=4\,\pi\,(1+\phi_{\rm s}) \int J(E) \sigma(E)\, dE\ ,
\end{equation}
where $\phi_{\rm s}$ accounts for secondary electrons and $\sigma(E)$ is the ionization cross section.
Following Cummings et al. (2016) we set  the integration limits to $3$~MeV and $10^7$~MeV and $\phi_{\rm s}=0.5$.
Moreover, we used the Spitzer \& Tomasko (1968) formula for the ionization cross section $\sigma(E)$. 

The CR nucleons (mainly protons) suffer energy losses by the combined effects of ionization and Coulomb losses, adiabatic
deceleration and pion production. For calculation of the local (solar neighborhood) CR density spectra of electrons and protons
we follow Pohl (1993). 
The equilibrium spectrum of CR protons is given by $n_{\rm p}=$
\begin{align}
  \label{eq:np}
  \begin{split}
    \xi_{\rm p} \frac{1}{q-1}\frac{5.5\times 10^{-3} (4.7 \times 10^8)^{q/2} T^{(2-q)/2}}{n_{\rm H}\,D_1 + D_2 + 5.55 \times 10^{-18} t_{\rm diff}^{-1}\,T^2}\ {\rm for}\ T<m_{\rm p}c^2 \\
    \xi_{\rm p} \frac{1}{q-1}\frac{1.3 \times 10^6 (9.4 \times 10^8)^q T^{-q}}{n_{\rm H} + D_2 + 2.8 \times 10^{-5} t_{\rm diff}^{-1}\,T^{1/2}}  \ {\rm for}\ T>m_{\rm p}c^2
  \end{split}
\end{align}
where $t_{\rm diff}$ is the diffusion timescale, $D_1=1+0.85\,n_{\rm e}/n_{\rm H}$, $D_2=1$, and $q=2.3$ for the CR protons and electrons.
Here $T$ is the kinetic energy, $E$ is the total energy, and $T=E$ for relativistic particles.
We use $q=2.3$ as expected for superbubbles created
by multiple SN remnants (Vieu et al. 2022).

For electrons one has to take into account ionization and Coulomb losses, non-thermal bremsstrahlung, adiabatic deceleration, inverse Compton losses, and
synchrotron emission (e.g., Lacki et al. 2010).
The equilibrium spectrum of CR electrons is given by
\begin{align}
  \label{eq:ne}
  \begin{split}
    n_{\rm e}=\xi_{\rm e} \frac{(m_{\rm e} c^2)^{(q-1)}}{(q-1)\,C_1} \frac{E^{(1-q)}}{(1+C_2E +C_3E^2+C_4 E^{3/2})}
  \end{split}
\end{align}
with $C_1=4.2 \times 10^7 n_{\rm H}$, $C_2=(8.57 \times 10^{-16}+6.34 \times 10^{-16})\,n_{\rm H}$, $C_3=6.4 \times 10^{-26} U+2.6 \times 10^{-25} B^2$,
and $C_4=2.2 \times 10^{-20}$ where $U$ is the galactic radiation field. The factors $C_1$ to $C_4$ were adapted to the values used by V22.

For the local ISM we assumed a density of $n_{\rm H}=0.8$~cm$^{-3}$, a magnetic field strength $B=6$~$\mu$G (Haverkorn 2015), which is close to
the value observed by the voyager spacecrafts ($B=5$~$\mu$G; Burlaga et al. 2023), a diffusion timescale $t_{\rm diff}=26$~Myr (Lacki et al. 2010),
and an electron density $n_{\rm e}=0.1\,n_{\rm H}$.

\begin{figure}
  \centering
  \resizebox{\hsize}{!}{\includegraphics{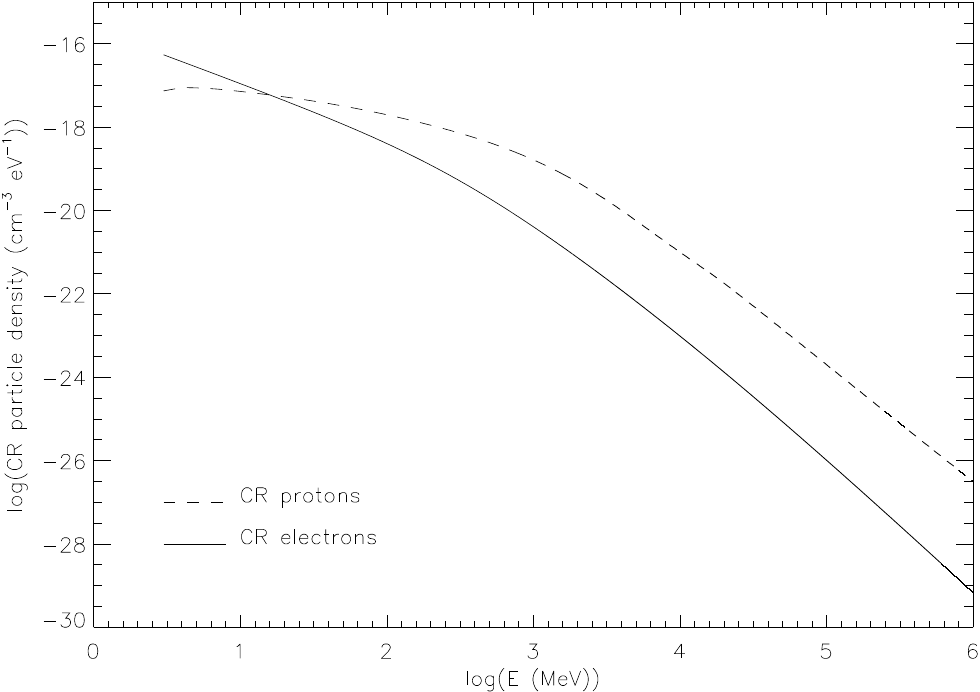}}
  \resizebox{\hsize}{!}{\includegraphics{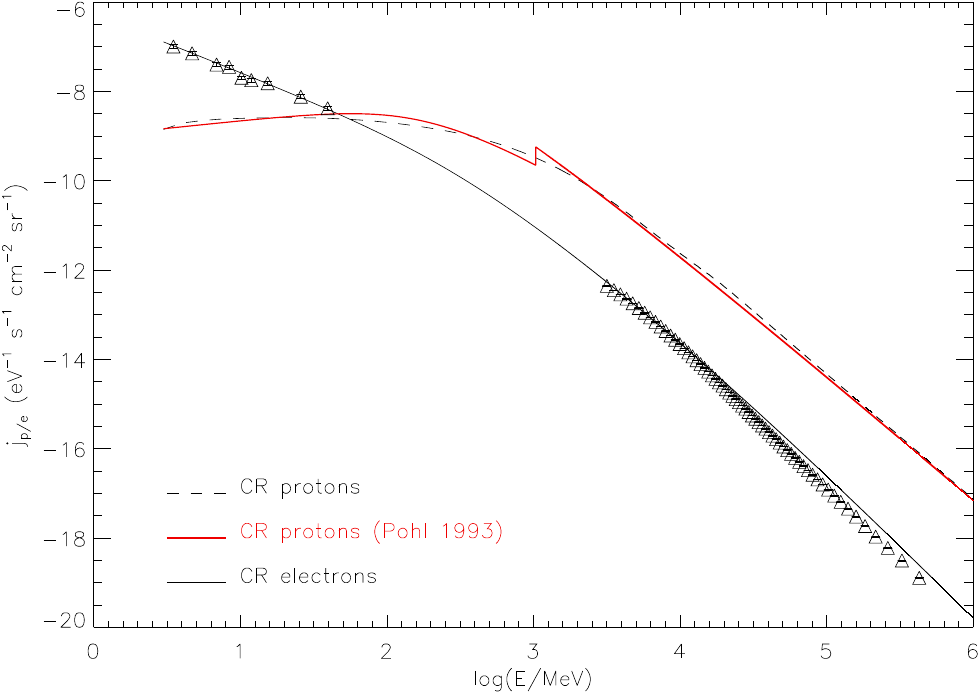}}
  \resizebox{\hsize}{!}{\includegraphics{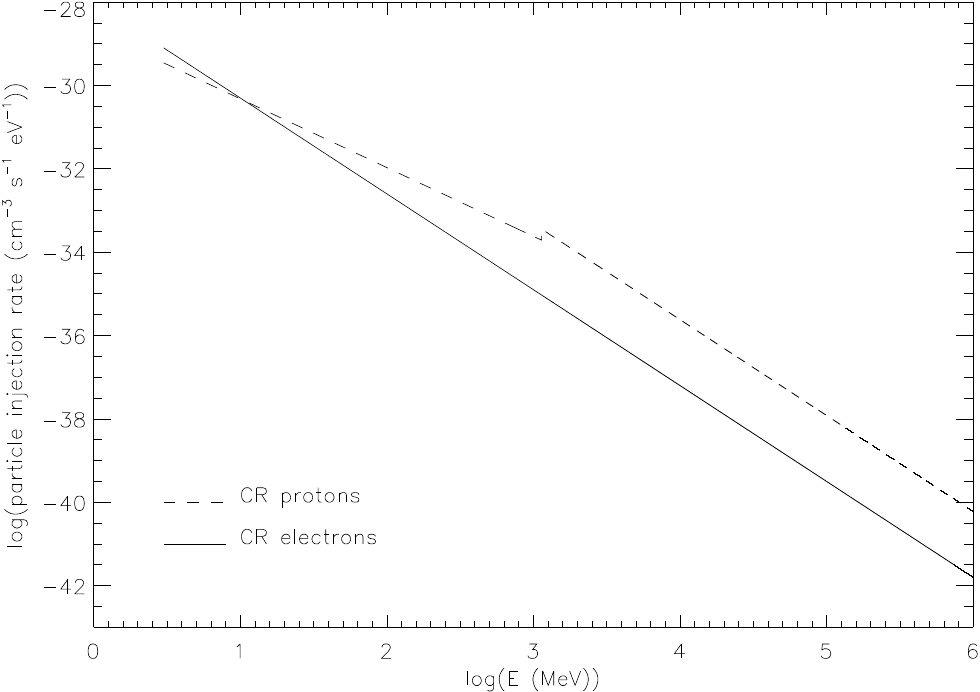}}
  \caption{Cosmic rays in the solar neighborhood. Upper panel: model CR particle density spectra.
    Middle panel: CR energy spectra; black solid line: CR electron model (Pohl 1993); red solid line: CR proton model (Pohl 1993);
    dashed line: GALPROP model fitted to observations by Cummings et al. (2016); triangles: Voyager~1 data (Cummings et al. 2016) and
    AMS data (Aguilar et al. 2019). Lower panel: CR injection spectra.
  \label{fig:cummings_norm_interpol_leakybox1a_final}}
\end{figure}

The normalization factors of proton and electron spectra, $\xi_{\rm p}$ and $\xi_{\rm e}$ were chosen such that the observed local CR intensities,
averaged over solid angle, $J(E)$, are recovered.
(middle panel of Fig.~\ref{fig:cummings_norm_interpol_leakybox1a_final};
Cummings et al. 2016, Aguilar et al. 2019). The resulting proton and electron spectra are presented in the upper panel of
Fig.~\ref{fig:cummings_norm_interpol_leakybox1a_final}. The corresponding injection spectra (Eqs.~3 and 4 of Pohl 1993)
are shown in the lower panel of Fig.~\ref{fig:cummings_norm_interpol_leakybox1a_final}.  The resulting total number 
rates of CR electrons and protons are the same as it is often assumed (e.g., Sect.~2 of Pohl 1993).

In the solar neighborhood, the CR energy density is $\epsilon_{\rm CR}=\int n(E)\,E\,dE\ = 0.9$~eV\,cm$^{-3}$, 
that of the magnetic field is also $\epsilon_{\rm B}=0.9$~eV\,cm$^{-3}$, following energy equipartition.
Assuming that CR ionization of atomic hydrogen due to nuclei with $Z>1$ amounts to $1.7$ times that due to protons
(Cummings et al. 2016), the total CR ionization rate is $\zeta_{\rm H}=1.6 \times 10^{-17}$~s$^{-1}$ (Cummings et al. 2016).
The CR ionization rate has been predicted to decrease with the cloud column density (e.g., Padovani et al. 2009).
This is because cosmic rays of low energy, which are most effective at ionization, lose all their energy due to ionization losses while
traversing large column densities of gas (e.g., Cravens \& Dalgarno 1978).
Padovani et al. (2022) presented three models for the decrease of the ionization rate with gas column density with
$\zeta_{{\rm H}_2} \propto N^{-\beta}$ where $N$ is the gas column density and $0.2 \le \beta \le 0.6$.  For simplicity, we adopted an exponent of $0.5$.

For the calculation of the cosmic ray ionization rate at a given distance $R$ from the center of a galaxy we set $n_{\rm H}$ to the
gas density of the analytical galaxy model (Sect.~\ref{sec:gasdisk}), use the interstellar radiation field $U$ and the magnetic field strength
of the galaxy model at $R$, and multiply $\xi_{\rm p}$ and $\xi_{\rm e}$ with $\dot{\Sigma}_*$ of the galaxy model.
We further set $t_{\rm diff}=\frac{1}{3} H\,v_{\rm turb}$ and $\zeta_{{\rm H}_2}=2.3/1.5\,\zeta_{\rm H}$ (Glassgold \& Langer 1974).
The final expression for the CR ionization rate of molecular hydrogen is
\begin{equation}
  \label{eq:crionf}
\zeta_{{\rm H}_2}=k\,\frac{\dot{\Sigma}_*}{\dot{\Sigma}_{*,{\rm local}}} 1.53\,\zeta_{\rm H}(n_{\rm H},B,U) \big(\sqrt{\Sigma_{\rm cl}/\Sigma}\big)^{-1}\ ,
\end{equation}
where $\dot{\Sigma}_{*,{\rm local}}=3\times 10^{-9}~{\rm M}_{\odot}{\rm pc}^{-2}{\rm yr}^{-1}$ is the local Galactic star formation rate per unit area (Elia et al. 2022).
$\Sigma_{\rm cl}$ and $\Sigma$ are the cloud and disk surface density. The local CR ionization rate (Cummings et al. 2016) corresponds to $k=1$. 

In our fiducial model we used a higher CR ionization rate ($k=3$) and 
also calculated a second model with $k=9$. $k=9$ approximates the value found
in the diffuse ISM (A$_{\rm V} < 0.3$) of the Galaxy (Indriolo et al. 2015, Neufeld \& Wolfire 2017).

\subsection{Radio continuum emission \label{sec:radcont}}

As stated in V22, the smallest scale on which the synchrotron--IR correlation holds is approximately the propagation length of CR
electrons, which is $\ga 0.5$~kpc at $\nu=5$~GHz and $\ga 1$~kpc at $\nu=1.4$~GHz in massive local spiral galaxies 
(e.g., Tabatabaei et al. 2013, Vollmer et al. 2020). Therefore, only the large-scale part of the gas disk model (Sect.~\ref{sec:gasdisk})
was used for the calculations of the model radio continuum emission.

Following V22, we assumed a stationary CR electron density distribution ($\partial n/\partial t$=0). 
The CR electrons are transported into the halo through diffusion or advection
where they lose their energy via adiabatic losses or where the energy loss through synchrotron emission is so small that the
emitted radio continuum emission cannot be detected. Furthermore, we assumed that the source term of CR electrons is
proportional to the SFR per unit volume $\dot{\rho}_*$. For the energy distribution of the CR electrons, the standard assumption is a power law
with index $q$, which leads to a power law of the radio continuum spectrum with index $-(q-1)/2$ (e.g., Beck 2015).
The details of the radio continuum model are described in Appendix~\ref{sec:radconta}.

\subsection{Molecular abundances and line emission \label{sec:molem}}

To determine chemical abundances, we used $7$ Nautilus grids for CR ionization rates of
$(1.3,\,3.0,\,6.5,\,30,\,130,\,650,\,6500) \times 10^{-17}$~s$^{-1}$ and varying time, gas density, and gas temperature.
The chemical abundances were calculated by interpolating the 4 dimensional Nautilus grid with the CR ionization rate of Eq.~\ref{eq:crionf}.

The line emission is calculated with RADEX using the inputs as calculated above.
The total emission is highly dependent on the gas mass surface densities (and hence filling factors). These come from the local gas densities and the prescription which decides whether gas is atomic or molecular based on the competition between the molecular formation time and the free-fall time.  Details are described in Sect. 2.1.1 and 2.4 of V17.

\section{Model results \label{sec:modelresults}}

The infrared, molecular line, and radio continuum luminosities were calculated with the analytical model presented in the previous section for each of the galaxies of the six samples
(local SF galaxies and starbursts, local and $z\sim 0.5$ LIRGs, high-z SF galaxies and starbursts). The molecular line emission was calculated by RADEX
with input parameters taken from the analytical model. Unlike V17, the CO, HCN, HCO$^+$ abundances were calculated with a variable CR ionization rate given by Eq.~\ref{eq:crionf}.

The model input parameters for each galaxy are the stellar surface density profile, the rotation curve, the 
Toomre $Q$ parameter, and the integrated SFR. All parameters are derived or constrained by observations (Leroy et al. 2008, Downes \& Solomon 1998, Genzel et al. 2010, Tacconi et al. 2013,
Freundlich et al. 2019, Fisher et al. 2019). 
We assumed constant rotation curves for the starburst galaxies and the high-z starforming galaxies.
To better reproduce the IR SEDs, we increased the star formation rates
of the PHIBSS2 galaxies by a factor of $1.5$ (Fig.~\ref{fig:IRspectra_phibss2_1}). To better reproduce the CO luminosities, we increased the stellar masses by the same factor, which led to increase of the rotation velocities
by a factor of $\sqrt{2}$ (Fig.~\ref{fig:plots_HCNCO_colum}). 
Both factors are within the uncertainties of the observational estimates via SED fitting
($0.3$~dex for the mass estimate; Roediger \& Courteau 2015) and combined UV and IR luminosities ($0.2$~dex for 
the star formation rate; Leroy et al. 2008). The use of a starburst attenuation law in the SED fitting can indeed underestimate the galaxy mass (e.g., Buat et al. 2019).
For internal consistency, the PHIBSS stellar masses were also increased by a factor of $1.5$.

\subsection{Metallicity, dust SED, TIR luminosity, and dust temperature}

Following V17, our leaky-box model uses an effective yield, meaning that enriched gas can escape and be accreted from the circumgalactic medium. This leads to a gas metallicity of
\begin{equation} \label{eq:zzodot}
Z/Z_{\odot}=(0.61~{\rm yr\,M_{\odot}pc^{-3}})/\alpha
\end{equation}
with
\begin{equation}
\label{eq:metulirgs}
\alpha=\frac{\big( \ln(\frac{\Sigma_{*}+\Sigma}{\Sigma})\big)^{-1}}{{\rm max}\big( (2 \times 10^9~{\rm yr}\,\frac{\dot{M_*}}{M_{\rm gas}})^{\frac{1}{3}},1.0\big)}\ {\rm yr\,M_{\odot}pc^{-3}}\ .
\end{equation}
This avoids severe underestimation of the  
metallicities of the ULIRG and high-z star-forming galaxy samples.
The oxygen abundance is then 12+log(O/H)=$\log(Z/Z_{\odot})+8.7$, with 12+log(O/H)=8.7 being the solar oxygen abundance (Asplund et al. 2005).

\begin{figure}
  \centering
  \resizebox{\hsize}{!}{\includegraphics{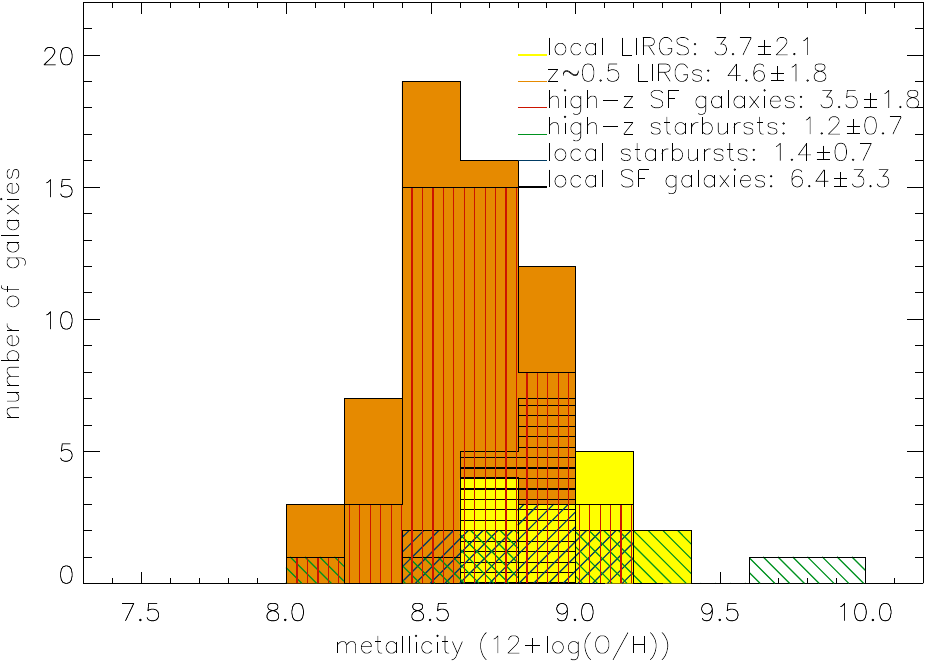}}
  \caption{Distribution of galaxy metallicities. Black solid line: local SF galaxies. Blue dotted line: local starbursts. Green dashed line:
    high-z starbursts. Red dash-dotted line: high-z SF galaxies. Yellow: local LIRGs. Orange: $z \sim 0.5$ LIRGs.
    Solar metallicity is 12+log(O/H)=8.7.
  \label{fig:plots_HCNCO_metal}}
\end{figure}

Fig.~\ref{fig:plots_HCNCO_metal} shows the metallicity distributions of our samples, with excellent agreement with V17 and observations for the local and high-z SF galaxies.  
However, the model metallicities of half of the starburst galaxies
show metallicities that are $2$-$4$ times higher than observed.
The mean model metallicity of the local LIRGs is 12+log(O/H)$= 9.0 \pm 0.2$, which is very close to the mean value of $6$ DYNAMO galaxies of
12+log(O/H)$\sim 9.1$ (Lenki\'c et al. 2021). The mean model metallicity of the $z \sim 0.5$ LIRGs is roughly solar, 12+log(O/H)$= 8.6 \pm 0.2$.
This is about a factor two lower than the mean gas metallicity found for $0.5 \leq z \leq 0.7$ galaxies with comparable stellar
mass and specific star formation rates (12+log(O/H)$\sim 8.9$; Guo et al. 2016).

Following V17, we extracted all available photometric data points for our galaxy samples from the CDS VizieR
database\footnote{\tt http://vizier.u-strasbg.fr/viz-bin/VizieR} for direct comparison between the model and observed dust SED. Since the flux densities
are determined within different apertures, we only take the highest flux densities for
a given wavelength range around a central wavelength $\lambda_0$ ($0.75 \leq \lambda/\lambda_0 \leq 1.25$).
In this way, only the outer envelope of the flux density distribution is selected. 
Since our dust model does not include stochastically heated small grains and PAHs, the observed IR flux densities
for $\lambda \la 50~\mu$m cannot be reproduced by the model.

We assumed a dust mass absorption coefficient of the following form:
\begin{equation}
\label{eq:kappa}
\kappa(\lambda)=\kappa_0\,(\lambda_0/\lambda)^{\beta}\ ,
\end{equation}
with $\lambda_0=250~\mu$m, $\kappa_0=0.48~{\rm m}^2{\rm kg}^{-1}$ (Dale et al. 2012), and a gas-to-dust ratio of $GDR=M_{\rm gas}/M_{\rm dust}=\frac{Z}{Z_{\odot}} \times 100$ 
(including helium; R{\'e}my-Ruyer et al. 2014). The exponent is $\beta=2$ for the starburst galaxies and $\beta=1.5$
for the starforming galaxies and LIRGs.

\begin{figure}
  \centering
  \resizebox{\hsize}{!}{\includegraphics{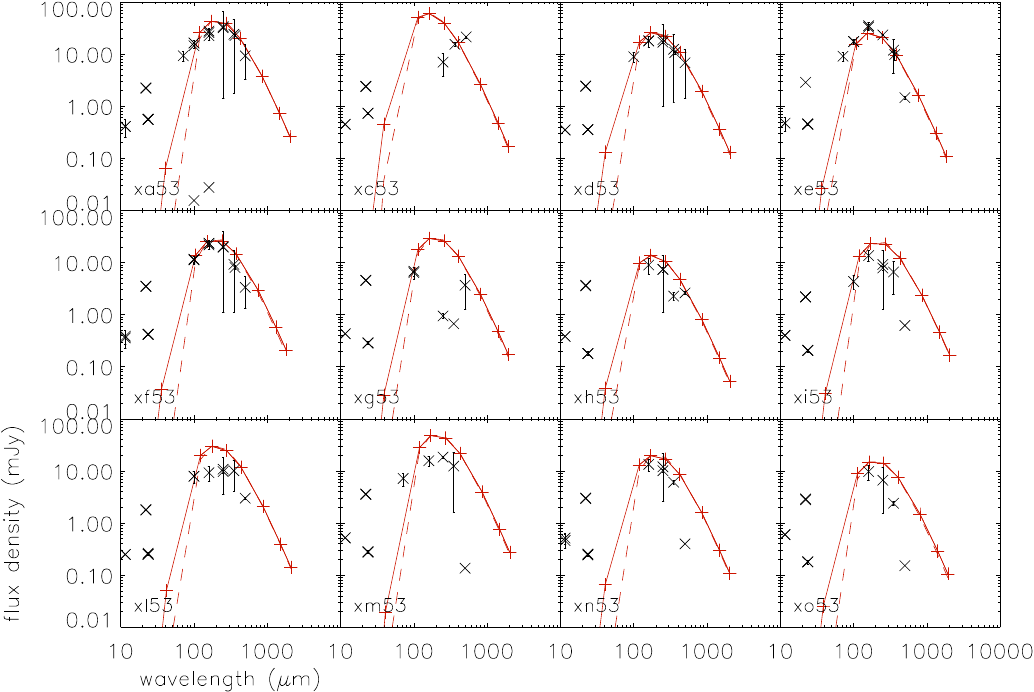}}
  \resizebox{\hsize}{!}{\includegraphics{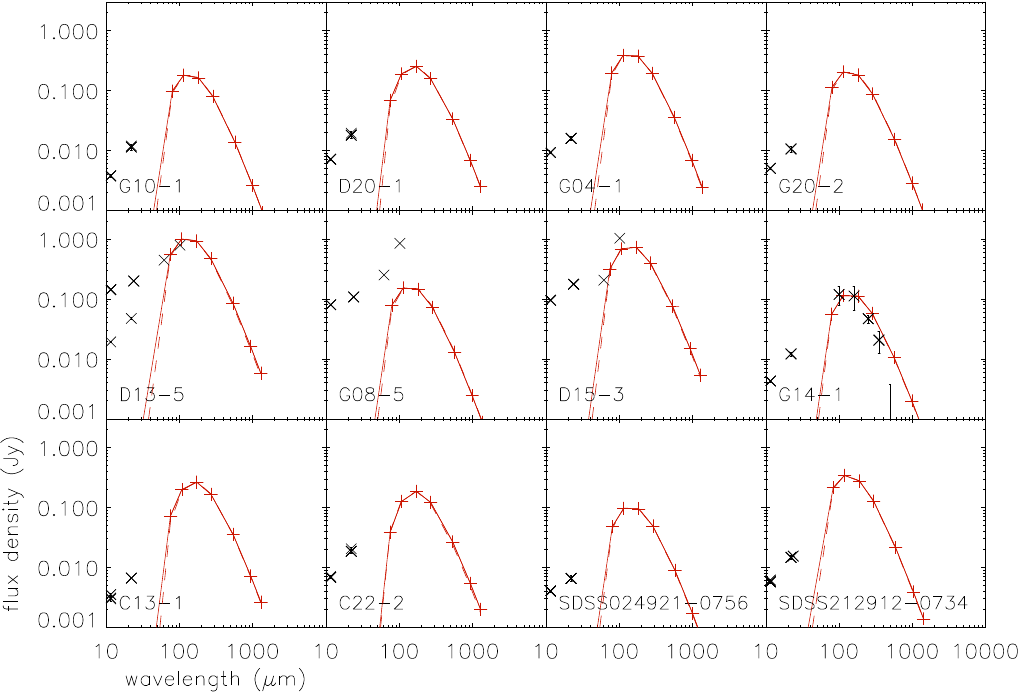}}
  \caption{Infrared spectral energy distributions. Red pluses and solid line: model SED.
    Red dashed line: modified Planck fit for temperature determination. 
    Black crosses: VizieR photometry. The errors bars are shown, if present in the VizieR tables, but often barely visible.
    Upper panel: $z \sim 0.5$ LIRGs. Lower panel: local LIRGs.
  \label{fig:IRspectra_phibss2_1}}
\end{figure}

The model dust IR SEDs of the local and $z \sim 0.5$ LIRGs are presented in Fig.~\ref{fig:IRspectra_phibss2_1} and Figs.~\ref{fig:IRspectra_phibss2_2}
to \ref{fig:IRspectra_dynamoe_2}.
Infrared flux densities at $\lambda > 100$~$\mu$m are only available for two DYNAMO galaxies and IRAS~08339+6517 (local LIRGs).
For these three galaxies the models reproduce the observations very well.
More data is available for the PHIBSS2 galaxies ($z \sim 0.5$ LIRGs) and the model dust IR SEDs reproduce the observations. For 80\% of the model galaxies, the fluxes are close the observed values.  For $8$ out of $59$ galaxies, the peak model flux is a factor of two higher than the observed SED and for two galaxies it is a factor two lower.

\begin{figure}
  \centering
  \resizebox{\hsize}{!}{\includegraphics{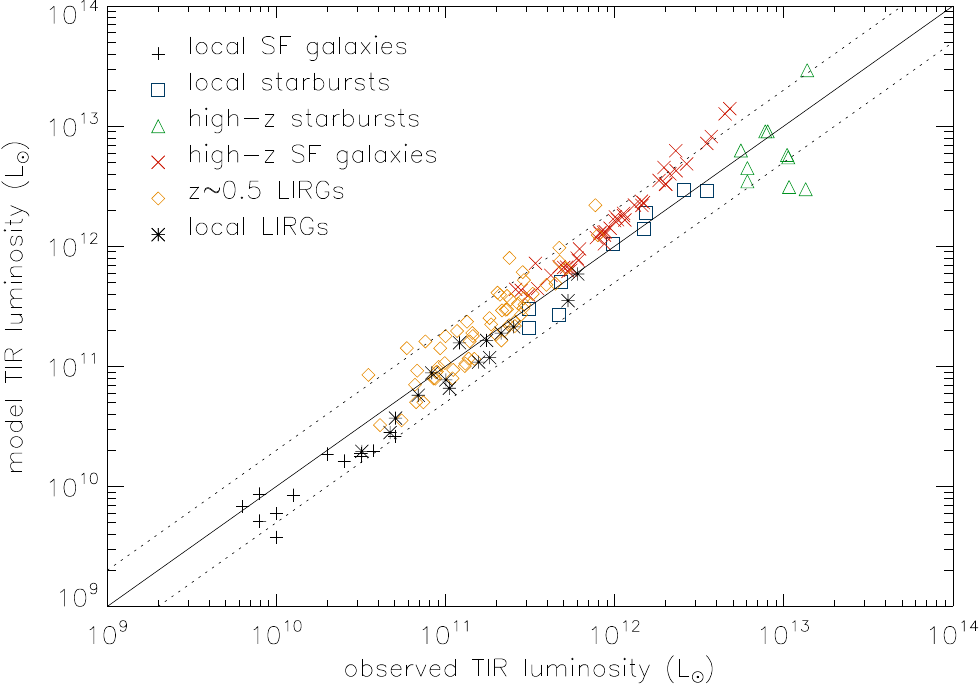}}
  \caption{Model total infrared luminosity as a function of the observed TIR luminosity of the galaxies.
    The lines correspond to an outlier-resistant linear bisector fit and its dispersion.
  \label{fig:plots_HCNCO_tirlum}}
\end{figure}

The model and observed $10~\mu$m to $1000~\mu$m luminosities (hereafter TIR) are presented
in the upper panel of Fig.~\ref{fig:plots_HCNCO_tirlum}. The TIR luminosities of the DYNAMO and PHIBSS2 galaxies were estimated
by $L_{\rm TIR}=10^{10} \times$SFR (Calzetti 2013).
The model and observed  TIR luminosities agree within a factor of two, with model  
local LIRGs slightly underestimated and $z \sim 0.5$ LIRGs are slightly overestimated (see upper panel of Fig.~\ref{fig:plots_HCNCO_tirlum}).

\begin{figure}
  \centering
  \resizebox{\hsize}{!}{\includegraphics{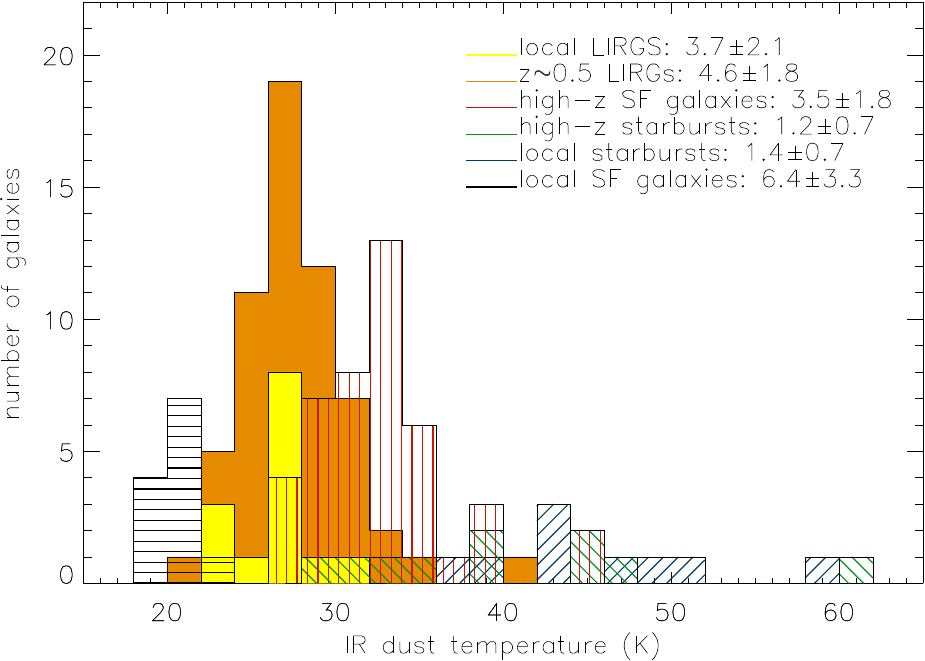}}
  \caption{Model distribution of dust temperatures. Black solid line: local SF galaxies. Blue dotted line: local starbursts.
    Green dashed line: high-z starbursts. Red dash-dotted line: high-z SF galaxies. Yellow: local LIRGs. Orange: $z \sim 0.5$ LIRGs.
  \label{fig:plots_HCNCO_tdust}}
\end{figure}

We fitted modified Planck functions with $\beta=1.5$ to the model dust IR SEDs to derive dust temperatures.
The modified Planck functions are shown as red dashed lines in Fig.~\ref{fig:IRspectra_phibss2_1} and Figs.~\ref{fig:IRspectra_phibss2_2}
to \ref{fig:IRspectra_dynamoe_2}.
The resulting distributions of dust temperatures for the different galaxy samples are presented in Fig.~\ref{fig:plots_HCNCO_tdust}.
The dust temperatures of the local and $z \sim 0.5$ LIRGs lie between $20$~K and $30$~K with a mean of $27$~K for both samples,
intermediate between those of the local and high-z SF galaxies. 

\subsection{Integrated CO, HCN($1-0$), and HCO$^+(1-0)$ fluxes}

\begin{figure}
  \centering
  \resizebox{\hsize}{!}{\includegraphics{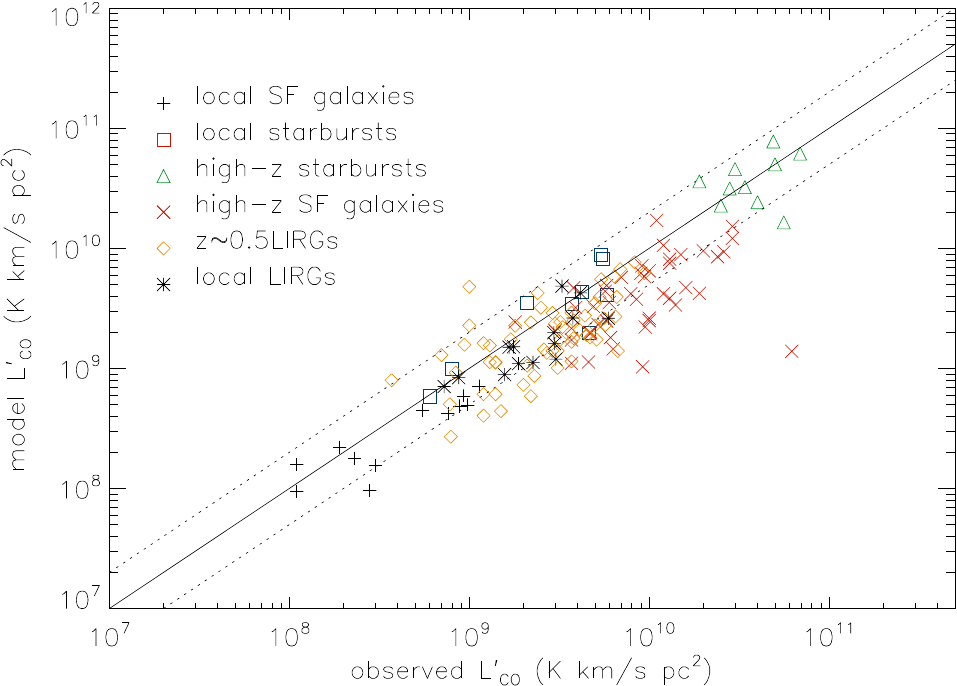}}
  \caption{Model CO luminosities as a function of the observed CO luminosities. The model CO transition corresponds to the observed transition.
    The lines correspond to the one-to-one relation with a scatter of a factor of two.
  \label{fig:plots_HCNCO_colum}}
\end{figure}

Model and observed CO luminosities are shown in Fig.~\ref{fig:plots_HCNCO_colum}: 
CO(2--1) for the local SF galaxies and LIRGs, CO(1--0) for local LIRGs and starbursts, and CO(3--2) for all high-z galaxies.  The transitions vary due to instruments and source redshifts.
The use of a non-constant ionization rate changes the CO emission, such that $L_{\rm CO}$ for high-z SF galaxies is $\sim 50$\,\% lower and $50$\,\% higher for high-z starbursts as compared to V17. The effect is lesser at low redshift.

\begin{figure}
  \centering
  \resizebox{\hsize}{!}{\includegraphics{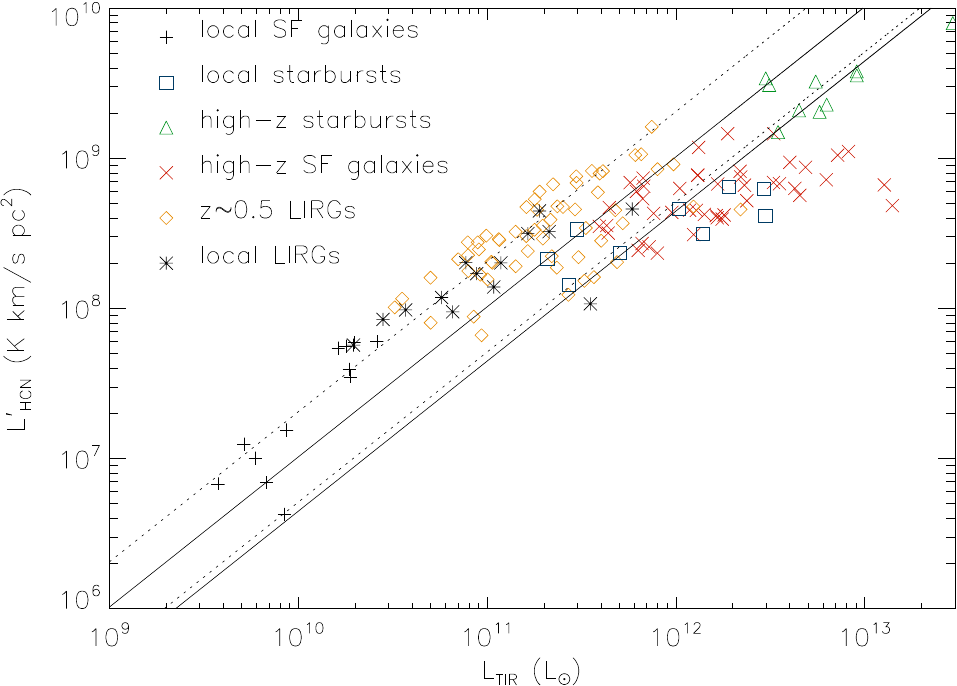}}
  \caption{Model HCN(1-0) luminosities as a function of the model TIR luminosities. The upper solid line corresponds to $L'_{\rm HCN}/L_{\rm TIR}=980$ 
  (Jimenez-Donaire et al. 2019), the dashed lines are offset by a factor of $0.5$ and $2$. 
  The lower solid line corresponds to a $2.3$ lower $L'_{\rm HCN}/L_{\rm TIR}$ as found for luminous IR 
  galaxies by Garcia-Burillo et al. (2012).
  \label{fig:plots_HCNCO_hcnlum}}
\end{figure}

\begin{figure}
  \centering
  \resizebox{\hsize}{!}{\includegraphics{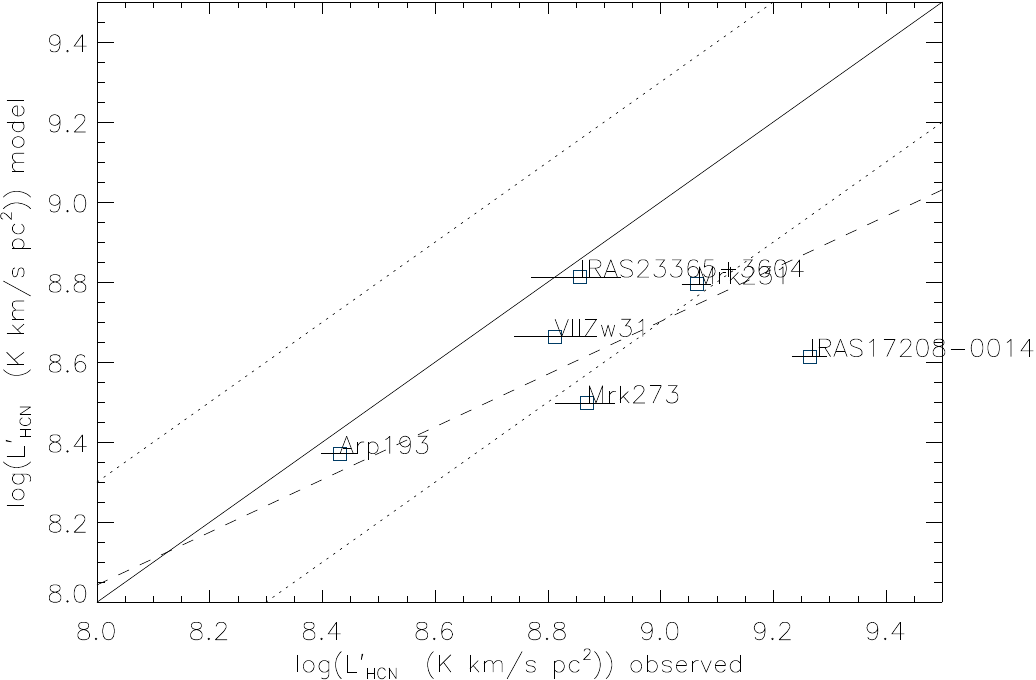}}
  \resizebox{\hsize}{!}{\includegraphics{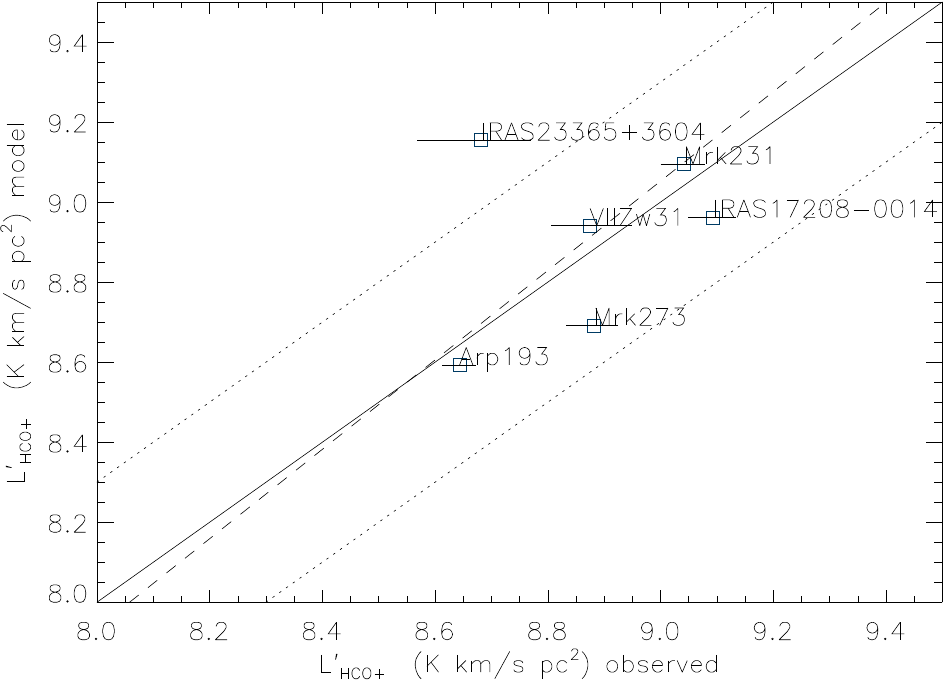}}
  \caption{Model HCN(1--0) (upper panel) and HCO$^+$(1-0) (lower panel) luminosity as a function of the observed HCN(1--0)/HCO$^+$(1-0) luminosity
    (Graci\'a-Carpio et al. 2008). The solid line corresponds to the one-to-one relation, the dotted lines to a dispersion of $0.3$~dex.
      The dashed lines correspond to a robust bisector fit.
  \label{fig:plots_HCNCO_hcn}}
\end{figure}

The model and observed $L_{\rm TIR}$--$L_{\rm HCN}$ relations are compared in Fig.~\ref{fig:plots_HCNCO_hcnlum}.
The model $L_{\rm HCN}$ of the local galaxies agree with V17 but 
the high-z galaxies are higher by factor of $1.5$ to $2$ than V17, primarily due to the variable CR rate (Eq.~\ref{eq:crionf}).
The model $L_{\rm TIR}$--$L_{\rm HCN}$ relations of the local SF and LIRGs as well as that of most $z \sim 0.5$ galaxies are 
a factor two higher than observed.
The low model $L_{\rm HCN}$ of high-z starburst galaxies are consistent with the results of Rybak et al. (2022) who only detected one out
of six $z \sim 3$ starburst galaxies in the HCN(1-0) line.
The model $L_{\rm HCN}$ of the local starbursts are broadly consistent with observations (upper panel of Fig.~\ref{fig:plots_HCNCO_hcn}).
Furthermore, the HCO$^+$(1-0) luminosities of the local starbursts are fully consistent with observations
(lower panel of Fig.~\ref{fig:plots_HCNCO_hcn}).

The new model, with a CR ionization rate calculated rather than injected, yields results as good (within less than a factor of two) as V17 and
has the advantage of revealing in which type of galaxy the CR ionization rate significantly modifies the
molecular line emission. For example, the model CO luminosities of the high-z SFR galaxies are about 50\,\% lower, whereas those of the high-z starbursts are about 50\,\% higher than those obtained by V17. Whereas the  agreement between the model and observed CO luminosities becomes somewhat worse for
the high-z starforming galaxies, it becomes somewhat better for the high-z starbursts.
V17 assumed a constant value of the CR ionization rate for each galaxy sample. A high CR ionization rate has the effect of
a high direct ionization of CO (see Sect.~\ref{sec:cosmicrays}) and thus a decrease of the CO abundance and line flux.
It thus appears that V17 under/overestimated the CR ionization rate in the   high-z SFR galaxies/starbursts, respectively.

\subsection{CO Spectral Line Energy Distributions}

\begin{figure}
  \centering
  \resizebox{\hsize}{!}{\includegraphics{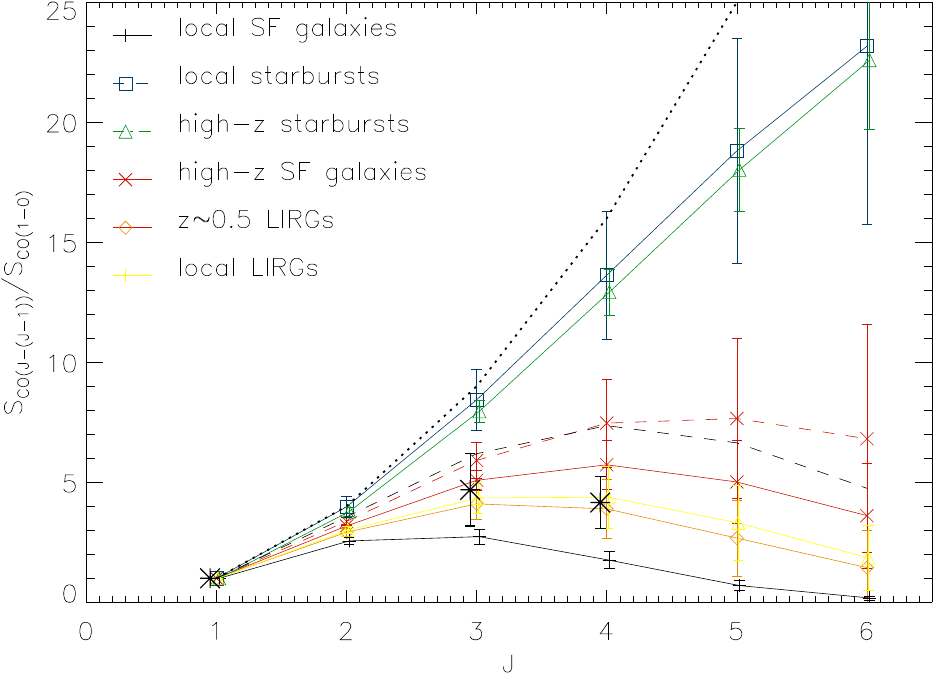}}
  \resizebox{\hsize}{!}{\includegraphics{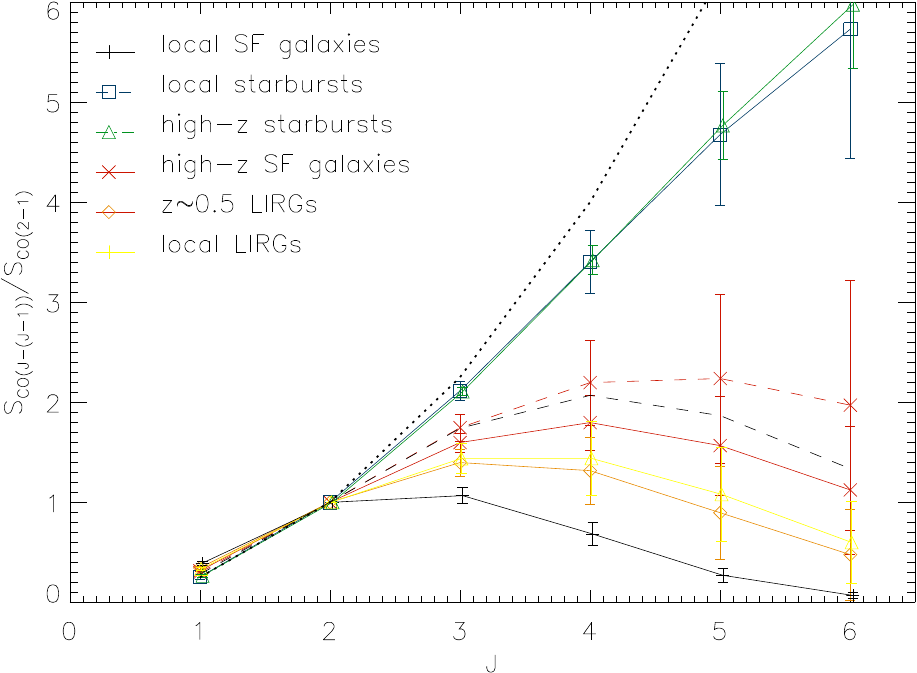}}
  \caption{Mean model CO spectral line energy distributions (SLEDs) of our galaxy samples.
    The dotted line corresponds to a constant brightness temperature.
    The high-z SF galaxy sample was divided into a $z \sim 1.5$ (solid red) and a $z > 2$ (dashed red) subsample.
    IRAS~08339+6517 is shown by a dashed black line.
  \label{fig:plots_HCNCO_cosed}}
\end{figure}

\begin{figure}
  \centering
  \resizebox{\hsize}{!}{\includegraphics{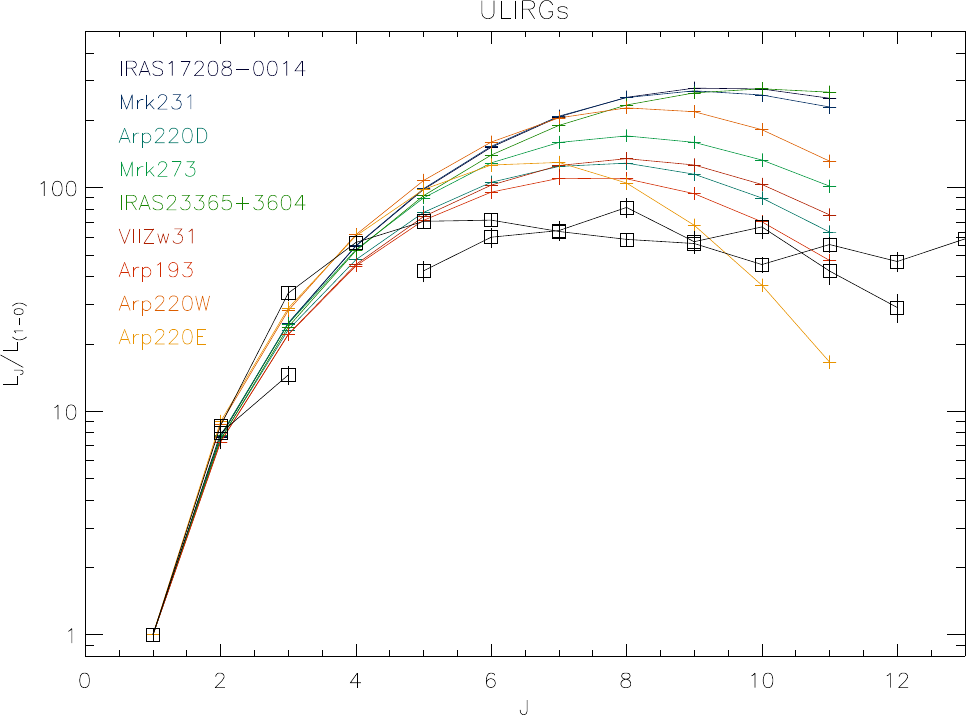}}
  \resizebox{\hsize}{!}{\includegraphics{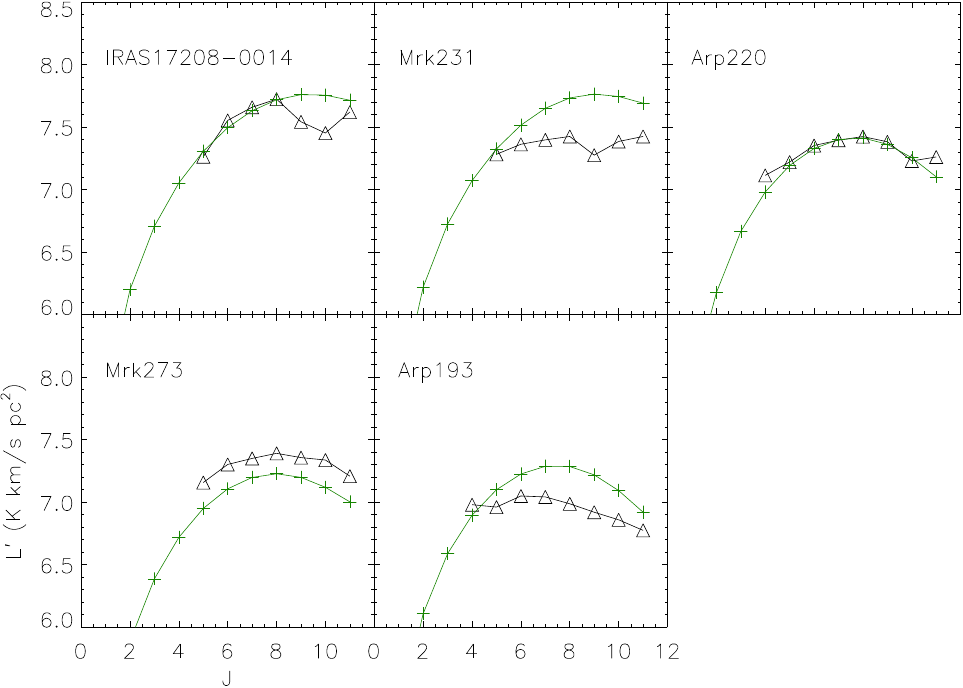}}
  \caption{CO SLEDs of local starburst galaxies (ULIRGs). Upper panel: the black boxes linked by black lines represent
    the observed mean CO SLEDs of ULIRGs with total infrared luminosities
    between $3 \times 10^{11}$ and $10^{12}$~L$_{\odot}$ and higher than $10^{12}$~L$_{\odot}$, respectively (from Kamenetzky et al. 2016).
    Lower panels: model (pluses) and observed CO SLEDs (Rosenberg et al. 2015; triangles).
  \label{fig:ulirg_ladder}}
\end{figure}

The CO SLEDs normalized to the CO(1-0) and CO(2-1) fluxes (expressed in Jy km s$^{-1}$) are presented in Fig.~\ref{fig:plots_HCNCO_cosed}.
For inner Milky Way,
the new mean model CO SLED reproduces the CO(3-2)/CO(2-1) line ratio (Fixsen et al. 1999) better than V17.
Kamenetzky et al. (2016) observed $J_{\rm upper} > 6$ transitions of CO in 
galaxies with TIR luminosities from $3 \times 10^{11}$ to $10^{12}$~L$_{\odot}$ and $L_{\rm TIR}>10^{12}$~L$_{\odot}$, shown with  model ULIRG CO SLEDs in Fig,~\ref{fig:ulirg_ladder}. 
The high-$J$ transitions have higher luminosities than observed, as in V17,
although shapes of the CO SLEDs are comparable.
Model and observed CO SLEDS of $5$ local ULIRGs (Rosenberg et al. 2015) are shown in the upper
panel of Fig.~\ref{fig:ulirg_ladder}. The CO(5-4) luminosities are well reproduced and the higher
CO transitions are reproduced within a factor of two except Mrk~231 where they are overestimated.  The lower transitions were not observed by Kamenetzky et al. (2016).

The CO model SLEDs of the high-z SF galaxies can be compared to the results of Boogaard et al. (2020) for $z \sim 1.5$ and Valentino et al. (2020) for $z > 2$.
The maximum of the mean CO SLED is observed at $J_{\rm up}=5$ at $z \sim 1.5$ and $J_{\rm up}=7$ at $z \sim 2.5$.
As in V17 the maximum of the model CO SLED is found at $J_{\rm up}=4$ at $z \sim 1.5$ and $J_{\rm up}=5$ at $z > 2$.
The model CO SLEDs normalized to the CO(1-0) and CO(2-1) fluxes are  compatible with the available observations.
The mean model CO SLEDs of 
LIRGs are identical within $1\sigma$ and the mean model SLED normalized to  CO(1-0) is consistent with the observations of $5$ DYNAMO galaxies by Lenki\'c et al. (2023).

\subsection{Integrated CO and HCN conversion factors}

\begin{figure}
  \centering
  \resizebox{\hsize}{!}{\includegraphics{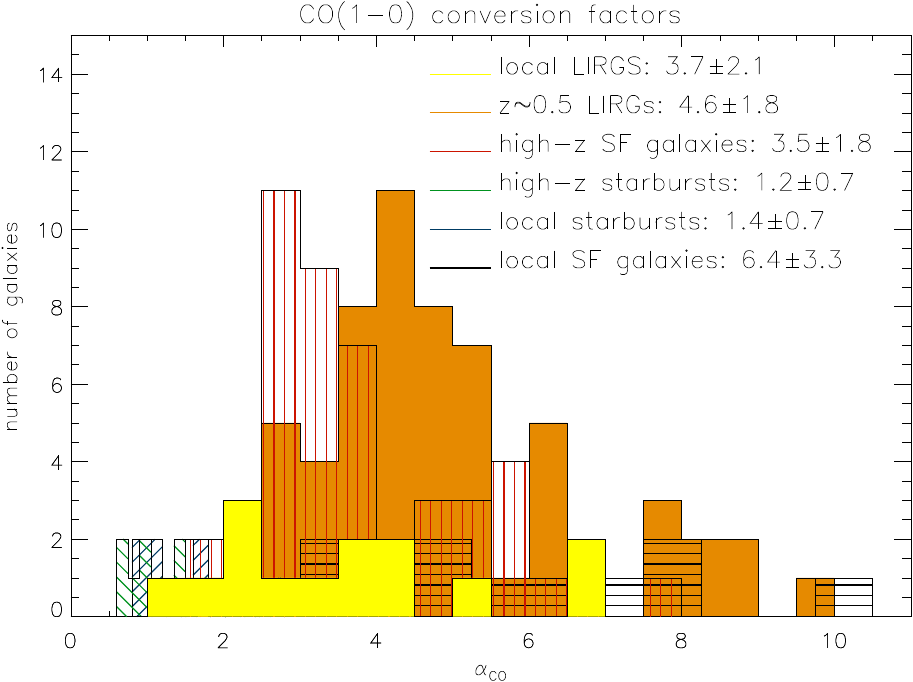}}
  \caption{CO(1--0) conversion factors for our galaxy samples in units of M$_{\odot}$(K\,km\,s$^{-1}$pc$^2$)$^{-1}$.
  \label{fig:plots_HCNCO_alphaco1}}
\end{figure}

Since the integrated H$_2$ mass and the line emission are calculated within the model (V17),
the integrated mass-to-light conversion factors can be determined (Fig.~\ref{fig:plots_HCNCO_alphaco1}).
The mean model conversion factor of local spirals 
($\langle \alpha_{\rm CO(1-0)} \rangle =6.4 \pm 3.3$~M$_{\odot}$(K\,km\,s$^{-1}$pc$^2$)$^{-1}$) 
is higher to but consistent within $1\sigma$ with the commonly-used value of $\alpha_{\rm CO(1-0)}=4.3$~M$_{\odot}$(K\,km\,s$^{-1}$pc$^2$)$^{-1}$ (Bolatto et al. 2013, Chiang et al. 2024).

The mean conversion factor of the local and high-redshift starburst galaxies 
($\langle \alpha_{\rm CO} \rangle \sim 1.3 \pm 0.4$~M$_{\odot}$(K\,km\,s$^{-1}$pc$^2$)$^{-1}$) is about twice that of observed high-z starburst galaxies 
($\alpha_{\rm CO}=0.8$~M$_{\odot}$(K\,km\,s$^{-1}$pc$^2$)$^{-1}$ (Downes \& Solomon 1998).
The model $\alpha_{\rm CO(1-0)}$ of high-z starforming galaxies is $\langle \alpha_{\rm CO} \rangle =3.5 \pm 1.8$~M$_{\odot}$(K\,km\,s$^{-1}$pc$^2$)$^{-1}$,close to the 
$\langle \alpha_{\rm CO} \rangle =3.7 \pm 2.1$~M$_{\odot}$(K\,km\,s$^{-1}$pc$^2$)$^{-1}$ for the LIRGs regardless of redshift,
similar to the standard MW value.

\begin{figure}
  \centering
  \resizebox{\hsize}{!}{\includegraphics{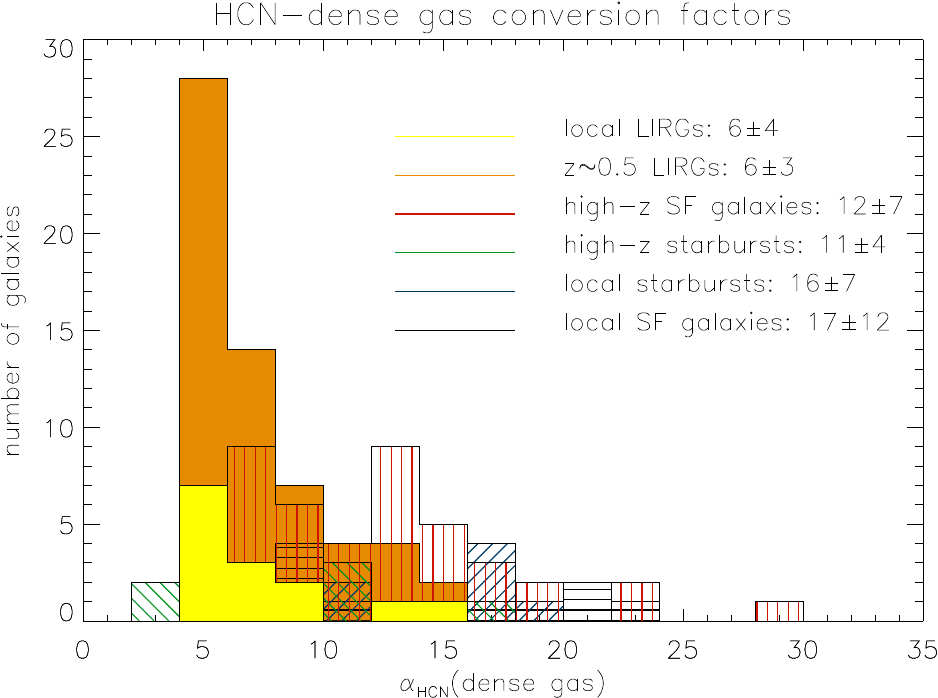}}
  \resizebox{\hsize}{!}{\includegraphics{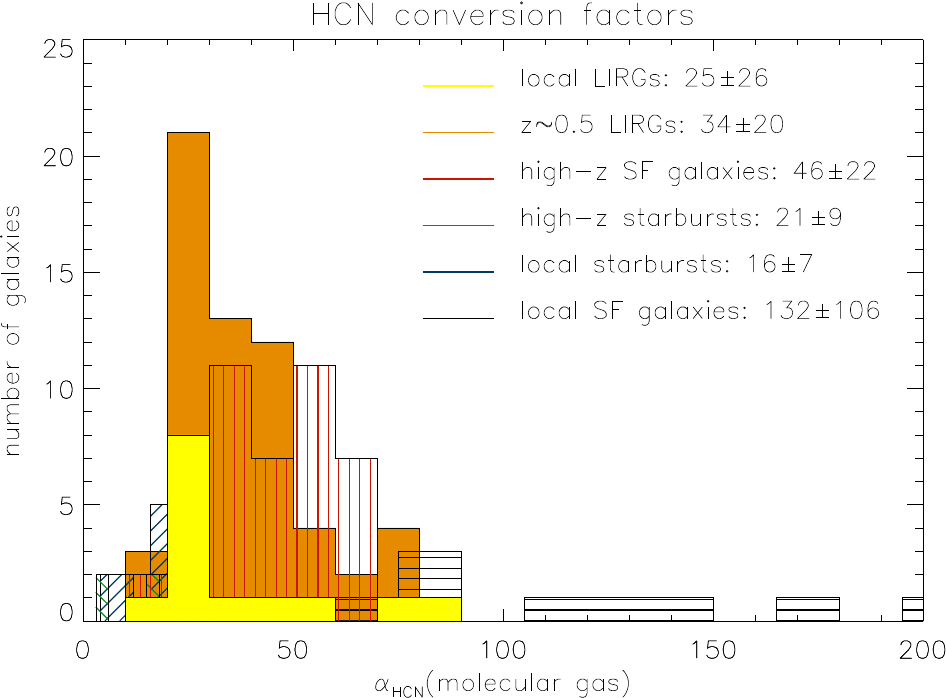}}
  \caption{HCN conversion factors in units of M$_{\odot}$(K\,km\,s$^{-1}$pc$^2$)$^{-1}$.
    Upper panel: HCN(1--0)--dense gas conversion factors; lower panel: HCN(1-0)--M$_{{\rm H}_2}$ conversion factors for our galaxy samples.
  \label{fig:plots_HCNCO_alphahcn}}
\end{figure}

The HCN conversion factor is usually interpreted as the HCN(1--0) flux (K\,km\,s$^{-1}$) divided by the column density of dense gas
($n\ge 3 \times 10^4$~cm$^{-3}$ e.g., Gao \& Solomon 2004), or equivalently the ratio of the HCN luminosity (K\,km\,s$^{-1}$pc$^2$) to dense gas mass (upper panel of Fig.~\ref{fig:plots_HCNCO_alphahcn}). In addition, the HCN(1--0)--$M_{\rm H_2}$ conversion factor (lower panel of Fig.~\ref{fig:plots_HCNCO_alphahcn}) might also be a useful quantity.
The canonical HCN(1--0)--dense gas conversion factor is $\alpha_{\rm HCN}=10$~M$_{\odot}$(K\,km\,s$^{-1}$pc$^{-1}$)$^{-1}$ (Gao \& Solomon 2004).

The model HCN(1--0)--dense gas conversion factors of the local and high-z starforming
and starburst galaxies are consistent with the canonical value. The HCN conversion 
factors of the local and $z \sim 0.5$ LIRGs are about a factor of two smaller than 
those of local and high-z starforming and starburst galaxies.

Whereas the HCN(1-0)--M$_{{\rm H}_2}$ conversion factors of the local and high-z starbursts and high-z SF galaxies are consistent within $1\sigma$ with those given
by V17, the HCN conversion factor of the local SF galaxies is twice that of V17.
Interestingly, the conversion factor of the local and $z \sim 0.5$ LIRGs are about the same and close to that of the high-z starbursts.

\subsection{Integrated radio continuum emission \label{sec:radcont1}}

\begin{figure*}
  \centering
  \resizebox{\hsize}{!}{\includegraphics{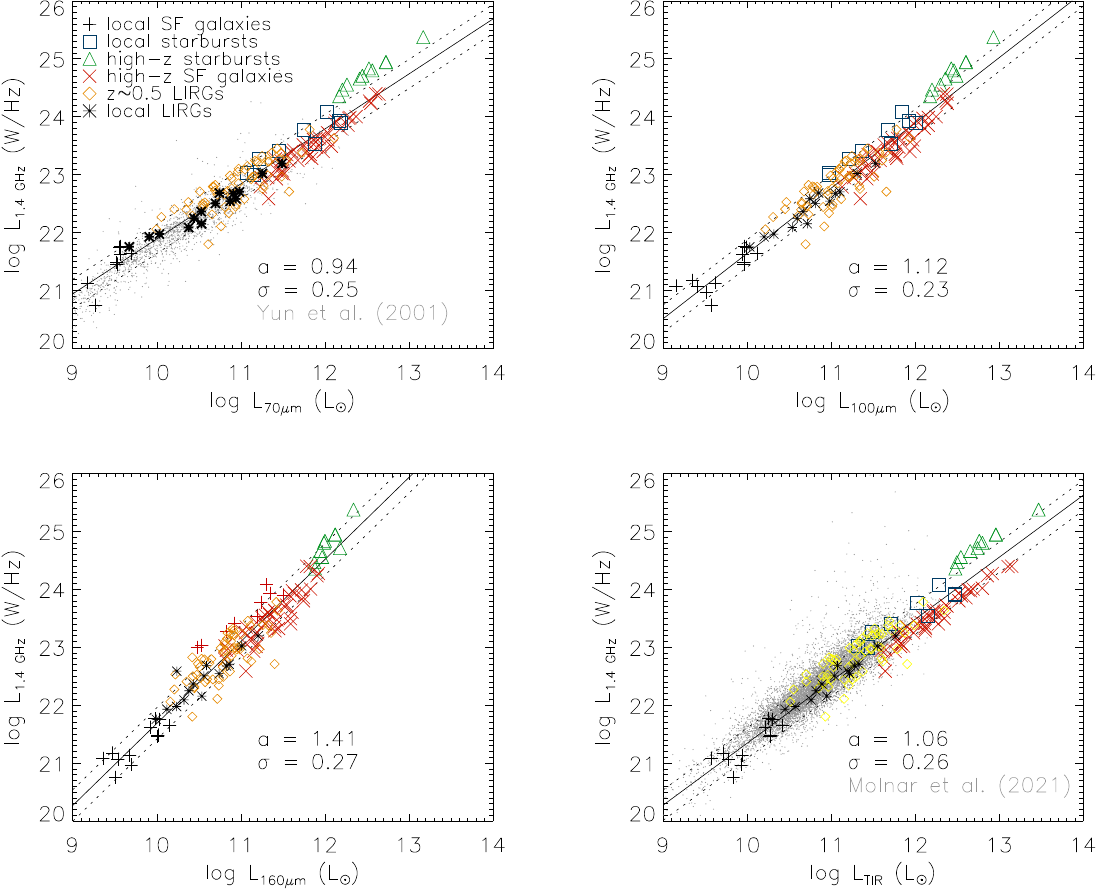}}
  \caption{IR-radio correlations. Upper left: $70$~$\mu$m - $1.4$~GHz correlation. Upper right: $100$~$\mu$m - $1.4$~GHz correlation.
    Lower left: $160$~$\mu$m - $1.4$~GHz correlation. Lower right: TIR - $1.4$~GHz correlation.
    Colored and black symbols show model galaxies. For a better visibility the $z \sim 0.5$ LIRGs are shown as yellow diamonds in the lower right panel.
    Black solid and dotted lines mark the model linear regression. Gray dots show observations.
  \label{fig:galaxies_FRC_vrotDifferentForPhibbs_4}}
\end{figure*}

The model radio continuum emission at $150$~MHz and $1.4$~GHz was calculated using the framework presented in Sect.~\ref{sec:radcont}.
The monochromatic ($70,\ 100,\ 160$~$\mu$m) and TIR--radio correlations of all six samples are shown in 
Fig.~\ref{fig:galaxies_FRC_vrotDifferentForPhibbs_4}.
The radio continuum luminosities at 1.4~GHz are consistent with those of V22 except for the high-z starburst galaxies whose
radio luminosities are about twice those of V22. The difference is caused by the different exponent of the CR energy distribution
(Eq.~\ref{eq:emissivity} with $q=2.3$ instead of $q=2$ in V22).
The local and $z \sim 0.5$ LIRGs nicely fall on and complement the correlations of the other galaxy samples already highlighted by V17.

We calculated the slopes and offsets of the correlation using an outlier-resistant bisector fit.
The results can be found together with the correlation scatter in Fig.~\ref{fig:galaxies_FRC_vrotDifferentForPhibbs_4}.
Furthermore, we used a Bayesian approach to linear regression with errors in both directions (Kelly 2007).
We assumed uncertainties on the TIR and radio luminosities of $0.2$~dex for the local galaxies and $0.3$~dex for the 
high-z galaxies.
In agreement with V22 the exponents derived from the bisector fits of the monochromatic IR--radio correlations increase
with increasing wavelength from $0.94$ at $70$~$\mu$m to $1.12$ at $100$~$\mu$m, and $1.41$ at $160$~$\mu$m.
The exponent of the TIR--radio correlation is $1.06$. These slopes are consistent with those derived with a Bayesian approach, the uncertainties
of which are about $0.04$. 

\begin{figure}
  \centering
  \resizebox{\hsize}{!}{\includegraphics{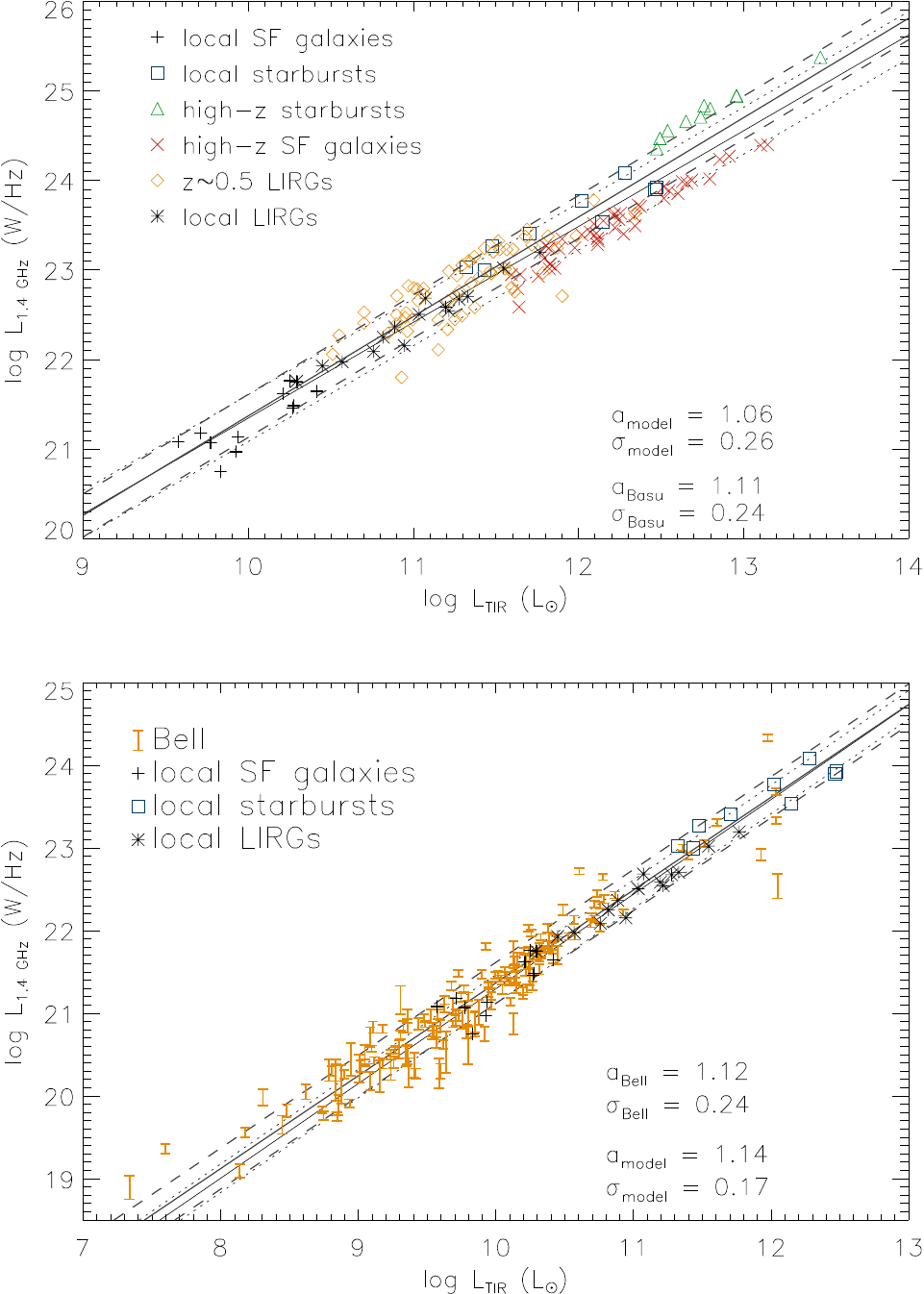}}
  \caption{TIR--$1.4$~GHz correlations. Symbols show model galaxies. Black solid and dotted lines show the
    model linear regression. Upper panel: Gray solid and dashed lines show the observed linear regression (Basu et al. 2015).
    Lower panel: Orange error bars show data from Bell (2003). Gray solid and dashed lines mark the observed linear regression (Bell 2003).
  \label{fig:galaxies_FRC_vrotDifferentForPhibbs_5}}
\end{figure}

Basu et al. (2015) studied the radio--TIR correlation in star-forming galaxies chosen from the PRism MUltiobject Survey up to 
redshift of $1.2$ in the XMM-LSS field, employing the technique of image stacking. These authors found a exponent of the TIR--$1.4$~GHz
correlation of  $1.11 \pm 0.04$. The lower panel of Fig.~\ref{fig:galaxies_FRC_vrotDifferentForPhibbs_5} shows 
the direct comparison between our model galaxies (local and high-redshift) and those of Basu et al. (2015).
The model and data show comparable exponents and scatters. 

Bell (2003) assembled a diverse sample of local galaxies from the literature with FUV, optical, IR, and 
radio luminosities and found a nearly linear radio--IR correlation. The upper panel of  
Fig.~\ref{fig:galaxies_FRC_vrotDifferentForPhibbs_5} shows the direct comparison between our local model galaxies
(spirals and low-z starbursts) and the compilation of Bell (2003). 
There is agreement between the slopes within $1 \sigma$ and between the offsets within $0.4 \sigma$ of the joint 
uncertainty of model and data.

\begin{figure}
  \centering
  \resizebox{\hsize}{!}{\includegraphics{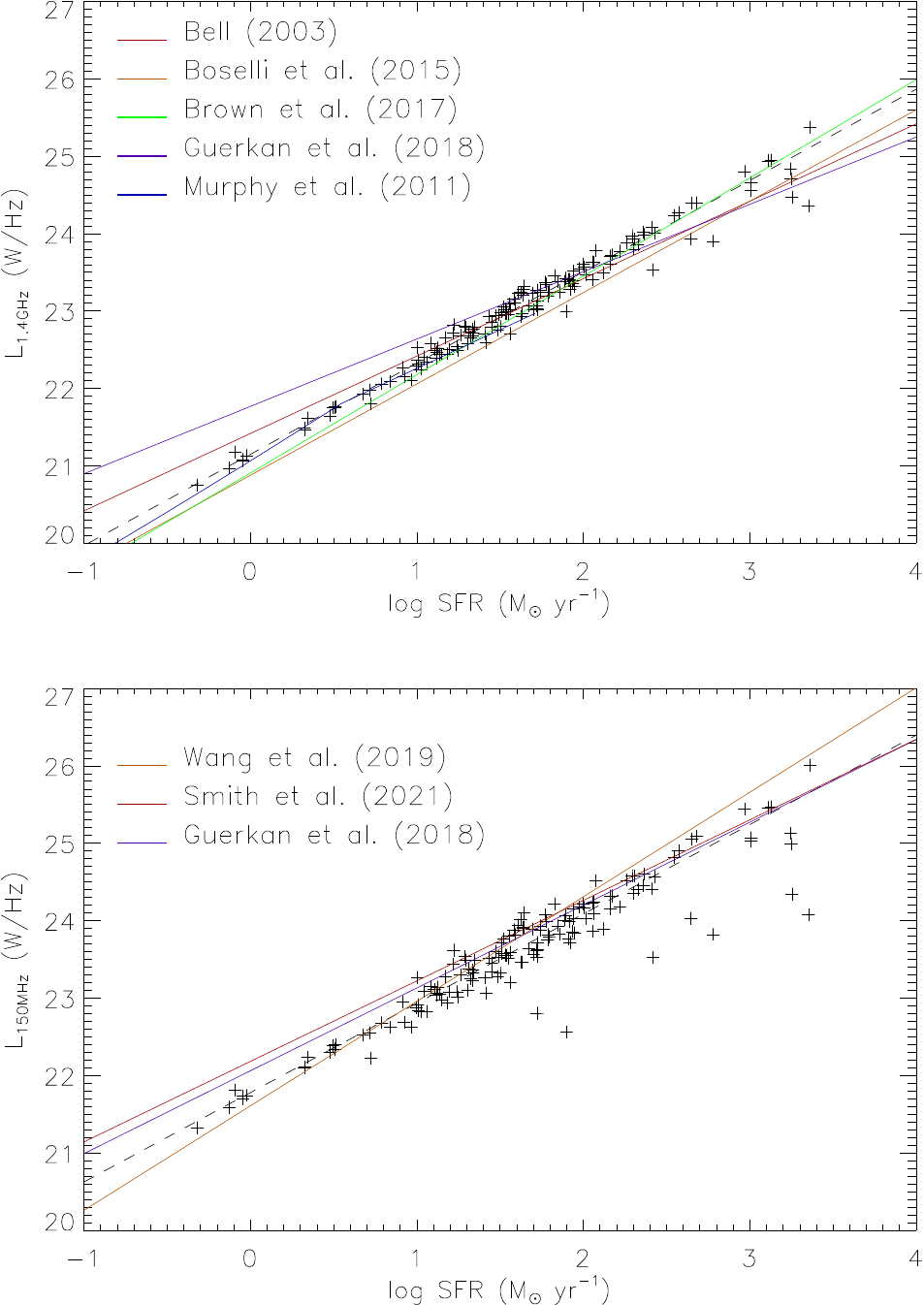}}
  \caption{Upper panel: SFR--$1.4$~GHz correlation. Lower panel: SFR--$150$~MHz correlation. 
    Colored lines show observed correlations. Plus symbols mark model galaxies. The black dashed lines in both panels correspond
      to outlier-resistant linear bisector fits.
  \label{fig:galaxies_FRC_vrotDifferentForPhibbs_6}}
\end{figure}

The resulting correlations between the $1.4$~GHz/$150$~MHz luminosity and the SFR are presented in Fig.~\ref{fig:galaxies_FRC_vrotDifferentForPhibbs_6}.
The model slopes of the log-log correlation are $1.07 \pm 0.04$ at $1.4$~GHz and $1.05 \pm 0.05$ at $150$~MHz.
Both correlations are thus slightly superlinear. Whereas the slope of the $1.4$~GHz correlation is consistent with,
the slope of the $150$~MHz correlation is slightly higher than that obtained by V17. Both model correlations
are consistent with existing estimates based on observations (Fig.~\ref{fig:galaxies_FRC_vrotDifferentForPhibbs_6}).

\section{Discussion \label{sec:discussion}}

Our modelling of the six galaxy samples at different redshifts reproduces the available observations within a factor of two.
The principal open parameter of the large-scale model of the turbulent star-forming galactic disk is the Toomre parameter, which
is expected to be not much higher than unity. The influence of the Toomre parameter $Q$ on the molecular line emission of the local
starburst galaxies is discussed in Sect.~\ref{sec:toomre}.

The model yields the gas mass, velocity dispersion, the turbulent driving length scale, and the gas viscosity $\nu=v_{\rm turb} l_{\rm driv}$.
The resulting star formation efficiencies are discussed with respect to the gas velocity dispersion in Sect.~\ref{sec:sfes}.
The radial gas transport within the galactic disks via the turbulent gas viscosity is discussed in Sect.~\ref{sec:viscosity}.

\subsection{The influence of the Toomre $Q$ parameter \label{sec:toomre}}

\begin{figure}
  \centering
  \resizebox{\hsize}{!}{\includegraphics{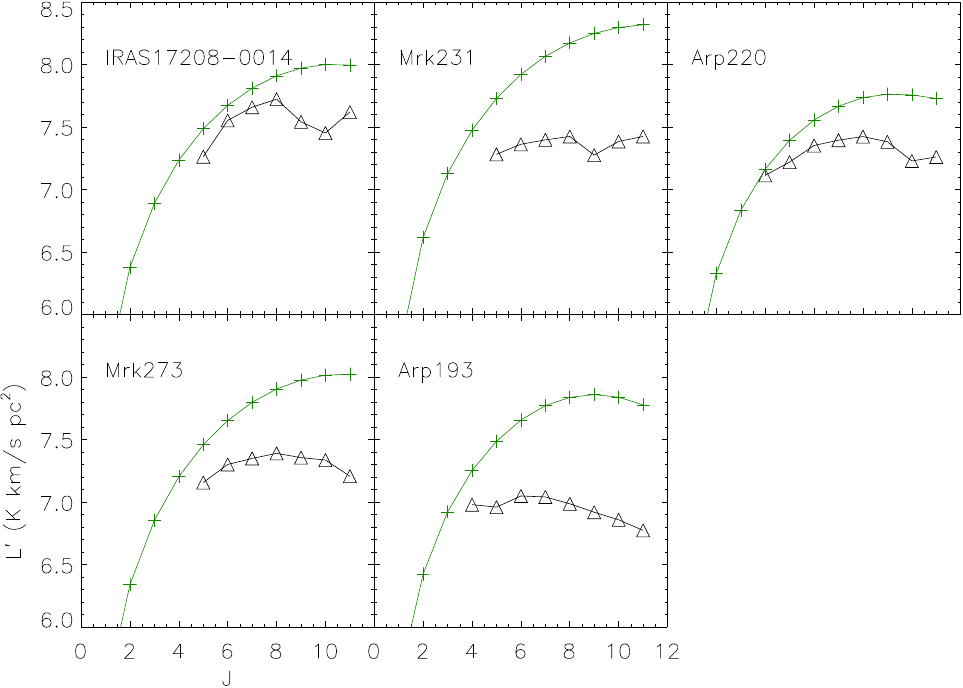}}
  \caption{CO SLEDs of the local starburst galaxies with $1.5$ to two times higher $Q$ values compared to Fig.~\ref{fig:ulirg_ladder}.
    Black triangles: Rosenberg et al. (2015). Green lines: models.
  \label{fig:ulirg_ladder_Q}}
\end{figure}

To investigate the influence of the Toomre $Q$ parameter on the CO SLEDs of the local starburst galaxies, we increased $Q$
by factors of $1.5$, $2$, $2$, $2$, and $1.5$. The resulting CO SLEDs are presented in Fig.~\ref{fig:ulirg_ladder_Q}.
The comparison with Fig.~\ref{fig:ulirg_ladder} shows that a higher $Q$ leads to a significantly steeper CO SLED.
We interpret this behavior as being caused by an increased gas velocity dispersion with increasing $Q$.
The higher velocity dispersion leads to a higher turbulent heating and thus higher gas temperatures.
The high gas temperature in turn lead to higher brightness temperatures of the high-J CO transitions.
The agreement between the model and observed SLEDs with higher values of $Q$ is much worse than that of
the fiducial values (Fig.~\ref{fig:ulirg_ladder}). This is strong support for low values of $Q$ in local starburst galaxies.

\subsection{Molecular gas depletion time and gas velocity dispersion \label{sec:sfes}}

Within our analytical model the only energy source for turbulence is stellar feedback (Sect.~\ref{sec:model}).
As stated in Sect.~\ref{sec:gasdisk}, our star formation prescription leads to a star formation rate per unit area, which is proportional to
the gas pressure (Fig.~\ref{fig:plots_1_phibss2_1}) as it is expected if star formation is pressure-regulated and feedback-modulated
(Ostriker \& Kim 2022). Since our model reproduces the molecular line and radio continuum emission of the six galaxy samples,
we conclude that the energy injection via stellar feedback is sufficient to maintain the observed level of turbulence in these galaxies.

\begin{figure}
  \centering
  \resizebox{\hsize}{!}{\includegraphics{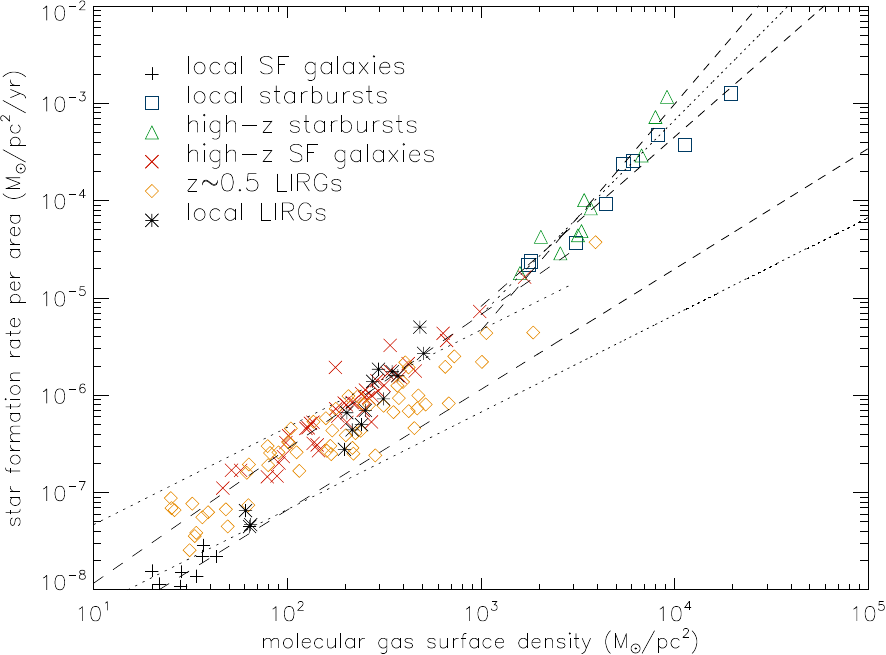}}
  \caption{Star formation rate per unit area as a function of the molecular gas surface density. The dotted lines are linear and square fits to guide
    the eye. The dashed lines are robust fits to the different galaxy samples.
  \label{fig:plots_1_phibss2_10}}
\end{figure}

The $\Sigma_{\rm SFR}-\Sigma_{\rm H2}$ correlation of our model galaxies is presented in
the upper panel of Fig.~\ref{fig:plots_1_phibss2_10}. 
For the area calculation, the stellar scalelength was adopted. The log-log relation cannot be fitted with a single slope or offset.
Despite similar $\Sigma_{\rm H2}$, the local LIRGs have SFRs up to $\sim 5$ times higher than the local SF galaxies. Perhaps surprisingly, the local and high-z starbursts and the local and $z \sim 0.5$ LIRGs have comparable star formation efficiencies. The slopes of the $log(\Sigma_{\rm SFR}) - log(\Sigma_{\rm H2})$ correlation are $1.2$, $1.7$, $2.0$, $1.1$, $1.4$, and $2.3$ respectively for local
SF galaxies and starbursts, local and $z \sim 0.5$ LIRGs, and high-z SF galaxies and starbursts (as derived by the IDL routine robust\_linefit). Only the local and high-z starbursts and the local LIRGs show
slopes higher than $1.5$. We interpret these steep slopes as a result of the compactness of these galaxies. Small galaxy sizes lead to deep potential wells, low dynamical times, and high gas densities.

\begin{figure}
  \centering
  \resizebox{\hsize}{!}{\includegraphics{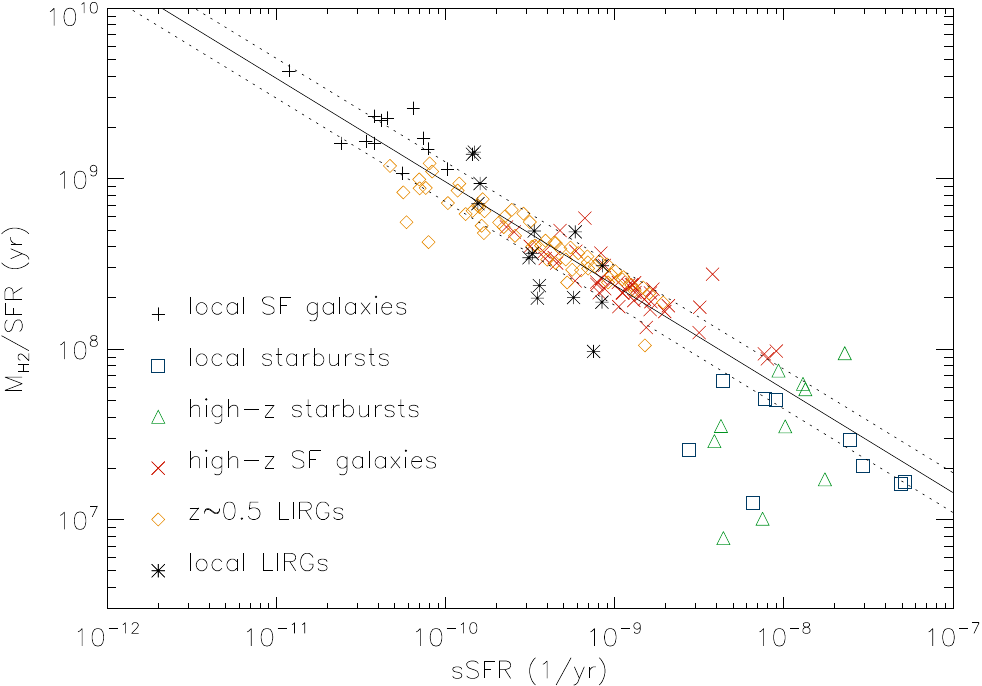}}
  \resizebox{\hsize}{!}{\includegraphics{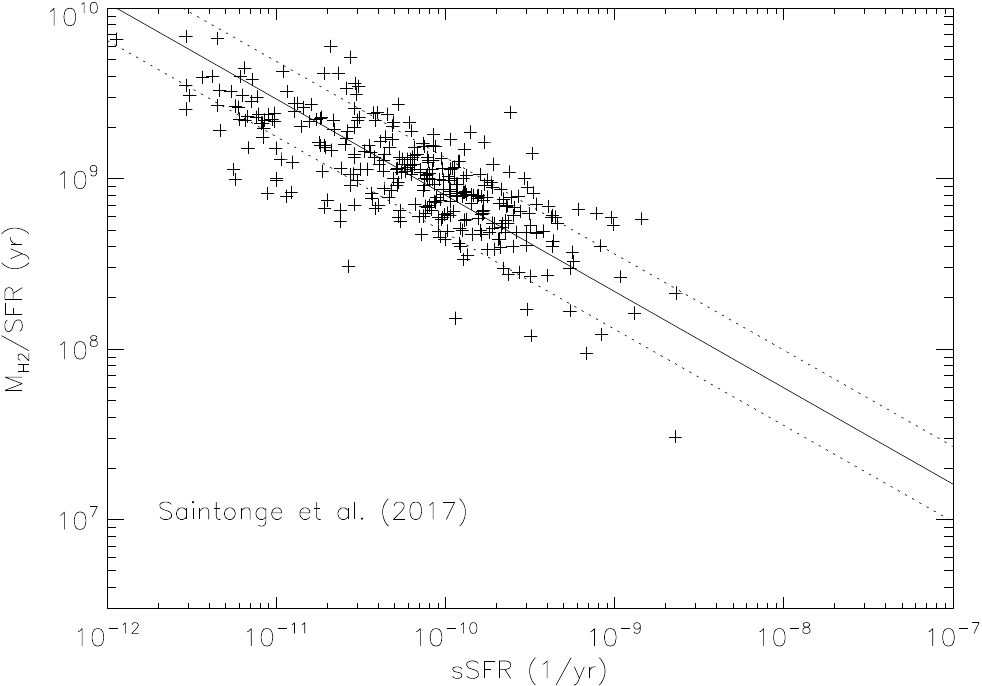}}
  \resizebox{\hsize}{!}{\includegraphics{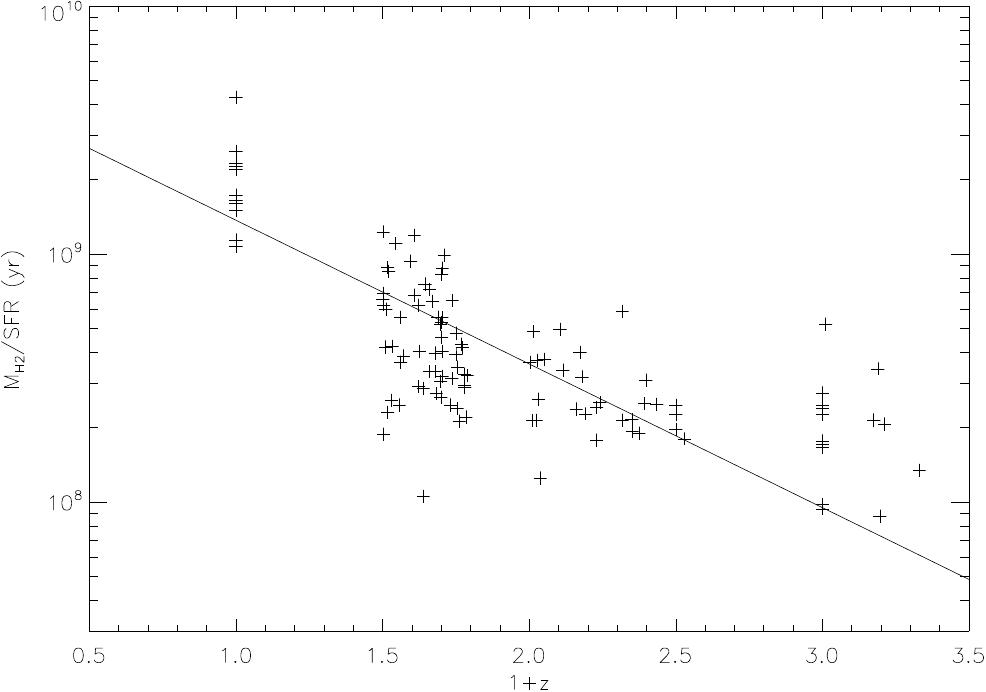}}
  \caption{Molecular gas depletion time. Upper panel:  as a function of the specific star formation rate for the model galaxies.
    Middle panel:  as a function of the specific star formation rate for the observations from Saintonge et al. (2017).
    Lower panel:  as a function of redshift for the model galaxies. The lines correspond to an outlier-resistant linear bisector fit and its dispersion.
  \label{fig:plots_1_phibss2_11}}
\end{figure}

The molecular gas depletion time as a function of the specific star formation rate of the model galaxies is presented in the upper panel of
Fig.~\ref{fig:plots_1_phibss2_11}. The slope of the log-log relation is $-0.61$, the scatter is $0.12$~dex. The local and high-z starbursts show by far the
largest scatter. The slope of the observed log-log relation ($-0.56$; Saintonge et al. 2017; middle panel of Fig.~\ref{fig:plots_1_phibss2_11}) is consistent with that of the model
relation. The observed scatter is a factor of two larger than the model scatter. 
We conclude that the tight observed $t_{\rm dep, H_{\rm 2}}$--sSFR relation is well reproduced by the SF galaxy models.
The molecular gas depletion time as a function of redshift is presented in the lower panel of  Fig.~\ref{fig:plots_1_phibss2_11}).
The depletion time decreases with redshift as $(1+z)^{-0.58}$ up to $z \sim 1.5$ and seems to become constant for $z \ga 2$.
The exponent of $-0.58$ is consistent with that found by Tacconi et al. (2018) for galaxies at $z < 0.7$ ($-0.62$; their Fig.~4).

\begin{figure}
  \centering
  \resizebox{\hsize}{!}{\includegraphics{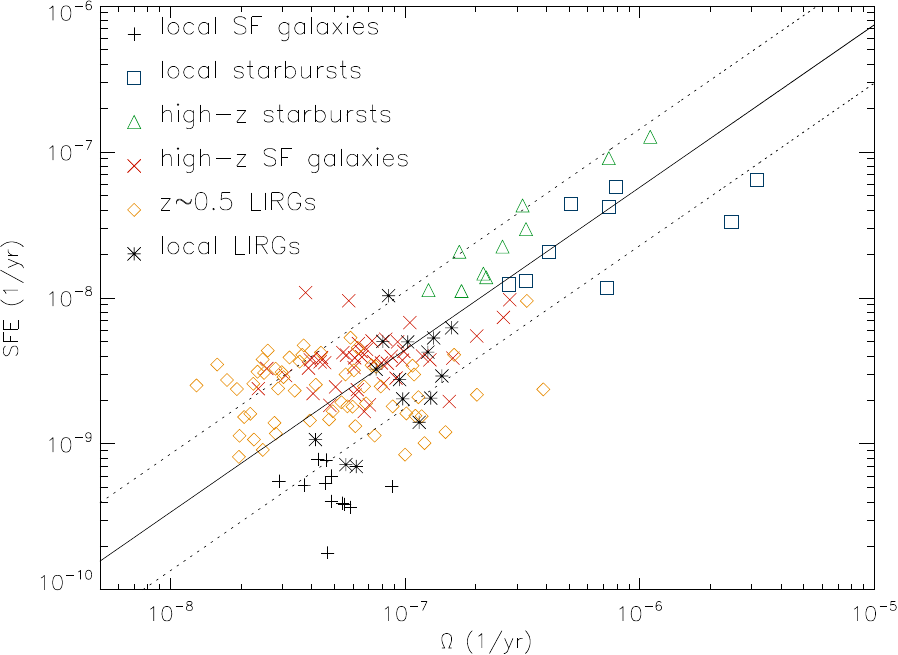}}
  \resizebox{\hsize}{!}{\includegraphics{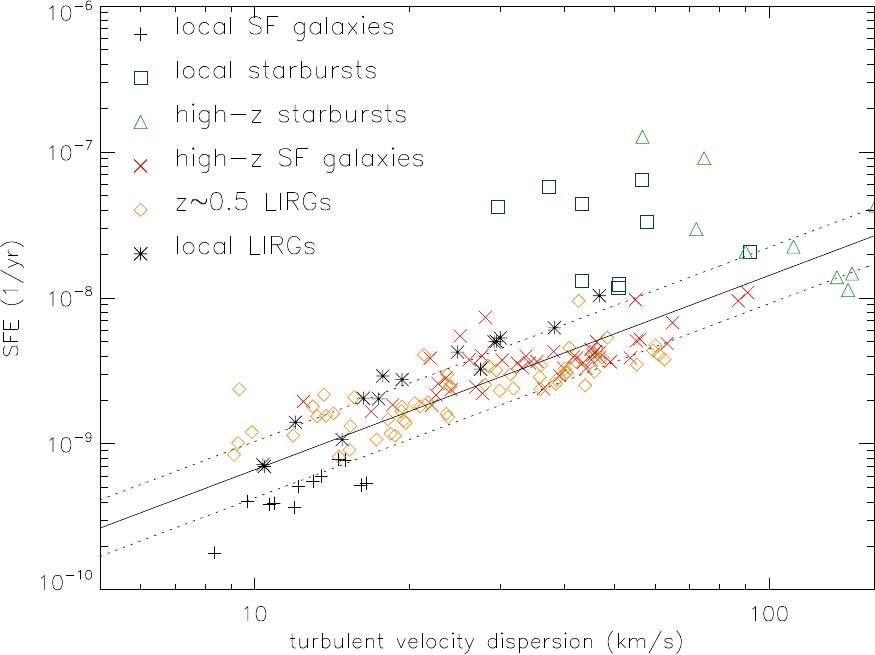}}
  \caption{Upper panel: star formation efficiency with respect to the molecular gas as a function of the angular velocity $\Omega$.
    The lines correspond to an outlier-resistant linear bisector fit and its dispersion.
    Lower panel: star formation efficiency with respect to the molecular gas as a function of the turbulent velocity dispersion. The fit only applies to the SF galaxies.
  \label{fig:plots_1_phibss2_6}}
\end{figure}

Silk (1997) suggested that the star formation rate and the gas mass are related by the dynamical time:
SFE$\propto \Omega^{-1}$. This relation with respect to the molecular gas is shown in the upper panel of Fig.~\ref{fig:plots_1_phibss2_6}.
The slope of the log-log relation is $1.1$. The $\Omega$-dependence captures the difference between the SF galaxies and starbursts.
However, the scatter of $0.4$ is quite large. And the $z \sim 0.5$ LIRGs do not follow the relation.
The $log(SFE) - log(V_{turb})$ relation has a slope of 1.3 and is considerably tighter ($0.20$~dex; see also Fisher et al. 2019).
Only the starburst galaxies do not follow this relation.

\begin{figure}
  \centering
  \resizebox{\hsize}{!}{\includegraphics{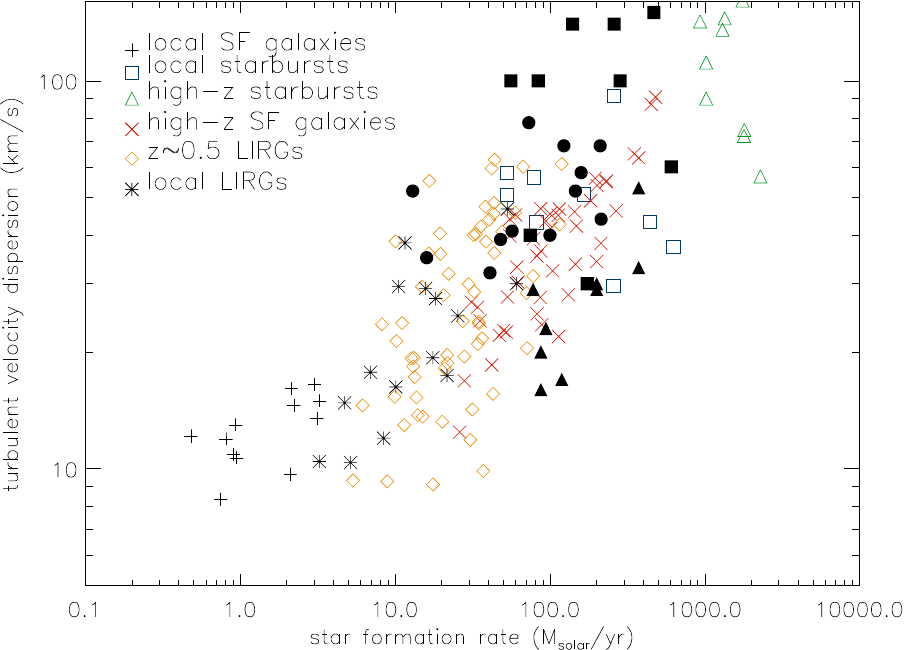}}
  \resizebox{\hsize}{!}{\includegraphics{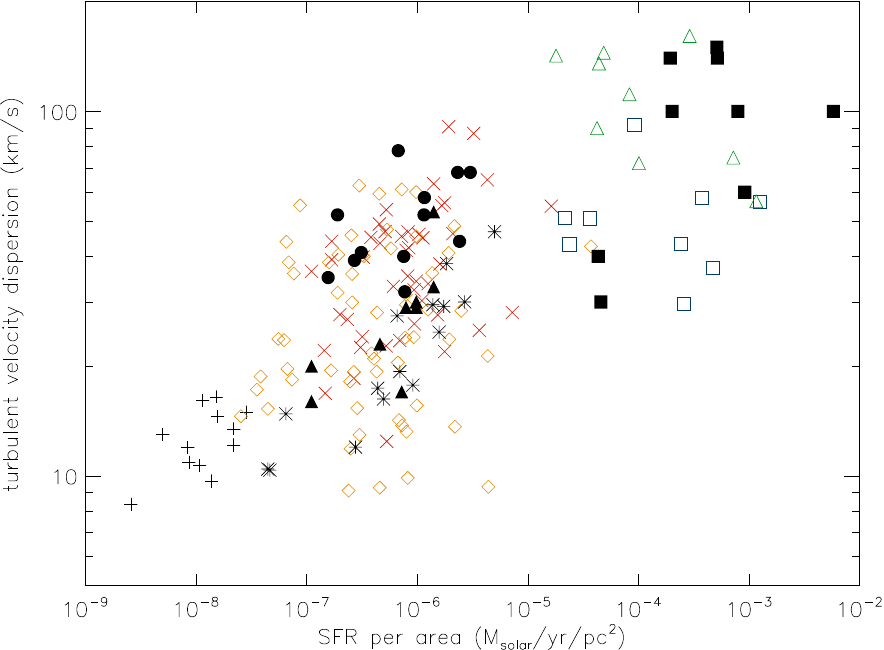}}
  \resizebox{\hsize}{!}{\includegraphics{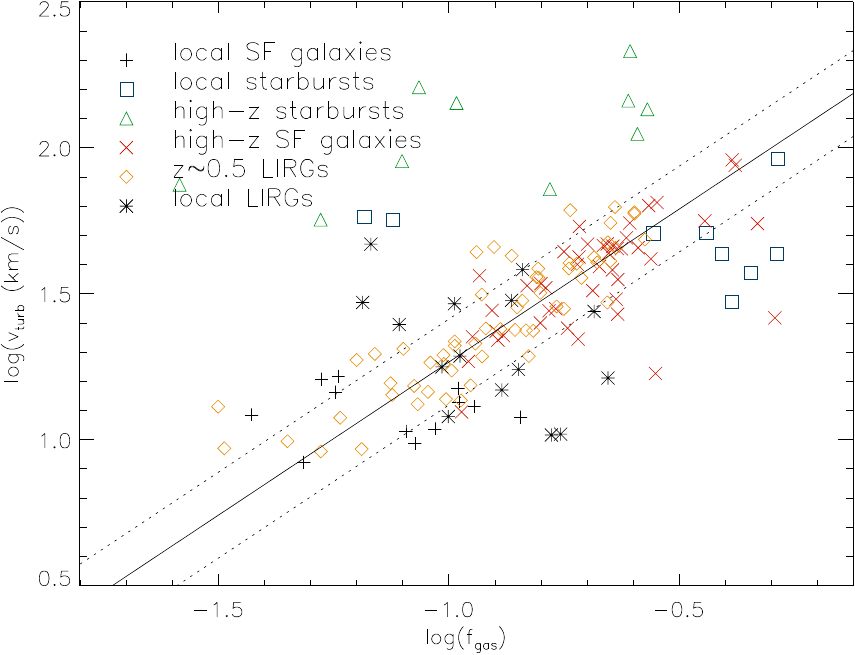}}
  \caption{Model turbulent velocity as a function of the star formation rate (upper panel), star formation rate per unit
    area (middle panel), and gas fraction (lower panel).
    Filled circles and boxes correspond to observed values (Downes \& Solomon 1998, Cresci et al. 2009, Tacconi et al. 2013,
    Girard et al. 2021, Nestor Shachar et al. 2023). The lines correspond to an outlier-resistant linear bisector fit and its dispersion.
  \label{fig:plots_1_phibss2_2}}
\end{figure}

The turbulent velocity dispersion increases with the star formation rate, star formation rate per unit area, and gas fraction
(Fig.~\ref{fig:plots_1_phibss2_2}). However, the scatter of these trends is large. The relation between the velocity dispersion
and the gas fraction of the SF galaxies has a scatter of $0.4$~dex (see also Girard et al. 2021).
The slope of the log-log relation is unity.

\begin{figure}
  \centering
  \resizebox{\hsize}{!}{\includegraphics{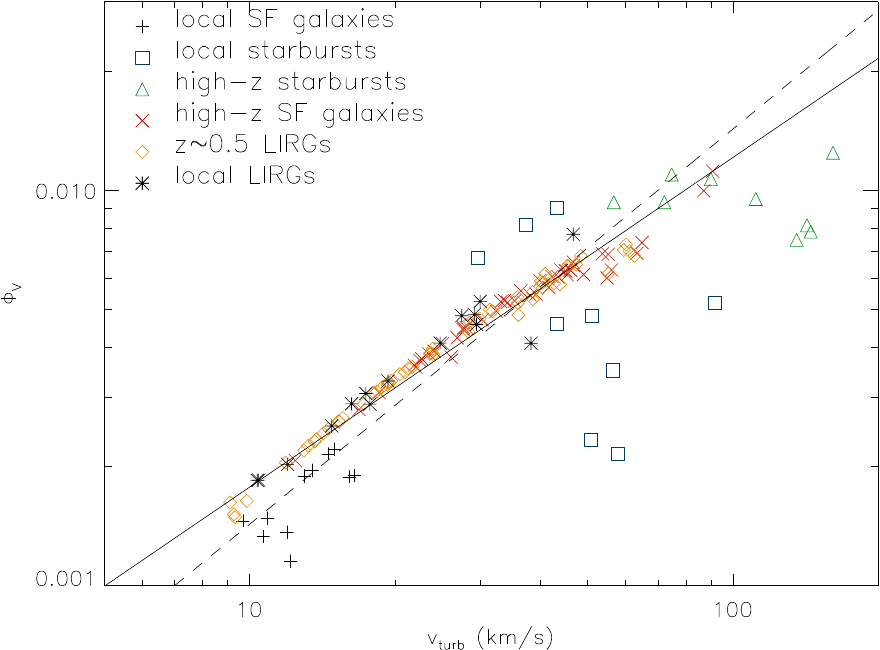}}
  \resizebox{\hsize}{!}{\includegraphics{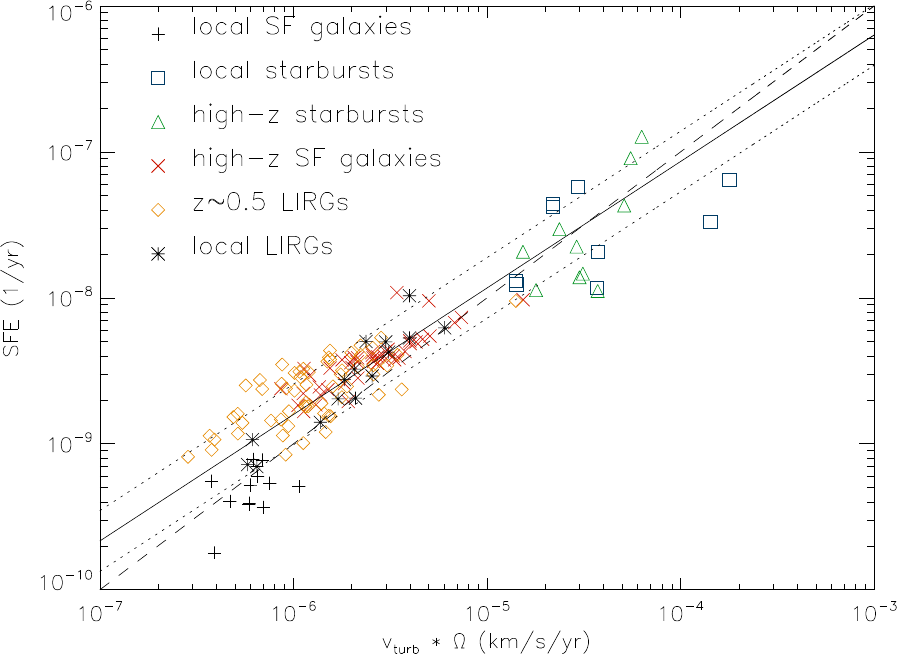}}
  \caption{Upper panel: inverse of the overdensity of selfgravitating clouds as a function of the turbulent velocity dispersion.
    The solid line corresponds to an outlier-resistant linear bisector fit, the dashed line to $\Phi_{\rm V}=v_{\rm turb}/(1\,{\rm km}\,{\rm s}^{-1})/7000$.
    Lower panel: star formation efficiency with respect to the molecular gas as function of ($v_{\rm turb} \times \Omega$).
    The solid and dotted lines correspond to an outlier-resistant linear bisector fit and its dispersion. The dashed line
    corresponds to SFE$=10^{-3}\,v_{\rm turb}\Omega$.
  \label{fig:plots_1_phibss2_8}}
\end{figure}

The $log(SFE) - log(v_{\rm turb} \Omega)$ correlation for the starburst galaxies yields the lowest scatter ($0.21$~dex)
The slope is $0.87$; (see lower panel of Fig.~\ref{fig:plots_1_phibss2_8}).
The reason for this behavior is the star formation prescription $\dot{\Sigma}_* \propto \Phi_{\rm V} \rho v_{\rm turb}$ (Eq.~\ref{eq:sfrrec}).
With $H \sim v_{\rm turb} \Omega^{-1}$ one obtains $\dot{\Sigma}_* \propto \Phi_{\rm V} \Sigma \Omega$.
With $\Phi_{\rm V} \propto v_{\rm turb}^{0.83}$ (upper panel of Fig.~\ref{fig:plots_1_phibss2_8}) one obtains the almost linear relation between the SFE
and ($v_{\rm turb} \times \Omega$).

\subsection{Gas viscosity \label{sec:viscosity}}

\begin{figure}
  \centering
  \resizebox{\hsize}{!}{\includegraphics{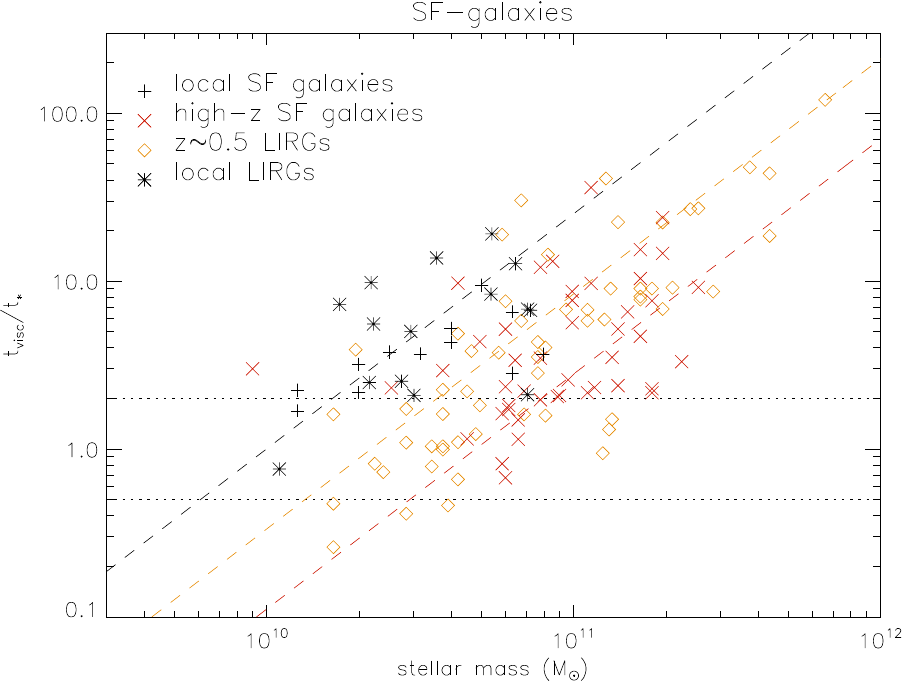}}
  \resizebox{\hsize}{!}{\includegraphics{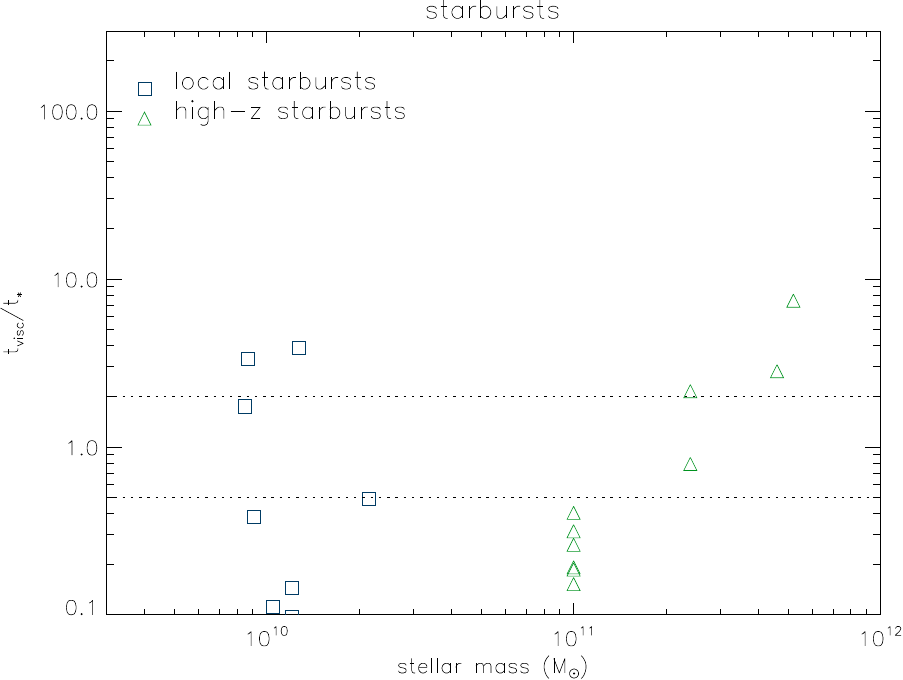}}
  \caption{The fraction $t_{\rm visc}/t_*$ as a function of the stellar mass of the model galaxies. Upper panel: SF galaxies.
    The $t_{\rm visc}/t_*$ fractions of the starforming galaxies should thus be taken as lower limits
    with an uncertainty of a factor of two towards higher values.
    The dashed lines are observed trends with stellar mass. Lower panel: starburst galaxies.
    \label{fig:plots_1_phibss2_14}}
\end{figure}

The turbulent viscosity of the gas is given by $\nu=v_{\rm turb}\,l_{\rm driv}$. To investigate the importance of radial gas accretion
with respect to the star formation rate, we evaluate the star formation timescale ($t_*=\Sigma/\dot{\Sigma}_*$) and the viscous
timescale ($t_{\rm visc}=R^2/\nu$) at the effective radius of each galaxy.
Fig.~\ref{fig:plots_1_phibss2_14} shows the model $t_{\rm visc}/t_*$ ratio vs. stellar mass for the SF galaxies (top) and starbursts (bottom).
Most starburst galaxies have $t_{\rm visc}/t_* \le 2$, meaning that the radial gas transport can approximately compensate the mass
loss via star formation. 

The situation is different for the SF galaxies (upper panel of Fig.~\ref{fig:plots_1_phibss2_14}). Only a minority of the SF galaxies have $t_{\rm visc}/t_* \le 2$.
The direct comparison with the spatially resolved models of the local SF galaxies (Liz\'ee et al. 2022) shows that our  $t_{\rm visc}/t_*$ fractions
can be underestimated by up to a factor of three. The $t_{\rm visc}/t_*$ fractions of the starforming galaxies should thus be taken as lower limits
with an uncertainty of a factor of two towards higher values. Our conclusions are not significantly affected by this uncertainty.
A trend appears where $t_{\rm visc}/t_*$ increases with stellar mass such that replenishment becomes more difficult at high mass, at least at the effective radius (see also Vollmer \& Leroy 2011).
Robust bisector fits of the log-log relations lead to slopes of $1.2$ for the local starbursts and LIRGs, $1.5$ for the $z \sim 0.5$ LIRGs, and
$1.3$ for the high-z SF galaxies. The trends with a log-log slope of $1.4$ are shown as dashed lines in the upper panel of Fig.~\ref{fig:plots_1_phibss2_14}.
In addition, there seems to be an offset related to redshift: the lines cross $t_{\rm visc}/t_*=2$ at higher stellar masses for increasing redshift:
$M(z=0) \sim 2 \times 10^{10}$~M$_{\odot}$, $M(z=0.5) \sim 4 \times 10^{10}$~M$_{\odot}$, and $M(z=1.5) \sim 8 \times 10^{10}$~M$_{\odot}$.
They cross $t_{\rm visc}/t_*=1$ at $M(z=0) \sim 1 \times 10^{10}$~M$_{\odot}$, $M(z=0.5) \sim 2 \times 10^{10}$~M$_{\odot}$, and $M(z=1.5) \sim 5 \times 10^{10}$~M$_{\odot}$.
This means that galaxies below these limiting masses can in principle compensate their gas consumption via star formation by radial viscous gas accretion.

\subsection{Cosmic ray ionization rate \label{sec:crir}}

There is a discrepancy of about a factor of ten between the measurements of low energy cosmic rays by Voyager~1 in the Solar neighborhood
($\zeta_{\rm H} \sim 2\,10^{-17}$~s$^{-1}$; Cummings et al. 2016) and the
astronomical inferences towards many lines of sight and at several Galactocentric radii ($\zeta_{\rm H} \sim 2\,10^{-16}$~s$^{-1}$, e.g., Indriolo et al. 2015, Neufeld \& Wolfire 2017).
New estimates of the gas density in diffuse molecular clouds led to a significantly smaller discrepancy (a factor of $\sim 3$; Obolentseva et al. 2024).
As described in Sect.~\ref{sec:cosmicrays} the CR ionization fraction was calculated for a given gas density,
radiation field, and magnetic field strength. The only free parameter is the overall normalization of the CR ionization rate
(parameter $k$ in Eq.~\ref{eq:crionf}). For $k=1$ the CR ionization rate measured in the Solar neighborhood is recovered for the conditions of the local ISM.
In the following we give estimates of $k$ based on the molecular line emission and radio continuum emission.

\begin{figure}
  \centering
  \resizebox{\hsize}{!}{\includegraphics{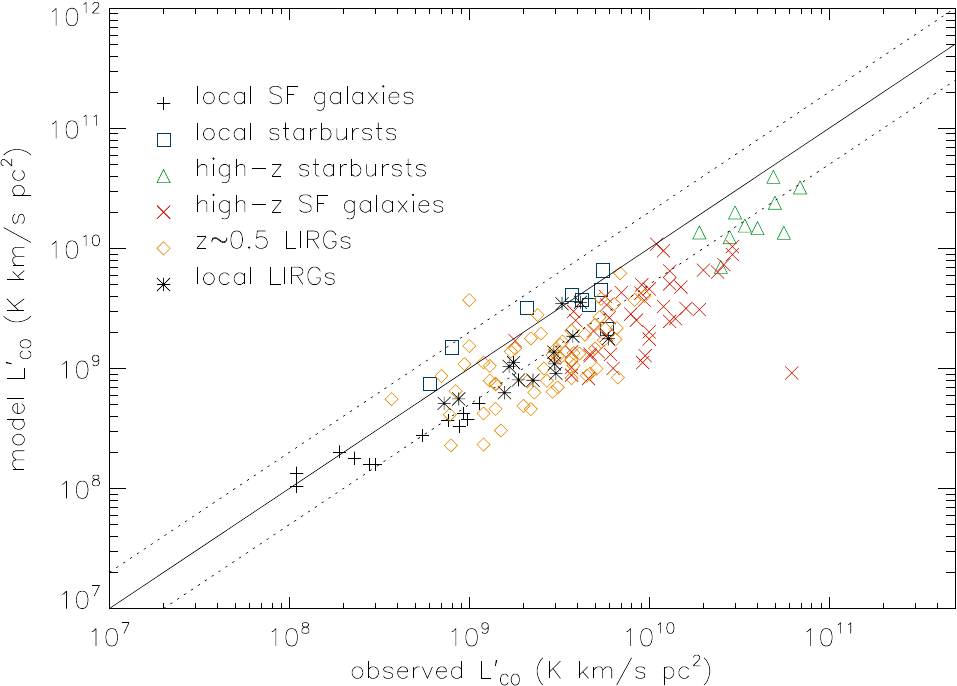}}
  \resizebox{\hsize}{!}{\includegraphics{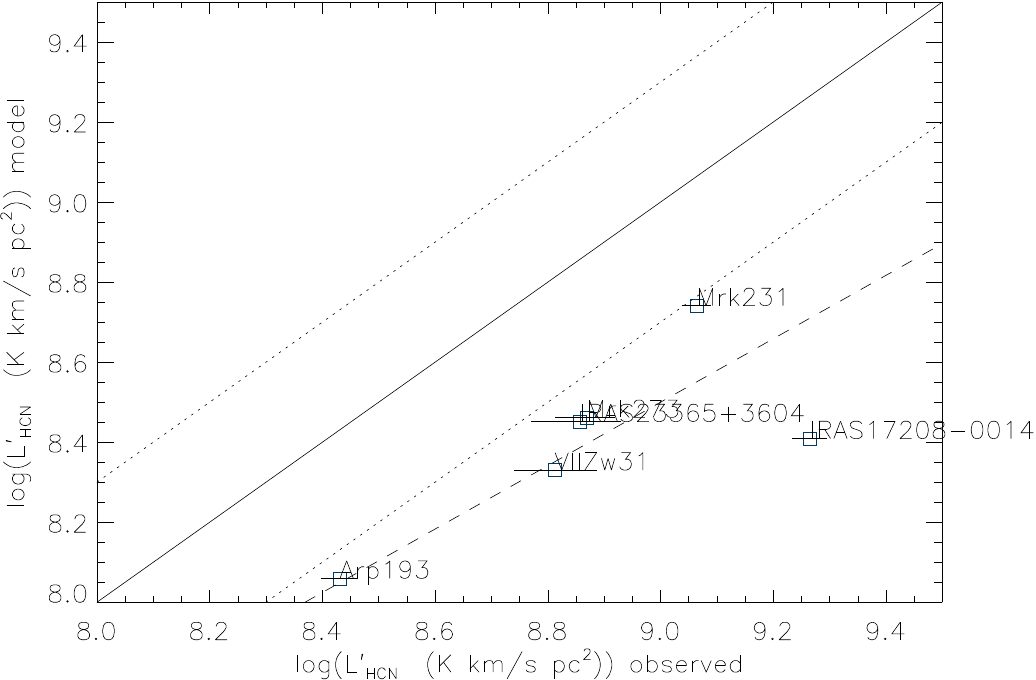}}
  \caption{Models with $k=9$ instead of $k=3$. Upper panel: model CO flux of the galaxies of the different samples as a function of the observed CO flux.
    Lower panel: model HCN(1-0) flux as a function of the observed HCN(1-0) flux for the local starburst galaxies.
  \label{fig:plots_HCNCO_cummings2}}
\end{figure}

The molecular line emission depends on the CR ionization rate, which is taken into account by Nautilus. 
In all previous sections we showed the results for $k=3$, which led to results closest to the available molecular line and radio continuum observations. 
Motivated by the observed CR ionization rate at  several Galactocentric radii we set $k=9$.
In this case the CO(1-0) and HCN(1-0) fluxes are significantly lower than it is observed (Fig.~\ref{fig:plots_HCNCO_cummings2}).
The $k=9$ model can thus be discarded.

Does $\zeta_{\rm H}$ really vary with $N_H$ (Eq.~\ref{eq:crionf}) ?
We calculated the $k=3$ model without the $N_H$ dependence of $\zeta_{\rm H}$ and
the CO(1-0) and HCN(1-0) fluxes become significantly lower than observed (Fig.~\ref{fig:plots_HCNCO_cummings2}).
We conclude that $\zeta_{\rm H}$ indeed varies with $N_H$.
\begin{figure}
  \centering
  \resizebox{\hsize}{!}{\includegraphics{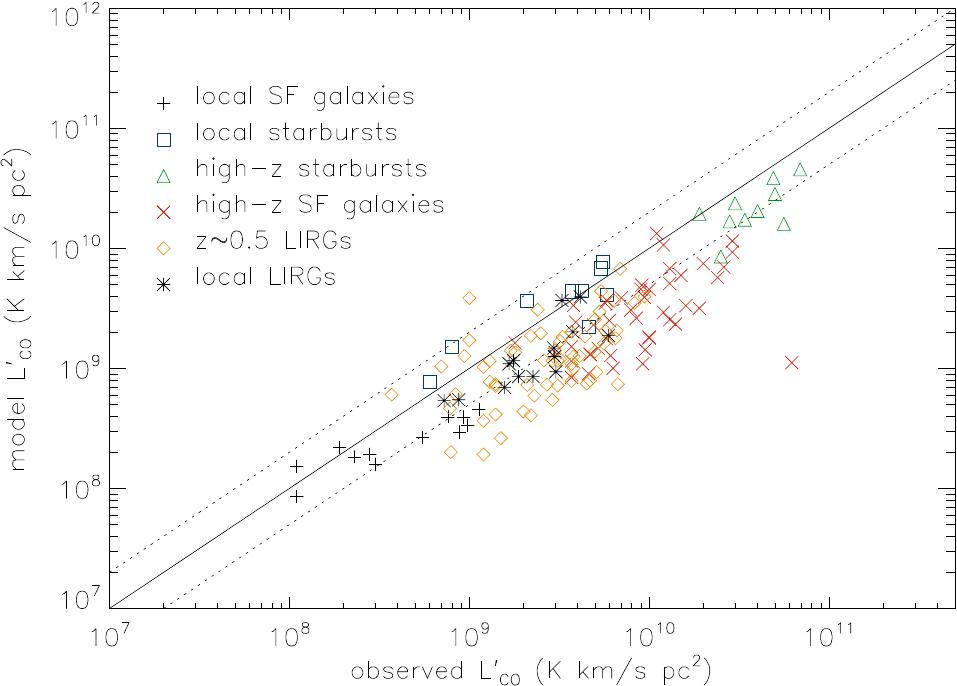}}
  \resizebox{\hsize}{!}{\includegraphics{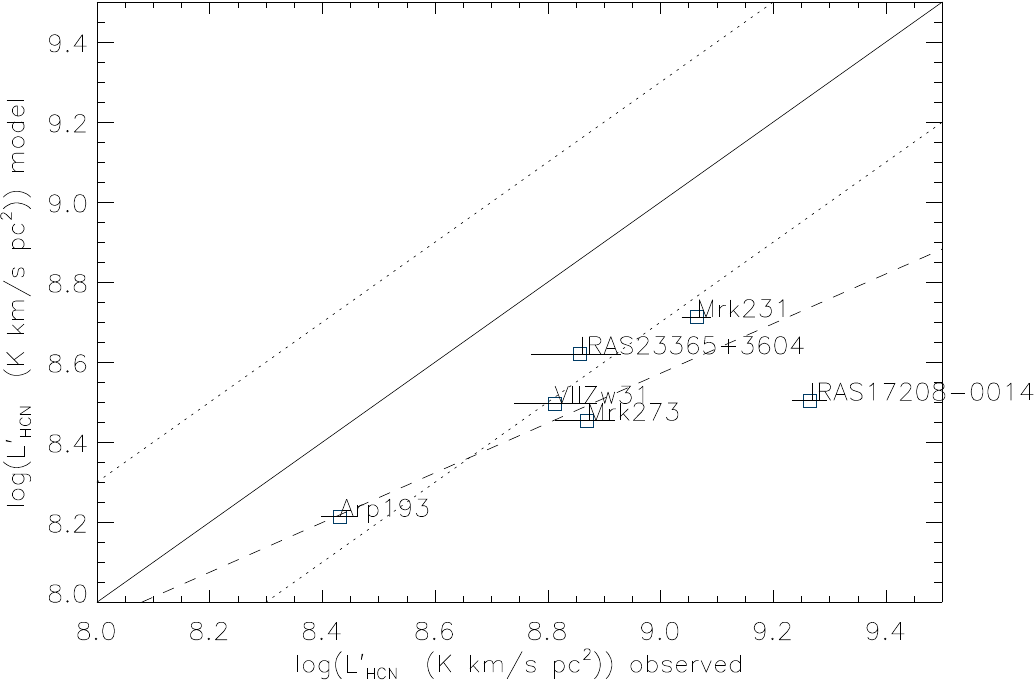}}
  \caption{Models without the dependence of $\zeta_{\rm H}$ on the gas column density. Upper panel: model CO flux of the galaxies of the different samples as a function of the observed CO flux.
    Lower panel: model HCN(1-0) flux as a function of the observed HCN(1-0) flux for the local starburst galaxies.
  \label{fig:plots_HCNCO_noPadovani}}
\end{figure}

On the other hand, the density of the CR electrons can be calculated from the radio continuum model (Sect.~\ref{sec:radcont}).
Cosmic ray electrons with energy $E$ emit most of their energy at the frequency $\nu_{\rm s}$ (e.g., Eq.~5 of V22) where
\begin{equation}
\label{eq:meanfreq}
\nu_{\rm s}=1.3 \times 10^{-2} \big(\frac{B}{10~\mu {\rm G}} \big) \big(\frac{E}{\rm GeV}  \big)^2 ~{\rm GHz}
\end{equation}
with $E=\gamma m_0 c^2$ for high Lorentz factors $\gamma$.
The Lorentz factor of electrons emitting at $1.4$~GHz $\gamma_{1.4}$ was calculated by inverting Eq.~\ref{eq:meanfreq}.
The CR electron density at a frequency of $1.4$~GHz was then calculated using Eq.~\ref{eq:emtsang}
with the model spectral index $\alpha_{12}$ calculated between $1$ and $2$~GHz
\begin{equation}
n_{\rm CRe,\ 1.4}=n_0\,\gamma_{1.4}^{\alpha_{12}}\ .
\end{equation}
To normalize the CR electron density at $1.4$~GHz to that based on the normalization found in the Solar neighborhood, we used Eq.~\ref{eq:ne} evaluated at
the energy corresponding to an emission frequency of $1.4$~GHz (Eq.~\ref{eq:meanfreq}) ($n_{\rm e,neighborhood}$) with the ISM conditions from the analytical model
(Sect.~\ref{sec:cosmicrays}). The final overdensity of CR electrons is $\tilde{k}=n_{\rm CRe,\ 1.4}/n_{\rm e,neighborhood}$.
Consistent results were found for frequencies of $15$ and $150$~MHz.

The visual inspection of the corresponding radial profiles $\tilde{k}(R)$ showed variations typically with a decreasing $\tilde{k}$ with
increasing radius. We calculated the mean of all $\tilde{k}(R)$ for the different galaxy samples. The results are presented in Table~\ref{tab:ktilde}.
\begin{table}
\begin{center}
\caption{Overdensity of CR electrons with respect to the Solar neighborhood. \label{tab:ktilde}}
\begin{tabular}{lcc}
\hline
sample & $\langle \log(\tilde{k}) \rangle$ & $\langle \tilde{k} \rangle$ \\
\hline
local SF galaxies & $0.70 \pm    0.29$ &    $5.0$ \\
local starbursts & $1.04 \pm 0.52$ & $11.1$ \\
local LIRGs & $0.91 \pm 0.39$ & $8.1$ \\
$z \sim 0.5$ LIRGs & $0.44 \pm 0.59$ & $2.7$ \\
high-z SF galaxies & $0.28 \pm 0.55$ & $1.9$ \\
high-z starbursts & $0.21 \pm 0.49$ & $1.6$ \\
\hline
\end{tabular}
\end{center}
\end{table}
The overdensities of CR electrons with respect to the Solar neighborhood range between $2$ and $10$. Based on Eq.~\ref{eq:neo} we assume
that $\tilde{k} \sim k$.

The mean of $\tilde{k}$ from the radio continuum observations is about $4$--$5$. This compares well to the value of $k=3$ found from the molecular line emission.
We thus conclude that the normalization of the CR ionization rate found for the different galaxy samples is about a factor of three to five higher
than the normalization for the Solar neighborhood. The mean yield of low energy CR particles for a given star formation rate per unit area is thus
about three to five times higher in external galaxies than the yield in the Solar neighborhood. This is two to three times less than the factor of ten discrepancy between
measurements of low energy cosmic rays by Voyager~1 in the Solar neighborhood and the astronomical inferences towards many lines of sight and at several Galactocentric radii.
It is in good agreement with the recent measurements of Obolentseva et al. (2024).
Given the large uncertainties of the measured CR ionization rates (Dalgrano 2006), we think that both findings broadly consistent.
It seems thus that the low energy CR particles found in the Solar neighborhood Voyager~1 are underabundant by about a factor of $5 \pm 4$ with respect to the mean
at the Solar radius. We can only speculate that this is caused by the 3D geometry of the magnetic field in the Local Bubble (Alves et al. 2018) leading to an increased escape
of CR particles and/or preventing them from entering the Solar neighborhood.

\section{Summary and conclusions \label{sec:conclusions}}

The theory of turbulent clumpy starforming gas disks (Vollmer \& Beckert 2003) provides the large-scale and small-scale properties of galactic gas disks. 
Within this model stellar feedback provides the energy injection to maintain turbulence. Stars are formed according to $\dot{\Sigma}_*=6.3\,\Phi_{\rm V}\rho v_{\rm turb}$
(Eq.~\ref{eq:sfrrec}) where the inverse of the overdensity of selfgravitating clouds $\Phi_{\rm V}$ is proportional to the gas velocity dispersion $v_{\rm turb}$
(upper panel of Fig.~\ref{fig:plots_1_phibss2_8}). This prescription leads to a star formation rate per unit area, which is proportional to the gas pressure
(Fig.~\ref{fig:plots_1_phibss2_1}) as it is expected if star formation is pressure-regulated and feedback-modulated (Ostriker \& Kim 2022).
Moreover, the radially-averaged star formation efficiencies per free-fall time range between $0.004$ and $0.075$, broadly within the range of the canonical value of
$\epsilon_{\rm ff} \sim 0.01$ (Krumholz et al. 2012).

We extended the work of V17 on local and high-z SF and starburst galaxies by (i) adding local and $z \sim 0.5$ LIRG samples, (ii) including RADEX, and (iii) adding
a CR ionization fraction, which is proportional to the star formation surface density (Eq.~\ref{eq:crionf}).
These newly added galaxy samples nicely fill the gap in total IR luminosity between the local and high-z SF galaxies.
The gas chemistry computed by the Nautilus code depends on the CR ionization rate, which was calculated following the analytical model of Pohl (1993)
using the local density, magnetic field strength, and radiation field of our analytical model. The injection rate of CR particles is proportional
to the star formation rate per unit area (Eq.~\ref{eq:crionf}).
In the present version of the code the molecular line emission are calculated by RADEX. The model reproduces the IR luminosities, CO, HCN, and HCO$^+$ line luminosities,
and the CO SLEDs
of the LIRGs (Fig.~\ref{fig:IRspectra_phibss2_1} to \ref{fig:ulirg_ladder}). We derived the model CO(1-0) and HCN(1-0) conversion factors
for all galaxy samples (Fig.~\ref{fig:plots_HCNCO_alphaco1} and \ref{fig:plots_HCNCO_alphahcn}).

Since the model yields the H$_2$ masses, the relation between the star formation per unit area and H$_2$ surface density (Kennicutt-Schmidt law) can be established.
The log-log relation cannot be fitted with a single line nor a single offset (Fig.~\ref{fig:plots_1_phibss2_10}). As observed, the SFE varies linearly with the sSFR (Fig.~\ref{fig:plots_1_phibss2_11}). Moreover, the log(SFE)--log($v_{\rm turb}$) relation has a small scatter of $0.20$~dex with an exponent of $1.3$
(lower panel of Fig.~\ref{fig:plots_1_phibss2_6}). The reason for this behavior is the $\Phi_{\rm V}$ is approximately proportional to $v_{\rm turb}$
(upper panel of Fig.~\ref{fig:plots_1_phibss2_8}). The lowest scatter of $0.21$~dex for the sample including the starburst galaxies is obtained for the log-log
relation between the SFE and the product of $v_{\rm turb}$ and $\Omega$ (lower panel of Fig.~\ref{fig:plots_1_phibss2_8}).

The model also yields the gas viscosity $\nu=v_{\rm turb} l_{\rm driv}$ and thus the viscous timescale $t_{\rm visc}$.
In addition to $t_{\rm visc}/t_*$ increasing with stellar mass, there seems to be an offset related to redshift.:
At $z=0$, $t_{\rm visc}/t_*=1$ at $M \sim 1 \times 10^{10}$~M$_{\odot}$, $M(z=0.5) \sim 2 \times 10^{10}$~M$_{\odot}$, and $M(z=1.5) \sim 5 \times 10^{10}$~M$_{\odot}$.
This means that galaxies more easily compensate their gas consumption by radial viscous gas accretion at higher redshifts.

As a second step, we calculated the radio continuum emission of the galaxies of the different samples following the framework of V22.
Compared to the previous work we used a spectral index of the injection CR electrons of $2.3$ instead of $2.0$. Such  a higher exponent is expected for superbubbles
created by multiple SN remnants (Vieu et al. 2022). The local and $z \sim 0.5$ LIRGs nicely fit into the existing radio-IR and radio-SFR relations
(Fig.~\ref{fig:galaxies_FRC_vrotDifferentForPhibbs_4} to \ref{fig:galaxies_FRC_vrotDifferentForPhibbs_6}).

Whereas the radio continuum emission is directly proportional to the density of CR electrons, the molecular line emission depends on the CR ionization rate via the
gas chemistry (Nautilus). Thus, both emission mechanisms depend on the density of CR particles. Since CR protons and electrons are provided by the same sources (mainly SN remnants),
the application of the radio continuum and molecular line emission models to observations should lead to coherent CR particle densities.
We showed in Sect.~\ref{sec:crir} that this is approximately the case and that the mean yield of low energy CR particles for a given star formation rate is about a factor of three to five higher than in the solar neighborhood.

\begin{acknowledgements}
  This research has made use of the VizieR catalogue access tool, CDS, Strasbourg, France (DOI : 10.26093/cds/vizier).
  The original description  of the VizieR service was published in 2000, A\&AS 143, 23.
  We would like to thank the referee for their careful reading and constructive comments.
\end{acknowledgements}

\clearpage

\begin{appendix}

\section{The model gas disk \label{sec:dgasdisk}}

\begin{figure}
  \centering
  \resizebox{\hsize}{!}{\includegraphics{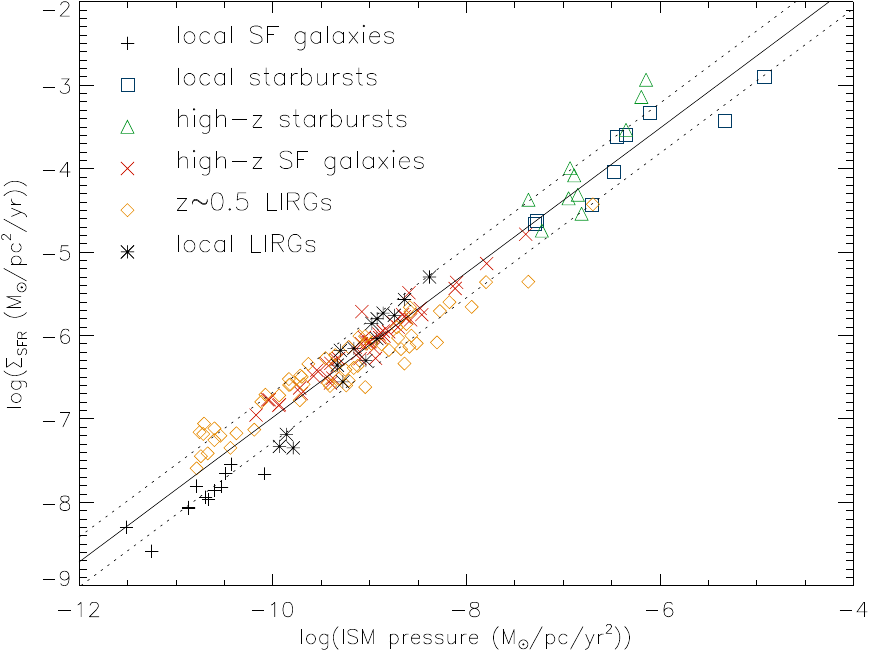}}
  \caption{Star formation rate as a function of ISM pressure for the six galaxy samples. The lines correspond to an outlier-resistant linear bisector fit and its dispersion.
    The slope is $0.87 \pm 0.02$.
  \label{fig:plots_1_phibss2_1}}
\end{figure}

In this section and the following, we briefly describe the V17 model with minor modifications
of the factors in Eq.~\ref{eq:tturbcl} to \ref{eq:sigmastar1}. The meaning of the different variables is given in Appendix~\ref{sec:variables}.
Following Vollmer \& Leroy (2011) and Vollmer et al. (2021), the SFR per unit volume is given by
\begin{equation}\label{eq:starform}
\dot{\rho}_{*} = \Phi_{\rm V} \frac{\rho}{t_{\rm ff}^{\rm l}}\ .
\end{equation}
where $\Phi_{\rm V}$ is the volume filling factor, $\rho$ is the large-scale density, $l$ and $\rho_{\rm l}=\rho/\Phi_{\rm V}$ the size and density of selfgravitating clouds,
and $t_{\rm ff}^{\rm l}=\sqrt{3\pi/(32\,G\,\rho_{\rm l})}$ the local free-fall time.
This prescription leads to a star formation rate per unit area which is proportional to the gas pressure $p=\rho v_{\rm turb}^2$ (Fig.~\ref{fig:plots_1_phibss2_1})
as expected if star formation is pressure-regulated and feedback-modulated (Ostriker \& Kim 2022).

The radially-averaged gas fraction converted to stars per free-fall time is $\epsilon_{\rm ff} = t_{\rm ff}/t_{\rm dep}$
where $t_{\rm ff}$ is calculated at the scale of $l_{\rm driv}$ (Krumholz \& McKee 2005), can be compared to the
value of $\epsilon_{\rm ff}=0.01$ determined by Krumholz et al. (2012).  
For our main sequence samples (ordered by increasing $z$), we find $\epsilon_{\rm ff} = 0.004 \pm 0.001, 0.010 \pm 0.006, 0.018 \pm 0.007$
and for the starburst sequence, $\epsilon_{\rm ff} = 0.022 \pm 0.007, 0.024 \pm 0.031, 0.075 \pm 0.041$ as described in Sect.~\ref{sec:samples}.
It is clear that while $\epsilon_{\rm ff}$ increases with redshift in both samples, the starburst galaxies convert their gas more quickly into stars.
Except for the high-z starbursts, these values are close to the Krumholz et al. (2012) prediction.

For self-gravitating clouds with a virial parameter of two (Bertoldi \& McKee 1992), the turbulent crossing time $t_{\rm turb,cl}$ equals
$1.4$ times the free-fall time
$t_{\rm turb,cl}$:
\begin{equation}
t_{\rm turb,cl} =  \frac{l_{\rm cl}}{2\,v_{\rm turb,cl}} = 1.4 \times t_{\rm ff,cl}=1.4 \times \sqrt{\frac{3 \pi \Phi_{\rm V}}{32 G \rho}}
\label{eq:tturbcl}
\end{equation}
where $l_{\rm cl}$ and $v_{\rm turb,cl}$ are the size and the turbulent 1D velocity dispersion of a single gas cloud, respectively. Following Larson's
law (Larson 1981), we can simplify the expression of the turbulent crossing time:
\begin{equation}
t_{\rm turb,cl}=\frac{l_{\rm cl}}{2\,v_{\rm turb,cl}} = \frac{l_{\rm driv}}{2\,v_{\rm turb} \sqrt{\delta}}\ .
\end{equation}
where $\delta$ is the scaling between the driving length scale and the
size of the largest self-gravitating structures, such as $\delta = l_{\rm driv}/l_{\rm cl}$.
The expression for the star formation rate then becomes
\begin{equation}
  \label{eq:sfrrec}
\dot{\rho}_*= 2.8\,\sqrt{\delta} \Phi_{\rm V} \rho \frac{v_{\rm turb}}{l_{\rm driv}}\ .
\end{equation}
and $\dot{\Sigma}_*=\dot{\rho}_* l_{\rm driv}$.

The star formation rate density is used to calculate the rate of energy injection by supernovae (SNe):
\begin{equation}
  \label{eq:energyflux}
  \frac{\dot{E}_{\rm SN}}{\Delta A}=\xi\,\dot{\Sigma}_{*} 
  = \xi\,\dot{\rho}_{*} l_{\rm driv}=\frac{1}{2} \Sigma \nu \frac{v_{\rm turb,3D}^{2}}{l_{\rm driv}^{2}}
  = \frac{3\,\sqrt{3}}{2} \Sigma \frac{v_{\rm turb}^3}{l_{\rm driv}}\ ,
\end{equation}
where $\Delta A$ is the unit surface element of the disk, $\dot{\Sigma}_{*}$ and $\dot{\rho}_{*}$ the star formation per unit area and unit volume,
$\Sigma$ the total gas surface density, $\nu$ the turbulent viscosity, $v_{\rm turb}$ the turbulent velocity of the gas, and the CO disk thickness
is assumed to be $l_{\rm driv}$. The turbulent viscosity is $\nu=\sqrt{3}v_{\rm turb} l_{\rm driv}$.
The energy injection rate is related to the turbulent
velocity dispersion and the driving scale of turbulence. These quantities in turn provide estimates of the clumpiness of gas in the disk (i.e., the
contrast between local and average density) and the rate at which viscosity moves matter inward.
The derived expressions for $\Phi_{\rm V}$ and $\dot{\Sigma}_*$ are
\begin{equation}
\Phi_{\rm V}=\frac{4.1\,G\,\rho\,l_{\rm driv}^2}{3\,\pi\,\delta\,v_{\rm turb}^2}
\end{equation}
and
\begin{equation}
\dot{\Sigma}_*=\frac{11.4\,G\,\rho^2 l_{\rm driv}^2}{3\,\pi\,\sqrt{\delta} v_{\rm turb}}\ .
\label{eq:sigmastar1}
\end{equation}
  
The model relies on several empirical calibrations: the relation between the stellar
velocity dispersion and the stellar disk scale length, the relationship between the star-formation rate (SFR) and the energy injected into
the ISM by SNe, and the characteristic time of H$_2$ formation, which is related to the gas metallicity and the gas density.
The fraction between the molecular and the total gas mass is governed by the turbulent crossing time $t^{\rm l}_{\rm turb}$ and the molecule formation timescale.
In addition, photodissociation of molecules is taken into account.

The model is divided into a large-scale and a small-scale part. The former yields the surface density, turbulent velocity, disk
height, and gas viscosity. The latter is relevant for self-gravitating gas clouds $(t^{\rm l}_{\rm ff} \le t^{\rm l}_{\rm turb})$.
The non-self-gravitating and self-gravitating clouds obey different scaling relations, which are set by observations.
For each gas density, the mass fraction of clouds of this density is characterized by a log-normal probability distribution function and the Mach number
(Padoan et al. 1997). The temperatures of the gas clouds are calculated via the equilibrium between turbulent mechanical and
cosmic ray heating and gas cooling via CO and H$_2$ line emission. The abundances of the different molecules are determined using the time-dependent
gas-grain code Nautilus (Hersant et al. 2009). The dust temperatures are calculated via the equilibrium between heating by the interstellar UV and optical
radiation field and cooling via infrared emission.

\section{The model radio continuum emission \label{sec:radconta}}

In this section, we briefly describe the parts of the model where nothing has changed with respect to what was done in V22.
Under the assumption described in Sect.~\ref{sec:radcont}, the synchrotron emissivity is given by the density per unit energy interval of the primary CR 
electrons $n_0 \propto \dot{\rho}_* t_{\rm eff}$ and
\begin{equation}
\label{eq:emissivity}
\epsilon_\nu {\rm d}\nu \propto \dot{\rho}_* t_{\rm eff} E^{-q} \frac{E}{t_{\rm sync}} {\rm d}E\ .
\end{equation}

According to V22, the effective lifetime of synchrotron-emitting CR electrons $t_{\rm eff}$ is given by
\begin{equation}
\frac{1}{t_{\rm eff}}=\frac{1}{t_{\rm sync}}+\frac{1}{t_{\rm diff}}+\frac{1}{t_{\rm brems}}+\frac{1}{t_{\rm IC}}+\frac{1}{t_{\rm ion}}\ .
\end{equation}
For the characteristic timescales, we follow the prescriptions of Lacki et al. (2010).
The diffusion timescale based on observations of beryllium isotope ratios at the solar circle (Connell 1998, Webber et al. 2003) is
\begin{equation}
\label{eq:tdiffe}
t_{\rm diff}=26/\sqrt{E/3 {\rm GeV}}~{\rm Myr},
\end{equation}
where the mean energy $E$ is calculated via the mean synchrotron frequency.
The characteristic time for bremsstrahlung is
\begin{equation} 
t_{\rm brems}=37\, (\frac{n}{{\rm cm}^{-3}})^{-1}~{\rm Myr},
\end{equation}
and that for inverse Compton energy losses is
\begin{equation}
t_{\rm IC}=180\, (\frac{B}{10~\mu{\rm G}})^{\frac{1}{2}}(\nu_{\rm GHz})^{-\frac{1}{2}}(\frac{U}{10^{-12}~{\rm erg\,cm}^{-3}})^{-1}~{\rm Myr}
,\end{equation}
where $U$ is the interstellar radiation field. The  
timescale of ionization-energy loss is
\begin{equation}
t_{\rm ion}=210\, (\frac{B}{10~\mu{\rm G}})^{-\frac{1}{2}}(\nu_{\rm GHz})^{\frac{1}{2}}(\frac{n}{{\rm cm}^{-3}})^{-1}\ .
\end{equation}
The magnetic field strength $B$ is calculated under the assumption of energy equipartition between the turbulent kinetic energy of the gas
and the magnetic field:
\begin{equation}
\label{eq:Bmag2}
\frac{B^2}{8 \pi}=\frac{1}{2}\rho v_{\rm turb}^2\ ,
\end{equation}
where $\rho$ is the total midplane density of the gas and $v_{\rm turb}$ its turbulent velocity dispersion.

Secondary CR electrons can be produced via collisions between the interstellar medium (ISM) and CR protons.
The proton lifetime to pion losses (Mannheim \& Schlickeiser 1994) is
\begin{equation}
t_{\pi}=50\,(\frac{n}{{\rm cm}^{-3}})^{-1}~{\rm Myr}.
\end{equation}
The effective lifetime of CR protons is given by
\begin{equation}
\frac{1}{t_{\rm eff,p}}=\frac{1}{t_{\rm wind}}+\frac{1}{t_{\rm diff,p}}\ ,
\end{equation}
where the proton diffusion timescale is four times shorter than the CR electron diffusion timescale (Appendix B3 of Werhahn et al. 2021a).
The CR electron secondary fraction is given by
\begin{equation}
\label{eq:secondaries}
\eta_{\rm sec}=\frac{1}{2} (1+\frac{t_{\pi}}{t_{\rm eff,p}})
\end{equation}
(Werhahn et al. 2021a).
In models that include CR electron secondaries, the CR electron density is multiplied by $(1+\eta_{\rm sec})$.

With $\nu=C B E^2$, the synchrotron emissivity of Eq.~\ref{eq:emissivity} becomes
\begin{equation}
\label{eq:emfinal}
\epsilon_{\nu}= \xi \dot{\rho}_* \frac{t_{\rm eff}(\nu)}{t_{\rm sync}(\nu)} B^{\frac{q}{2}-1} \nu^{-\frac{q}{2}}\ .
\end{equation}
The constant is $C=e/(2 \pi m_{\rm e}^2 c^2)$.
With the CR electron density $n_0$ and
\begin{equation}
\label{eq:synch}
t_{\rm sync}=\frac{E}{b(E)} \simeq 4.5 \times 10^7 \big( \frac{B}{10~\mu{\rm G}} \big)^{-3/2} \big(\frac{\nu}{\rm GHz}\big)^{-1/2}~{\rm yr}\ .
\end{equation}
the classical expression 
$\epsilon_{\nu} \propto n_0 B^{(q+1)/2} \nu^{(1-q)/2}$ is recovered.
The factor $\xi$ was chosen such that the radio--IR correlations measured by Yun et al. (2001) and Molnar et al. (2021) are 
reproduced within $2 \sigma$ (Fig.~\ref{fig:galaxies_FRC_vrotDifferentForPhibbs_4}).
As in Sect.~\ref{sec:cosmicrays} we assumed $q=2.3$.
The gas density $\rho$, turbulent gas velocity dispersion $v_{\rm turb}$, and interstellar radiation field $U$ are 
directly taken from the analytical model of Sect.~\ref{sec:gasdisk}.

Following Tsang (2007) and Beck \& Krause (2005), the synchrotron emissivity is given by 
\begin{equation}
\label{eq:emtsang}
\epsilon_{\nu=}=a(s) \alpha_{\rm f} n_0 h \nu_{\rm L} \big(\frac{\nu}{\nu_{\rm L}}\big)^{-(q-1)/2}\ ,
\end{equation}
with $a(s)=3^{q/2}/(4 \pi(q+1))\,\Gamma((3s+19)/12)\,\Gamma((3s-1)/12)$, $\alpha_{\rm f}$ is the fine structure constant,
$h$ the Planck constant, and $\nu_{\rm L}$ the Larmor frequency.

The synchrotron luminosity was calculated via 
\begin{equation}
\label{eq:lnu8}
L_{\nu}=8 \pi^2 \int \epsilon_{\nu} \frac{(1-\exp(-\tau))}{\tau}\,H\,R\,{\rm d}R\ ,
\end{equation}
where $\tau=\tau_{\rm ff}+\tau_{\rm sync}$ is the optical depth caused by free-free and synchrotron self-absorption.

For the free-free absorption we used
\begin{equation} 
\tau_{\rm ff}=4.5 \times 10^{-9} (\frac{n}{{\rm cm}^{-3}})\,(\frac{l_{\rm driv}}{1~{\rm pc}})\,\nu_{\rm GHz}^{-2.1}\ ,
\end{equation}
where the height of the star-forming disk is assumed to be of the order of the turbulent driving length scale.
For the optical depth of synchrotron self-absorption, we used the formalism described by Tsang (2007).

Optically thin thermal emission was added according to the recipe of Murphy et al. (2012)
\begin{equation}
\big(\frac{L_{\nu}^{\rm ff}}{{\rm erg}\,s^{-1} {\rm Hz}^{-1}}\big)= 2.33 \times 10^{27} \big( \frac{T_{\rm e}}{10^4\ {\rm K}} \big)^{0.45} \big( \frac{\nu}{\rm GHz} \big)^{-0.1} \big(\frac{\rm SFR}{{\rm M}_{\odot}{\rm yr}^{-1}}\big)
\end{equation}
with an electron temperature of $T_{\rm e}=8000$~K.

For the galaxies at high redshifts, the inverse Compton (IC) losses from the cosmic microwave background (CMB) are taken into account
via the IC equivalent magnetic field:
\begin{equation}
U(z)=U+\frac{(3.25~\mu{\rm G} (1+z)^2)^2}{8 \pi}\ .
\end{equation}

\section{Model variables \label{sec:variables}}

\begin{table*}
\begin{center}
\caption{Model Disk Parameters.\label{tab:parameters}}
\begin{tabular}{lll}
\hline\hline
Parameter & Unit & Explanation \\
\hline
$G=5 \times 10^{-15}$ & pc$^{3}$yr$^{-1}$M$_{\odot} ^{-1}$ & gravitation constant \\
$\bf Q$ & & \bf Toomre parameter \\
$R$ & pc &galactocentric radius \\
$H$ & pc & thickness of the gas disk \\
$l_{\rm cl}$ & pc & cloud size \\
$v_{\rm rot}$ & pc\,yr$^{-1}$ & \it rotation velocity \\
$\Omega=v_{\rm rot}/R$ & yr$^{-1}$ & \it angular velocity \\
$\Phi_{\rm V}$ & & volume-filling factor \\
$\rho$ & M$_{\odot}$pc$^{-3}$ & disk midplane gas density\\
$\rho_{\rm cl}=\rho/\Phi_{\rm V}$ & M$_{\odot}$pc$^{-3}$ & cloud density \\
$\dot{\rho}_{*}$ & M$_{\odot}$pc$^{-3}$yr$^{-1}$ & star-formation rate \\
$\Sigma$ & M$_{\odot}$pc$^{-2}$ & gas surface density \\
$\Sigma_{*}$ & M$_{\odot}$pc$^{-2}$ & \it stellar surface density \\
$\dot{\Sigma}_{*}$ & M$_{\odot}$pc$^{-2}$yr$^{-1}$ & \it star-formation rate \\
$\xi=9.2 \times 10^{-8}$ & pc$^2$yr$^{-2}$ & constant relating SN energy input to SF \\
$\bf \dot{M}$ & M$_{\odot}$yr$^{-1}$ & \bf disk mass accretion rate (radial, within the disk) \\
$v_{\rm turb}$ & pc\,yr$^{-1}$ & gas turbulent velocity dispersion \\
$v_{\rm rad}$ & pc\,yr$^{-1}$ & gas radial velocity \\
$v_{\rm disp}^{*}$ & pc\,yr$^{-1}$ & \it stellar vertical velocity dispersion \\
$\nu$ & pc$^{2}$yr$^{-1}$ & viscosity \\
$f_{\rm mol}=\Sigma_{\rm H_{2}}/(\Sigma_{\rm HI}+\Sigma_{\rm H_{2}})$ &  & molecular fraction \\
$\alpha$ & yr\,M$_{\odot}$pc$^{-3}$ & constant of molecule formation timescale \\
$l_{\rm driv}$ & pc & turbulent driving length scale \\
$\bf \delta=5$ & & {\bf scaling between the driving length scale and the size of the} \\
& & {\bf  largest self-gravitating structures} \\
$SFE=\dot{\Sigma}_{*}/\Sigma$ & yr$^{-1}$ & star-formation efficiency \\
$t_{\rm ff,cl}$ & yr & cloud free fall timescale at size $l$ \\
$t_{\rm turb,cl}$ & yr & cloud turbulent timescale at size $l$ (turbulent crossing time) \\
$\epsilon_{\rm ff}$ &  & star formation efficiency per free-fall time $\epsilon_{\rm ff}=\dot{\Sigma}_*/\Sigma_{{\rm H}_2}\times t_{\rm ff}(l_{\rm driv})$ \\
\hline
\end{tabular}
\begin{tablenotes}
      \item {\bf boldface}: free parameters; {\it italic}: parameters determined from observations.
    \end{tablenotes}
\end{center}
\end{table*}

\begin{table*}
\begin{center}
\caption{CR Model Parameters.\label{tab:parameters1}}
\begin{tabular}{lll}
\hline\hline
Parameter & Unit & Explanation \\
\hline
$c$ & cm s$^{-1}$ & speed of light \\
$m_{\rm e/p}$ & g & electron/proton mass \\
$E$ & eV & CR particle energy \\
$T$ & eV & CR kinetic energy \\
$n(E)$ & cm$^{-3}$ & CR particle density \\
$n_{\rm e/p}$ & cm$^{-3}$ & CR electron/proton particle density \\
$n_{\rm H}$ & cm$^{-3}$ & hydrogen density \\
$J(E)$ & eV cm s$^{-1}$ & CR mean flux or spectrum \\
$\zeta_{\rm H,H_2}$ & s$^{-1}$ & CR ionization rate of atomic/molecular hydrogen \\
$\phi_{\rm s}$ & & CR secondary fraction \\
$\sigma(E)$ & CR ionization cross section \\
$\epsilon_{\rm CR}$ & eV cm$^{-3}$ & CR kinetic energy density \\
$\epsilon_{\rm B}$ & eV cm$^{-3}$ & magnetic energy density \\
$t_{\rm diff}$ & yr & CR diffusion timescale \\
$U$ & Habing radiation field G$_0$ & interstellar radiation field \\
$B$ & G & magnetic field strength \\
$\xi_{\rm p/e}$ & & normalization of the CR proton and electron densities \\
$k$ & & normalization factor of the local ionization fraction \\
\hline
\end{tabular}
\end{center}
\end{table*}

\begin{table*}
\begin{center}
\caption{Radio Continuum Model Parameters.\label{tab:parameters2}}
\begin{tabular}{lll}
\hline\hline
Parameter & Unit & Explanation \\
\hline
$h$ & erg s & Planck constant \\
$\nu$ & Hz & frequency \\
$\nu_{\rm L}$ & Hz  & Larmor frequency \\
$q$ & & power law index of the CR electron density distribution $n_{\rm e}(E) \propto E^{-q}$ \\
$\epsilon_{\nu}$ & erg s$^{-1}$ cm$^{-2}$ Hz$^{-1}$ sr$^{-1}$ & synchrotron emissivity \\
$L_{\rm nu}$ & erg s$^{-1}$ & synchrotron luminosity \\
$t_{\rm sync}$ & yr & synchrotron timescale \\
$t_{\rm brems}$ & yr & bremsstrahlung timescale \\
$t_{\rm IC}$ & yr & inverse Compton timescale \\
$t_{\rm ion}$ & yr & ionization-energy loss timescale \\
$t_{\pi}$ & yr & proton lifetime to pion losses timescale \\
$t_{\rm wind}$ & yr & Galactic wind advection timescale \\
$\alpha_{\rm f}$ & & fine structure constant \\
$\tau$ & & optical depth \\
$T_{\rm e}$ & K & thermal electron temperature \\
\hline
\end{tabular}
\end{center}
\end{table*}

\section{PHIBSS2 and DYNAMO galaxies \label{sec:phibss2a}}
  
\begin{table*}
  \tabcolsep=0.15cm
  \small
\begin{center}
\caption{PHIBSS2 galaxies from Freundlich et al. (2019).\label{tab:gphibss2}}
\begin{tabular}{lcccccccccc}
  \hline
Galaxy Name & short & RA & DEC & z & log(M$_*$) & SFR & $R_{\rm eff}^{\rm disk}$ & inc & $v_{\rm max}$ & log($L_{\rm CO(2-1)}$) \\
& name & & & & (M$_{\odot}$) & (M$_{\odot}$yr$^{-1}$) & (kpc) & ($^{\circ}$) & (km\,s$^{-1}$) & (K\,km\\
 & & & & & & & & & & s$^{-1}$pc$^2$) \\
\hline
zCOSMOS 822872 & xa53 & 10:02:02.09 & +02:09:37.40 & 0.7000 & 11.46 & 47.3 &  4.1 & 63 & 374 & 9.97 \\
zCOSMOS 805007 & xc53 & 10:00:58.20 & +01:45:59.00 & 0.6227 & 10.92 & 47.1 &  2.1 & 58 & 336 & 9.00 \\
zCOSMOS 822965 & xd53 & 10:01:58.73 & +02:15:34.20 & 0.7028 & 10.95 & 39.5 &  6.1 & 34 & 190 & 9.82 \\
zCOSMOS 811360 & xe53 & 10:01:00.74 & +01:49:53.00 & 0.5297 & 10.36 & 25.5 &  3.4 & 33 & 149 & 9.67 \\
zCOSMOS 834187 & xf53 & 09:58:33.86 & +02:19:50.90 & 0.5020 & 11.08 & 18.6 &  5.2 & 56 & 268 & 9.75 \\
zCOSMOS 800405 & xg53 & 10:02:16.78 & +01:37:25.00 & 0.6223 & 11.20 & 21.0 &  2.7 & 34 & 401 & 9.70 \\
zCOSMOS 837919 & xh53 & 10:01:09.67 & +02:30:00.70 & 0.7028 & 10.73 & 18.2 &  2.2 & 37 & 225 & 9.23 \\
zCOSMOS 838956 & xi53 & 10:00:24.70 & +02:29:12.10 & 0.7026 & 11.46 & 20.3 &  4.4 & 40 & 480 & 9.38 \\
zCOSMOS 824759 & xl53 & 10:00:28.27 & +02:16:00.50 & 0.7506 & 11.23 & 28.6 &  2.6 & 43 & 463 & 9.72 \\
zCOSMOS 810344 & xm53 & 10:01:53.57 & +01:54:14.80 & 0.7007 & 11.64 & 23.9 &  2.7 & 67 & 538 & 9.73 \\
zCOSMOS 839268 & xn53 & 10:00:11.16 & +02:35:41.60 & 0.6967 & 11.04 & 24.2 &  3.9 & 38 & 287 & 9.57 \\
zCOSMOS 828590 & xo53 & 10:02:51.41 & +02:18:49.70 & 0.6077 & 11.40 & 11.7 &  3.4 & 77 & 475 & 9.52 \\
zCOSMOS 838696 & xq53 & 10:00:35.69 & +02:31:15.60 & 0.6793 & 10.92 & 26.9 &  9.9 & 70 & 168 & 9.49 \\
zCOSMOS 816955 & xr53 & 10:01:41.85 & +02:07:09.80 & 0.5165 & 11.28 & 14.5 &  9.5 & 44 & 248 & 9.26 \\
zCOSMOS 823380 & xt53 & 10:01:39.31 & +02:17:25.80 & 0.7021 & 11.04 & 22.7 &  3.7 & 53 & 275 & 9.64 \\
zCOSMOS 831385 & xu53 & 10:00:40.37 & +02:23:23.60 & 0.5172 & 10.28 & 28.0 &  3.8 & 61 & 139 & 9.46 \\
zCOSMOS 850140 & xv53 & 10:01:43.66 & +02:48:09.40 & 0.6248 & 10.80 & 23.1 &  2.7 & 68 & 237 & 9.81 \\
zCOSMOS 824627 & xw53 & 10:00:35.52 & +02:16:34.30 & 0.7503 & 10.40 & 13.7 &  2.7 & 29 & 157 & 9.11 \\
zCOSMOS 831870 & l14co001 & 10:00:18.91 & +02:18:10.10 & 0.5024 & 10.18 & 29.0 &  1.8 & 30 & 139 & 9.48 \\
zCOSMOS 831386 & l14co004 & 10:00:40.29 & +02:20:32.60 & 0.6885 & 10.45 &  8.8 &  2.2 & 61 & 175 & 9.15 \\
zCOSMOS 838945 & l14co007 & 10:00:25.18 & +02:29:53.90 & 0.5015 & 10.71 &  4.1 &  6.2 & 77 & 163 & 9.15 \\
zCOSMOS 820898 & l14co008 & 09:58:09.07 & +02:05:29.76 & 0.6081 & 10.94 & 13.9 &  3.7 & 56 & 229 & 9.61 \\
zCOSMOS 826687 & l14co009 & 09:58:56.45 & +02:08:06.72 & 0.6976 & 10.45 & 21.4 &  3.9 & 56 & 160 & 9.43 \\
zCOSMOS 839183 & l14co011 & 10:00:14.30 & +02:30:47.16 & 0.6985 & 10.41 & 29.3 &  4.8 & 46 & 160 & 9.65 \\
zCOSMOS 838449 & l14co012 & 10:00:45.53 & +02:33:39.60 & 0.7007 & 10.59 & 10.0 &  1.0 & 40 & 283 & 9.41 \\
 AEGIS 30084 & xa54 & 14:19:17.33 & +52:50:35.30 & 0.6590 & 11.11 & 51.7 &  4.0 & 32 & 292 & 9.81 \\
 AEGIS 24556 & xd54 & 14:19:46.35 & +52:54:37.20 & 0.7541 & 10.36 & 28.9 &  2.7 & 33 & 155 & 9.58 \\
 AEGIS 25608 & xe54 & 14:19:35.27 & +52:52:49.90 & 0.5090 & 10.40 & 11.0 &  5.9 & 73 & 153 & 9.34 \\
 AEGIS 32878 & xf54 & 14:19:41.70 & +52:55:41.30 & 0.7683 & 10.71 & 19.9 &  4.3 & 44 & 200 & 9.51 \\
 AEGIS 3654 & xg54 & 14:20:13.43 & +52:54:05.90 & 0.6593 & 11.15 & 14.4 &  7.2 & 54 & 265 & 9.70 \\
 AEGIS 30516 & xh54new & 14:19:45.42 & +52:55:51.00 & 0.7560 & 10.28 & 13.1 &  3.5 & 38 & 144 & 8.92 \\
 AEGIS 23488 & l14eg006 & 14:18:45.52 & +52:43:24.10 & 0.5010 & 10.48 &  7.4 &  5.6 & 53 & 161 & 9.08 \\
 AEGIS 21351 & l14eg008 & 14:19:39.46 & +52:52:33.60 & 0.7315 & 10.94 & 79.5 &  5.1 & 50 & 247 & 9.92 \\
 AEGIS 31909 & l14eg009 & 14:20:04.88 & +52:59:38.84 & 0.7359 & 10.04 &  9.9 &  1.7 & 40 & 138 & 9.49 \\
 AEGIS 4004 & l14eg010 & 14:20:22.80 & +52:55:56.28 & 0.6702 & 10.74 &  9.3 &  1.7 & 32 & 273 & 9.11 \\
 AEGIS 6274 & l14eg011 & 14:20:26.20 & +52:57:04.85 & 0.5705 & 10.73 & 25.7 &  6.2 & 55 & 195 & 9.57 \\
 AEGIS 6449 & l14eg012 & 14:19:52.95 & +52:51:11.06 & 0.5447 & 11.04 &  9.1 &  7.0 & 65 & 228 & 9.08 \\
 AEGIS 9743 & l14eg014 & 14:20:33.58 & +52:59:17.46 & 0.7099 & 10.93 &  5.9 &  1.8 & 22 & 280 & 8.97 \\
 AEGIS 26964 & l14eg015 & 14:20:45.61 & +53:05:31.18 & 0.7369 & 10.97 & 13.4 &  2.0 & 36 & 345 & 9.00 \\
 AEGIS 34302 & l14eg016 & 14:18:28.90 & +52:43:05.28 & 0.6445 & 10.60 &  6.6 &  2.3 & 38 & 190 & 9.15 \\
 GOODS-N 21285 & xa55 & 12:36:59.92 & +62:14:50.00 & 0.7610 & 10.45 & 44.7 &  3.3 & 56 & 163 & 9.46 \\
 GOODS-N 6666 & xb55 & 12:36:08.13 & +62:10:35.90 & 0.6790 & 10.65 & 23.1 &  1.7 & 29 & 262 & 9.34 \\
 GOODS-N 19725 & xc55 & 12:36:09.76 & +62:14:22.60 & 0.7800 & 10.66 & 29.1 &  2.3 & 47 & 219 & 9.76 \\
 GOODS-N 12097 & xd55 & 12:36:21.04 & +62:12:08.50 & 0.7790 & 10.49 & 21.6 &  2.0 & 55 & 204 & 9.51 \\
 GOODS-N 19815 & xe55 & 12:36:11.26 & +62:14:20.90 & 0.7720 & 10.52 & 14.8 &  4.3 & 78 & 172 & 9.32 \\
 GOODS-N 7906 & xf55 & 12:35:55.43 & +62:10:56.80 & 0.6382 & 10.04 & 11.1 &  5.5 & 80 & 118 & 9.08 \\
 GOODS-N 19257 & xg55 & 12:37:02.93 & +62:14:23.60 & 0.5110 & 10.58 &  8.5 &  2.8 & 59 & 183 & 9.36 \\
 GOODS-N 16987 & xh55 & 12:37:13.87 & +62:13:35.00 & 0.7784 & 10.20 & 13.0 &  4.0 & 40 & 136 & 9.30 \\
 GOODS-N 10134 & xl55 & 12:37:10.56 & +62:11:40.70 & 0.7880 & 10.51 & 22.2 &  4.1 & 49 & 171 & 9.58 \\
 GOODS-N 30883 & l14gn006 & 12:36:34.41 & +62:17:50.50 & 0.6825 & 10.40 & 23.8 &  3.1 & 74 & 157 & 9.71 \\
 GOODS-N 939 & l14gn007 & 12:36:32.38 & +62:07:34.10 & 0.5950 & 10.87 &  8.9 &  7.7 & 46 & 234 & 9.57 \\
 GOODS-N 11532 & l14gn008 & 12:36:07.83 & +62:12:00.60 & 0.5035 & 10.28 &  5.5 &  4.6 & 30 & 140 & 9.18 \\
 GOODS-N 25413 & l14gn018 & 12:36:31.66 & +62:16:04.10 & 0.7837 & 10.40 & 32.8 &  2.0 & 36 & 168 & 9.40 \\
 GOODS-N 8738 & l14gn021 & 12:36:03.26 & +62:11:10.98 & 0.6380 & 10.71 & 76.9 &  0.7 & 47 & 311 & 9.84 \\
 GOODS-N 11460 & l14gn022 & 12:36:36.76 & +62:11:56.09 & 0.5561 & 10.11 &  6.8 &  0.6 & 24 & 133 & 8.57 \\
 GOODS-N 36596 & l14gn025 & 12:37:13.99 & +62:20:36.60 & 0.5320 & 10.65 &  3.5 &  0.4 & 60 & 262 & 8.89 \\
 GOODS-N 21683 & l14gn032 & 12:37:16.32 & +62:15:12.30 & 0.5605 & 11.11 &  7.6 &  2.5 & 28 & 304 & 8.85 \\
 GOODS-N 1964 & l14gn033 & 12:36:53.81 & +62:08:27.70 & 0.5609 & 10.04 &  6.7 &  4.8 & 38 & 116 & 8.90 \\
 GOODS-N 33895 & l14gn034 & 12:36:19.68 & +62:19:08.10 & 0.5200 & 10.87 &  8.7 &  5.3 & 64 & 213 & 9.83 \\
\hline
\hline
\end{tabular}
\begin{tablenotes}
  The table displays the original values from Freundlich et al. (2019) except for the effective radius of the stellar disk
  $R_{\rm eff}^{\rm disk}$ and the maximum rotation velocity
  $v_{\rm max}$, which is based on the total stellar mass and the I-band surface brightness profile (Sect.~\ref{sec:rotcurv}).
  The SFRs are estimated following a combination of UV and IR luminosities. The stellar masses are
  derived from SED fitting.
  For our model calculations the stellar masses and star formation rates were multiplied by a factor of $1.5$,
  the maximum rotation velocity by a factor of $\sqrt{1.5}$.
  The SIMBAD designations of AEGIS and GOODS-N objets are [SWM2014] AEGIS and [SWM2014] GOODS-N. 
\end{tablenotes}
\end{center}
\end{table*}

\begin{table*}
\begin{center}
\caption{DYNAMO galaxies.\label{tab:gdynamo}}
\begin{tabular}{lcccccccc}
\hline
Galaxy Name & RA & DEC & z & log(M$_*$) & SFR & $l_*$ & $v_{\rm flat}$ & $L_{\rm CO}$ \\
& & & & (M$_{\odot}$) & (M$_{\odot}$yr$^{-1}$) & (kpc)  & (km\,s$^{-1}$) & (K\,km\,s$^{-1}$pc$^2$) \\
\hline
G10-1 & 10 21 42.47 & +12 45 18.73 & 0.1437 & 10.44 & 15.7 &  1.2 & 117 &  9.89 \\
D20-1 & 20 52 09.08 & -00 30 39.32 & 0.0705 & 10.47 &  4.7 &  3.4 & 134 &  9.58 \\
G04-1 & 04 12 19.71 & -05 54 48.70 & 0.1298 & 10.81 & 21.3 &  2.8 & 260 & 10.41 \\
G20-2 & 20 44 02.91 & -06 46 57.91 & 0.1411 & 10.33 & 18.2 &  2.1 & 150 &  9.86 \\
D13-5 & 13 30 07.03 & +00 31 52.48 & 0.0754 & 10.73 & 17.5 &  2.0 & 180 & 10.11 \\
G08-5 & 08 54 18.74 & +06 46 20.57 & 0.1322 & 10.24 & 10.0 &  1.8 & 220 &  9.99 \\
D15-3 & 15 34 35.40 & -00 28 44.54 & 0.0671 & 10.73 &  8.3 &  2.2 & 240 & 10.11 \\
G14-1 & 14 54 28.31 & +00 44 34.49 & 0.1323 & 10.35 &  6.9 &  1.1 & 150 &  9.84 \\
C13-1 & 13 26 39.42 & +01 30 01.35 & 0.0788 & 10.55 &  5.1 &  4.2 & 223 &  9.91 \\
C22-2 & 22 39 49.34 & -08 04 18.05 & 0.0712 & 10.34 &  3.2 &  3.4 & 200 &  9.50 \\
SDSS 024921-0756 & 02 49 21.42 & -07 56 58.67 & 0.1530 & 10.48 & 10.5 &  1.1 &  84 & 10.12 \\
SDSS 212912-0734 & 21 29 12.14 & -07 34 57.69 & 0.1840 & 10.85 & 53.0 &  1.3 & 105 & 10.26 \\
SDSS 013527-1039 & 01 35 27.11 & -10 39 38.55 & 0.1270 & 10.85 & 25.3 &  1.6 & 190 & 10.22 \\
SDSS 033244+0056 & 03 32 44.77 & +00 58 42.08 & 0.1820 & 10.86 & 60.5 &  1.9 & 239 & 10.16 \\
\hline
IRAS08339+6517 & 08 38 23.09 & +65 07 16.05 & 0.0191 & 10.04 & 12.1 &  1.0 & 150 &  9.32 \\
\hline
\end{tabular}
\begin{tablenotes}
  SFRs were calculated from the emission-line ﬂux using the extinction-corrected H$\alpha$ line luminosity. The stellar masses were derived from the 4000-\r{A} break strength,
   the Balmer absorption-line index H$\delta_{\rm A}$ (Kauffmann et al. 2003).
  The SIMBAD designation of the first 10 sources is [GGM2014]. 
\end{tablenotes}
\end{center}
\end{table*}

\begin{figure*}
  \centering
  \resizebox{14cm}{!}{\includegraphics{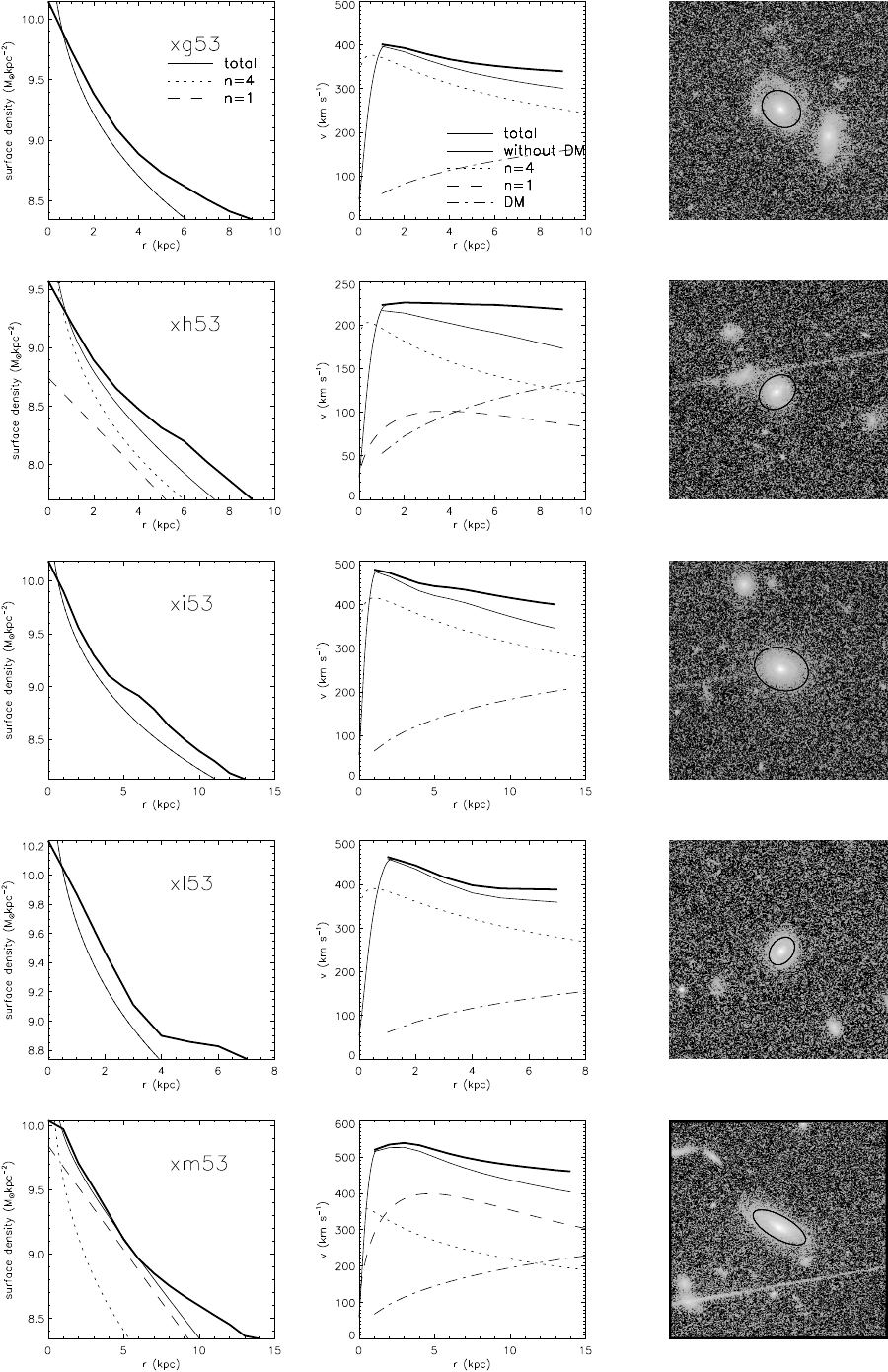}}
  \caption{Fig.~\ref{fig:profiles_phibss2_1} continued.
  \label{fig:profiles_phibss2_2}}
\end{figure*}
\begin{figure*}
  \centering
  \resizebox{14cm}{!}{\includegraphics{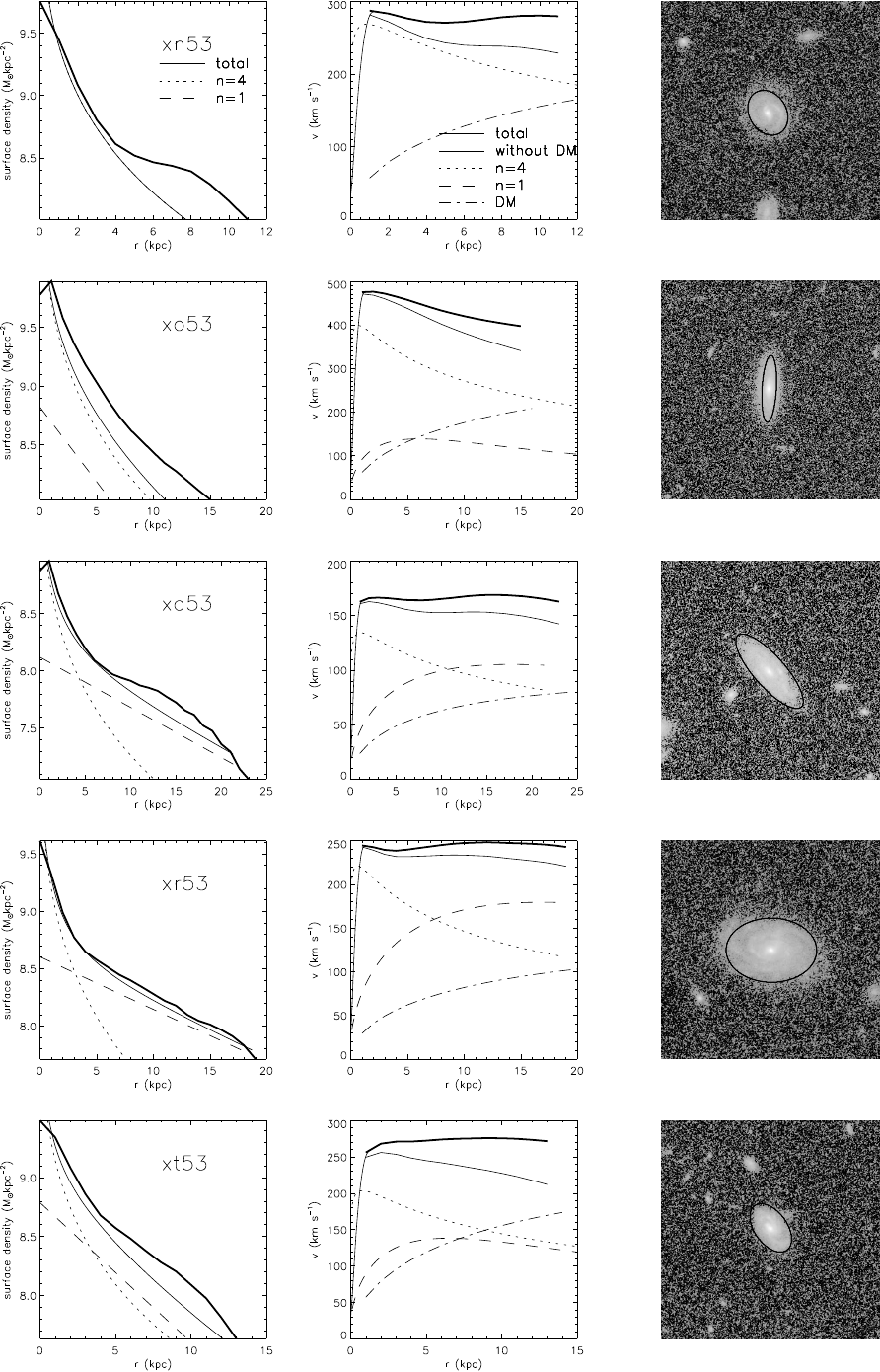}}
  \caption{Fig.~\ref{fig:profiles_phibss2_1} continued.
  \label{fig:profiles_phibss2_3}}
\end{figure*}
\begin{figure*}
  \centering
  \resizebox{14cm}{!}{\includegraphics{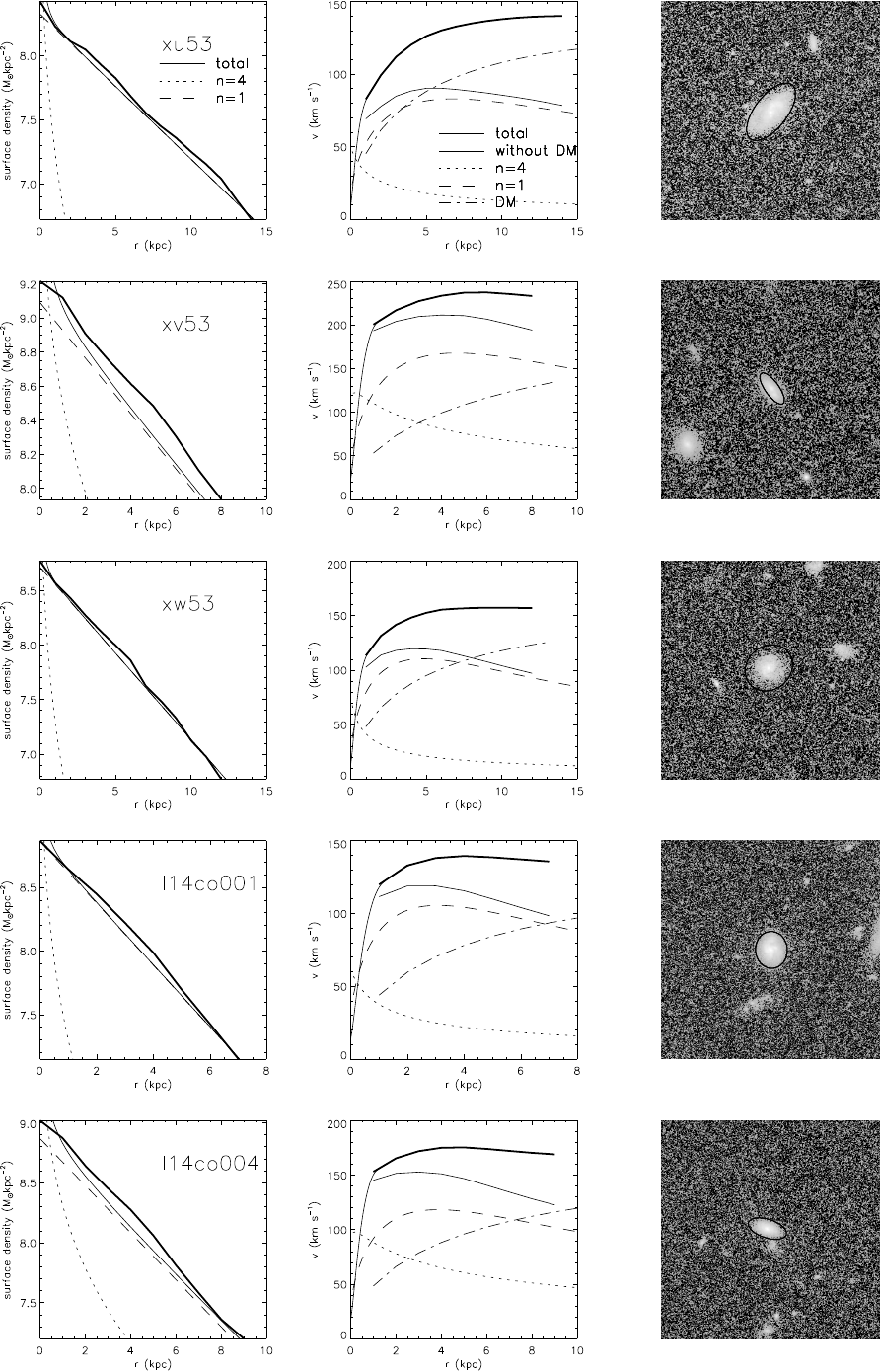}}
  \caption{Fig.~\ref{fig:profiles_phibss2_1} continued.
  \label{fig:profiles_phibss2_4}}
\end{figure*}
\begin{figure*}
  \centering
  \resizebox{14cm}{!}{\includegraphics{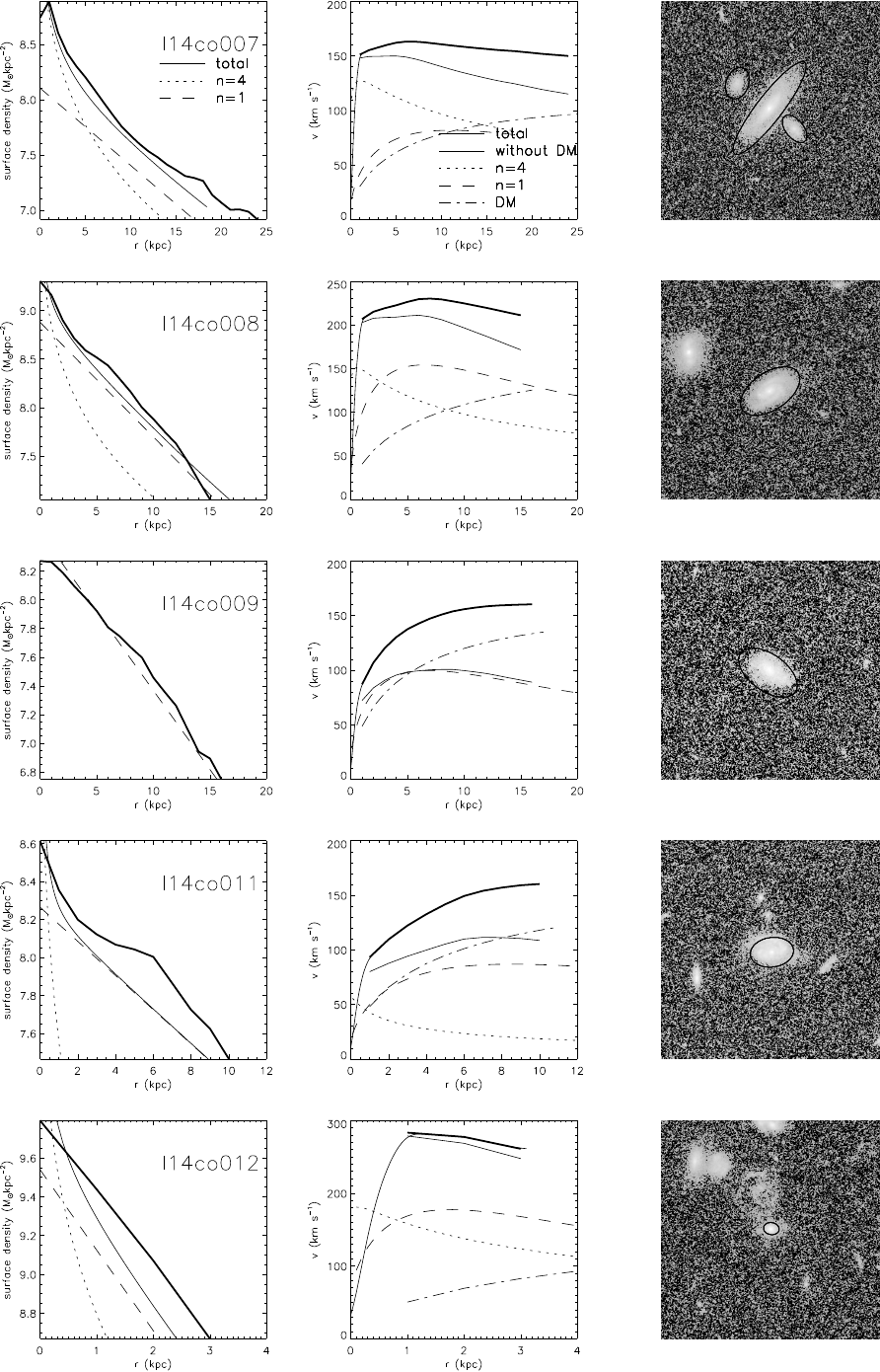}}
  \caption{Fig.~\ref{fig:profiles_phibss2_1} continued.
  \label{fig:profiles_phibss2_5}}
\end{figure*}
\begin{figure*}
  \centering
  \resizebox{14cm}{!}{\includegraphics{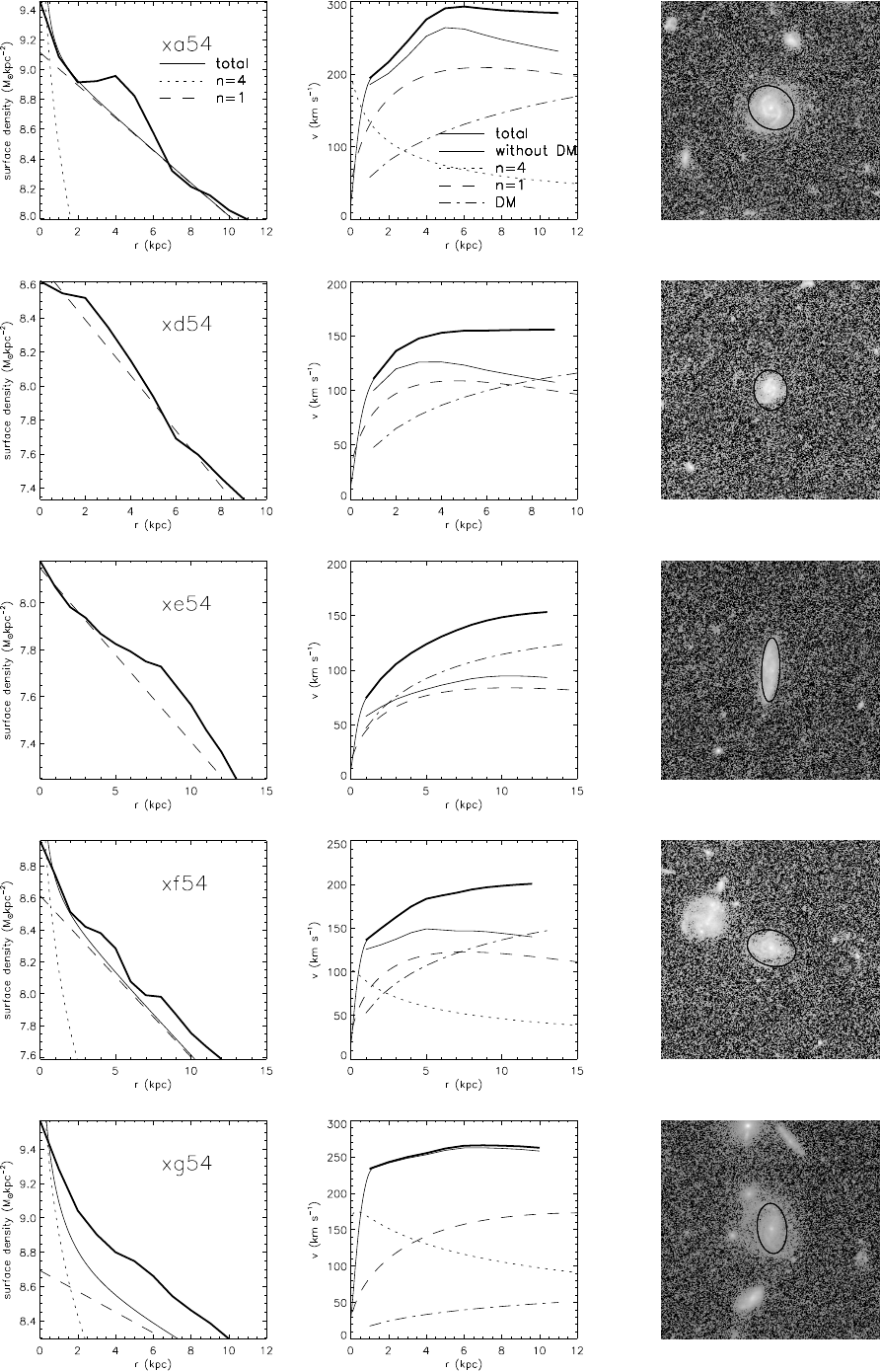}}
  \caption{Fig.~\ref{fig:profiles_phibss2_1} continued.
  \label{fig:profiles_phibss2_6}}
\end{figure*}
\begin{figure*}
  \centering
  \resizebox{14cm}{!}{\includegraphics{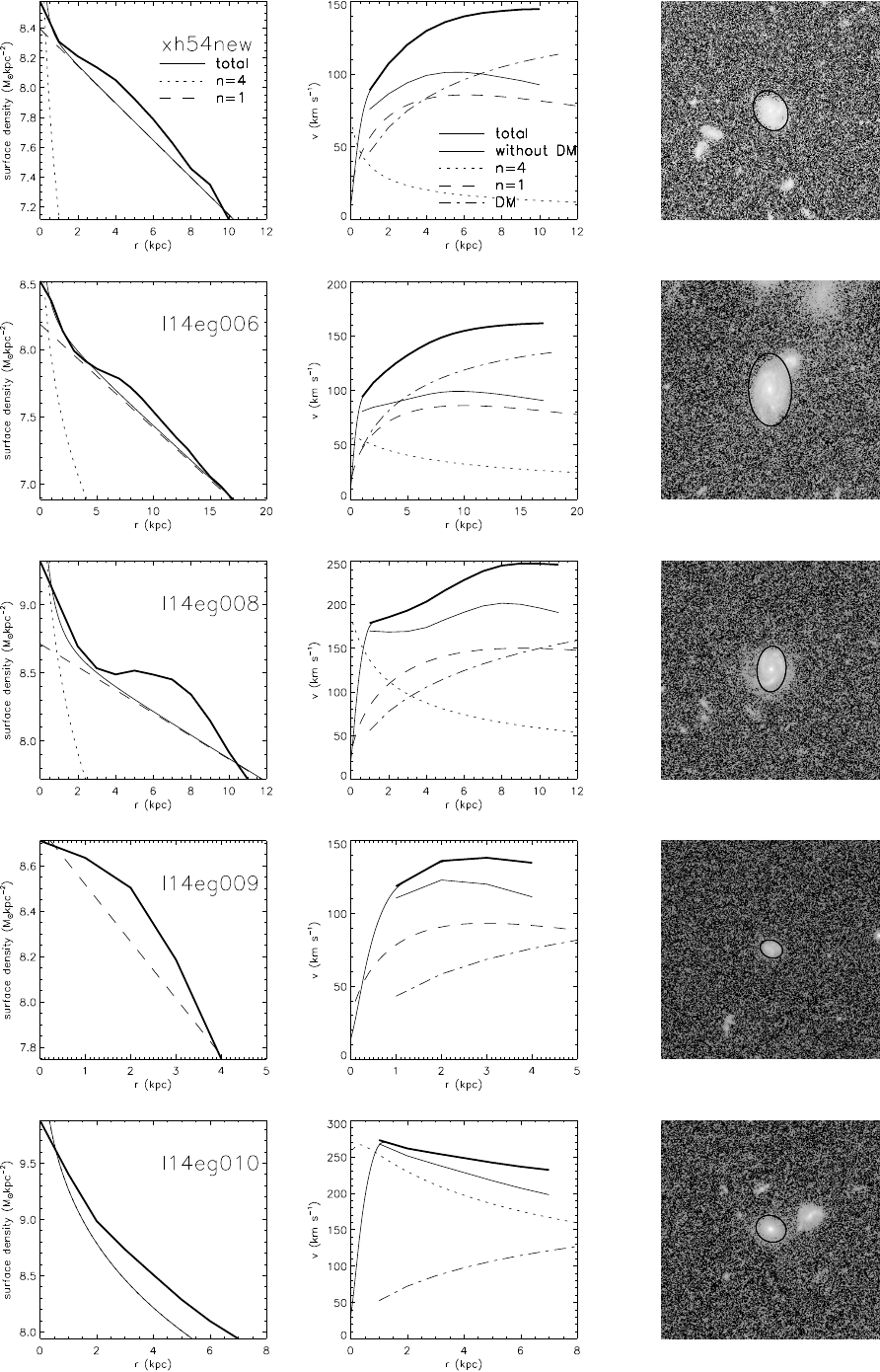}}
  \caption{Fig.~\ref{fig:profiles_phibss2_1} continued.
  \label{fig:profiles_phibss2_7}}
\end{figure*}
\begin{figure*}
  \centering
  \resizebox{14cm}{!}{\includegraphics{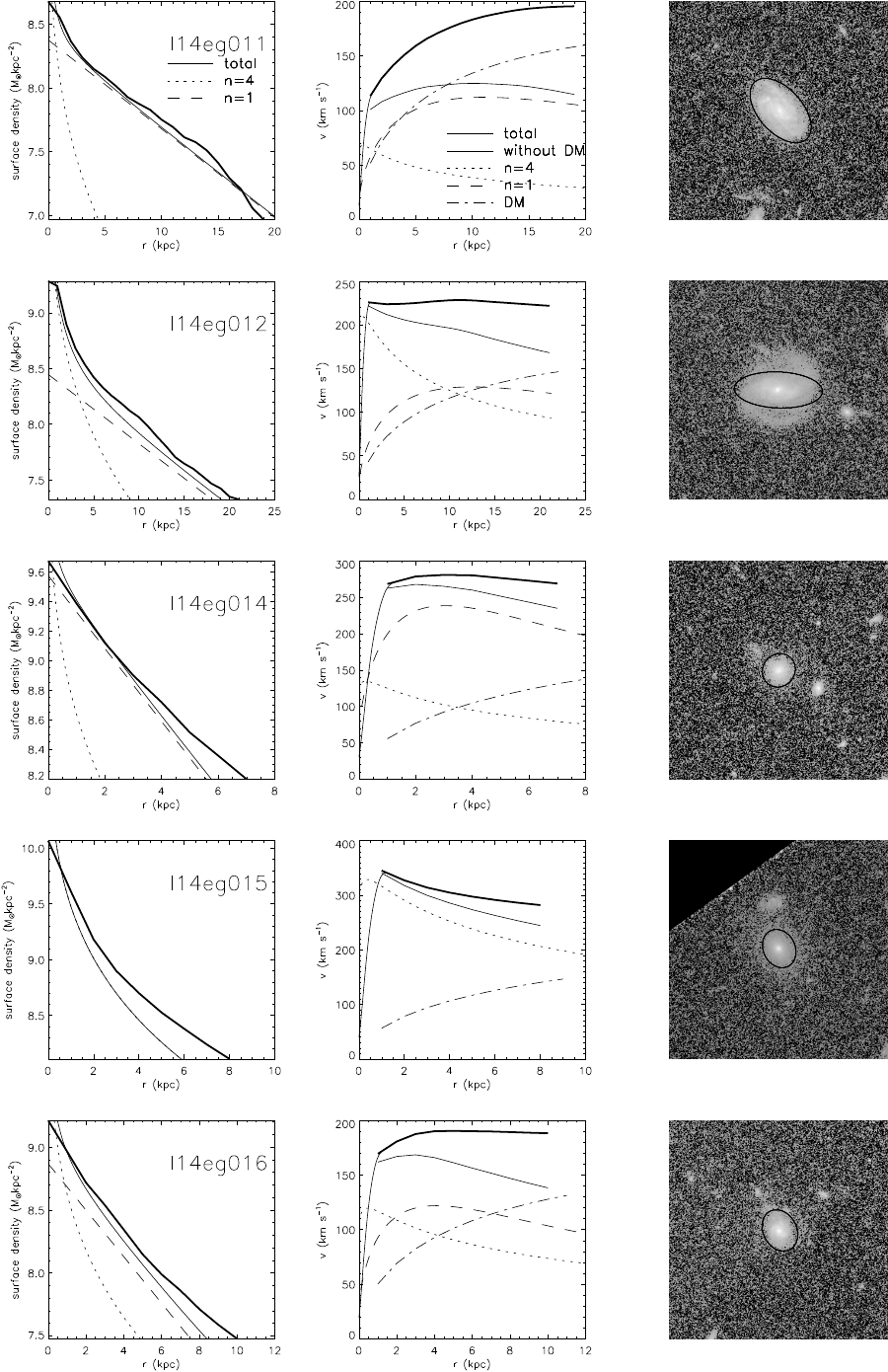}}
  \caption{Fig.~\ref{fig:profiles_phibss2_1} continued.
  \label{fig:profiles_phibss2_8}}
\end{figure*}
\begin{figure*}
  \centering
  \resizebox{14cm}{!}{\includegraphics{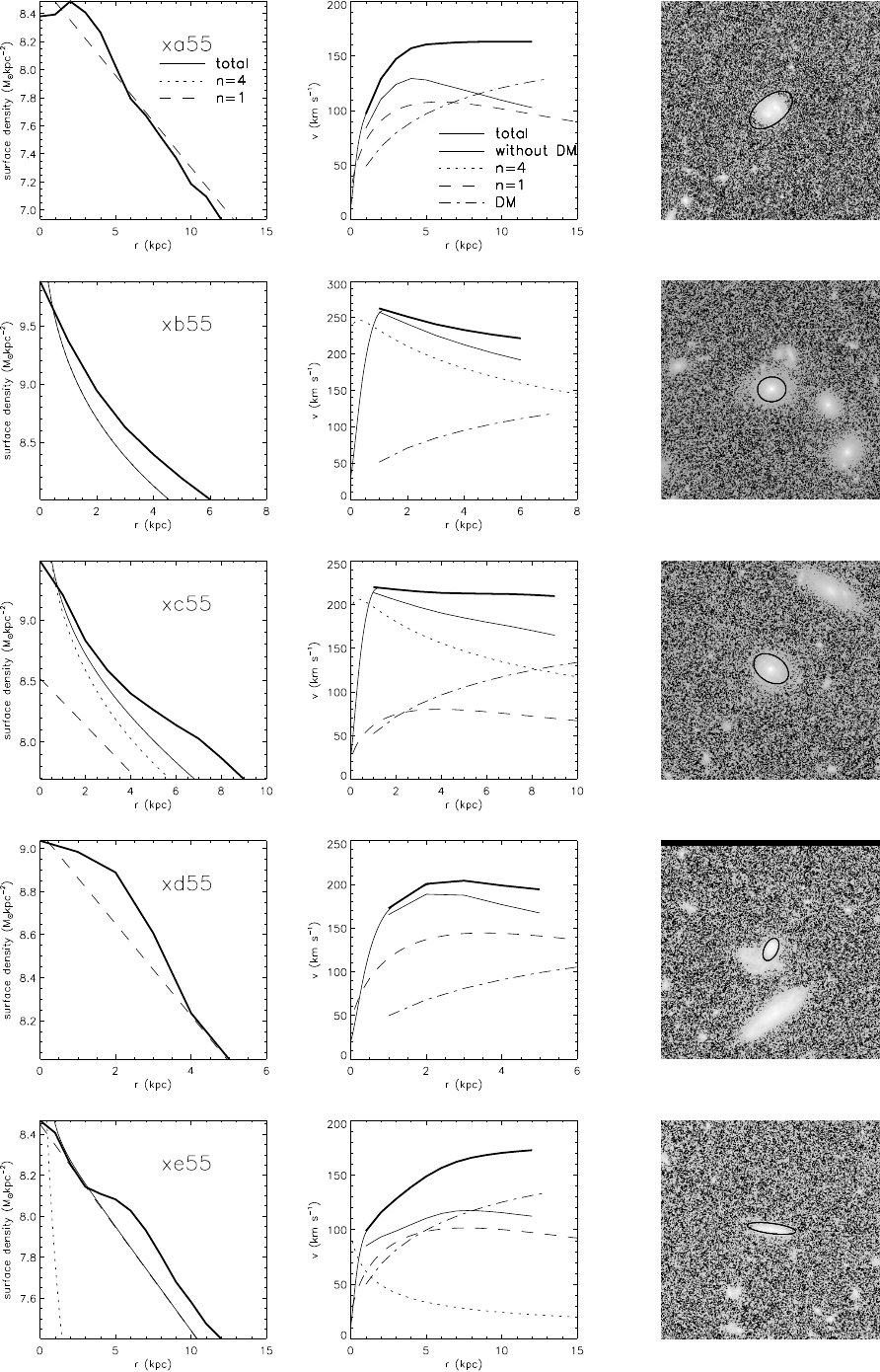}}
  \caption{Fig.~\ref{fig:profiles_phibss2_1} continued.
  \label{fig:profiles_phibss2_9}}
\end{figure*}
\begin{figure*}
  \centering
  \resizebox{14cm}{!}{\includegraphics{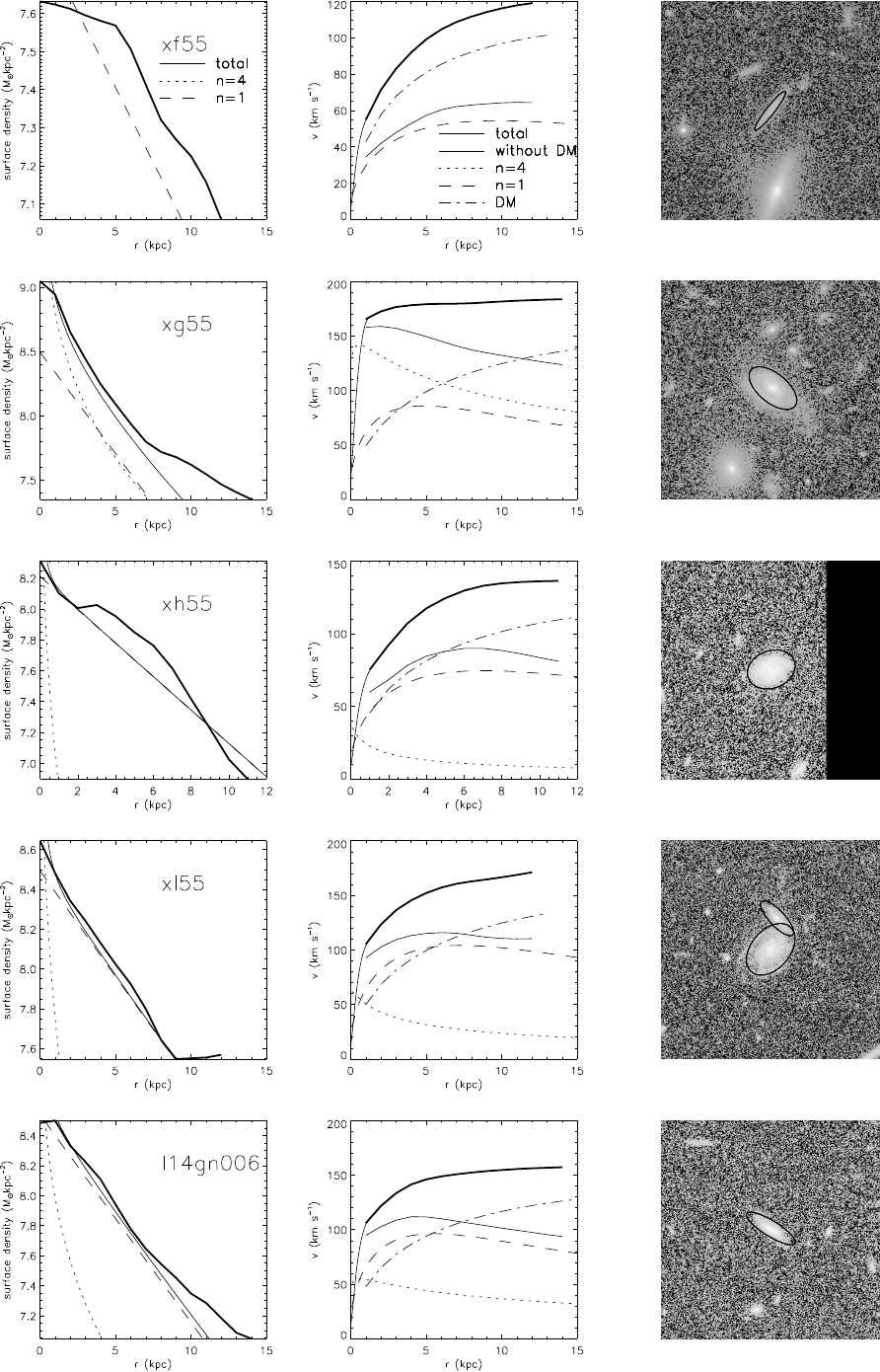}}
  \caption{Fig.~\ref{fig:profiles_phibss2_1} continued.
  \label{fig:profiles_phibss2_10}}
\end{figure*}
\begin{figure*}
  \centering
  \resizebox{14cm}{!}{\includegraphics{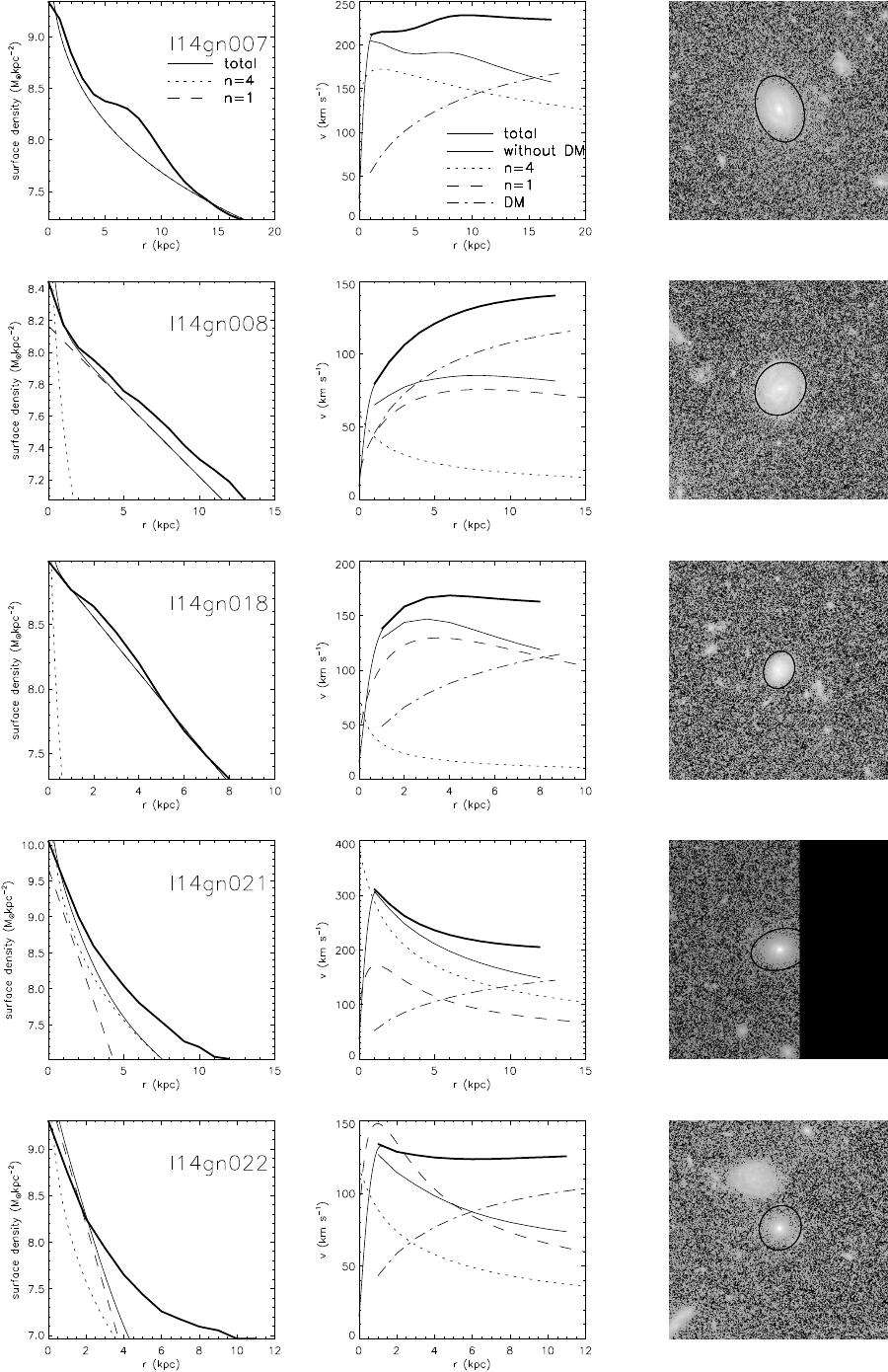}}
  \caption{Fig.~\ref{fig:profiles_phibss2_1} continued.
  \label{fig:profiles_phibss2_11}}
\end{figure*}
\begin{figure*}
  \centering
  \resizebox{14cm}{!}{\includegraphics{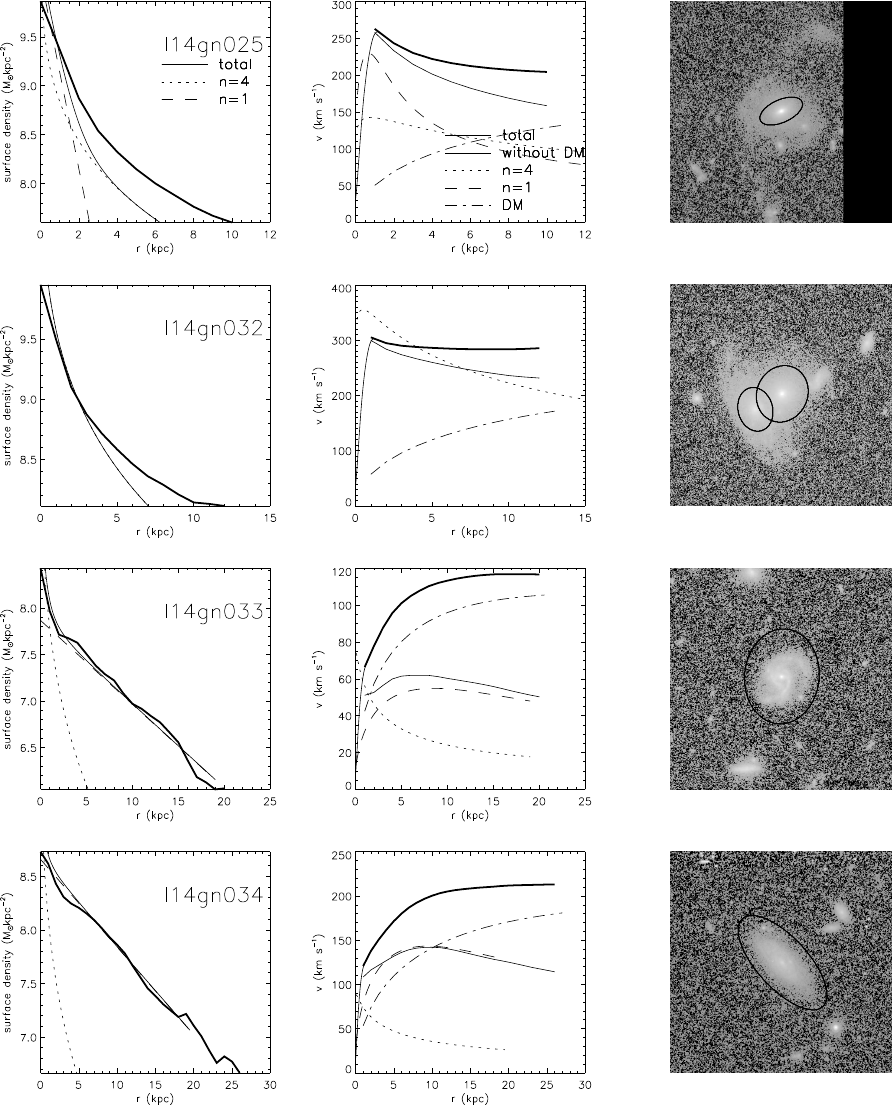}}
  \caption{Fig.~\ref{fig:profiles_phibss2_1} continued.
  \label{fig:profiles_phibss2_12}}
\end{figure*}

\begin{figure*}
  \centering
  \resizebox{\hsize}{!}{\includegraphics{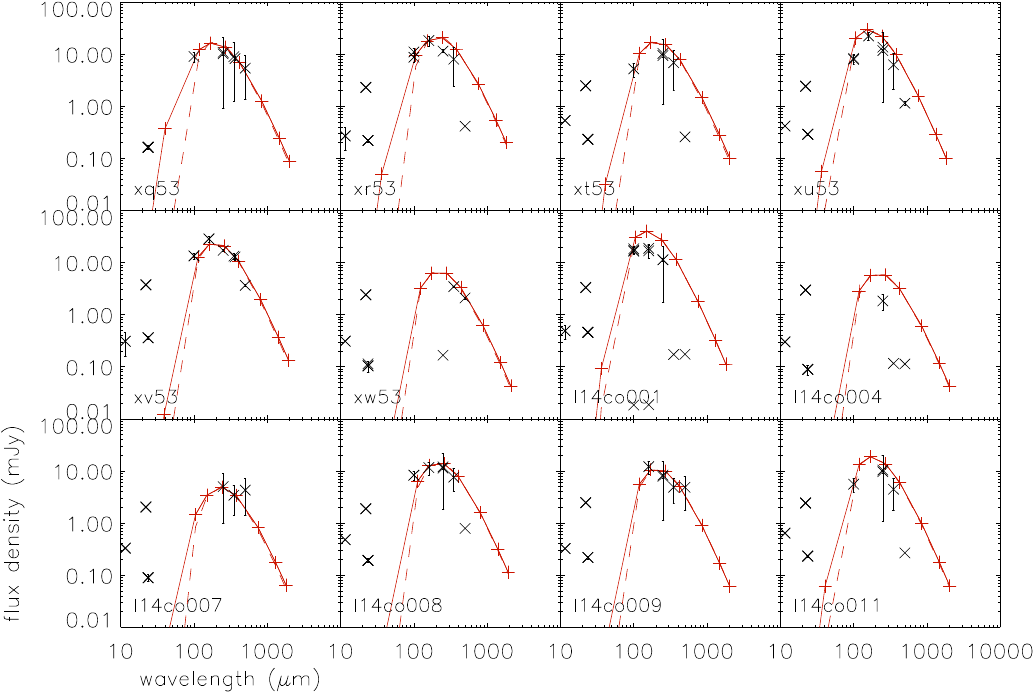}\includegraphics{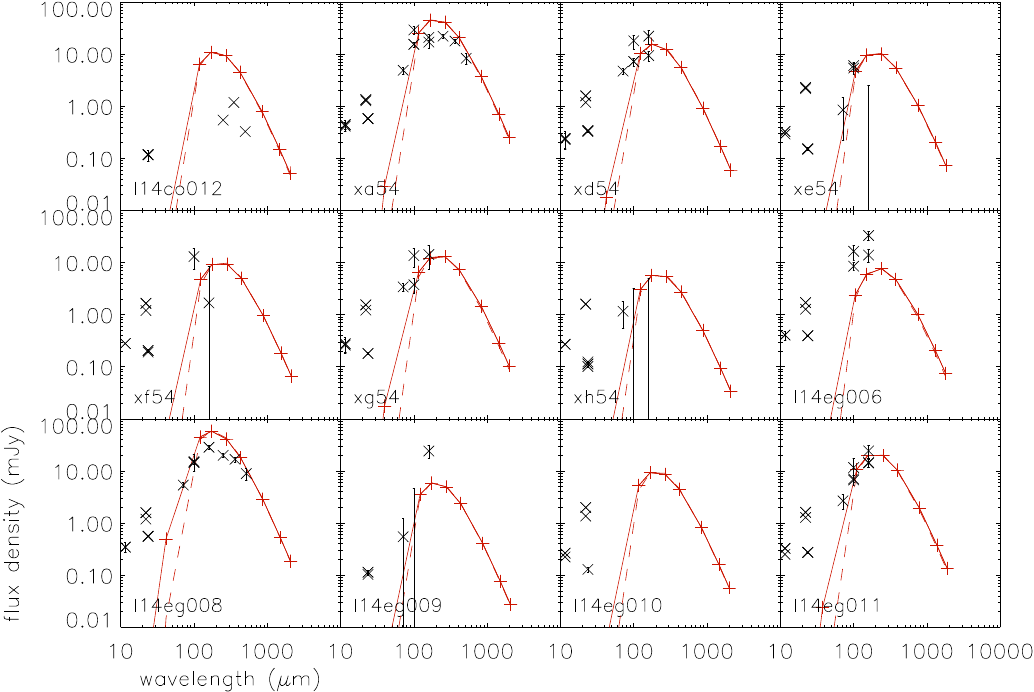}}
  \caption{IR SEDs of $z \sim 0.5$ LIRGs. Fig.~\ref{fig:IRspectra_phibss2_1} continued.
  \label{fig:IRspectra_phibss2_2}}
\end{figure*}

\begin{figure*}
  \centering
  \resizebox{\hsize}{!}{\includegraphics{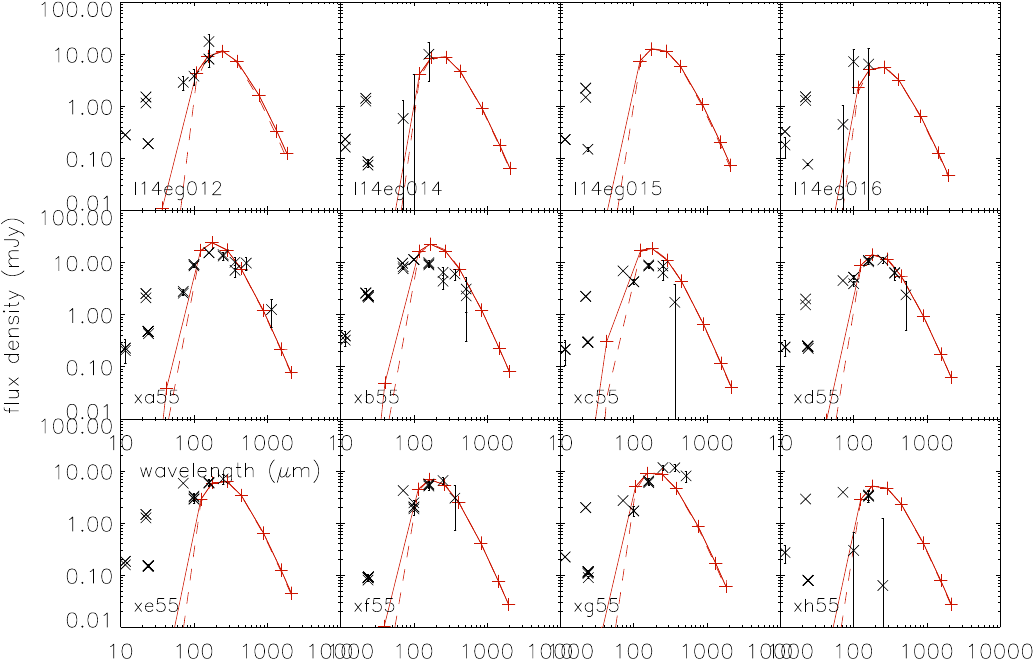}\includegraphics{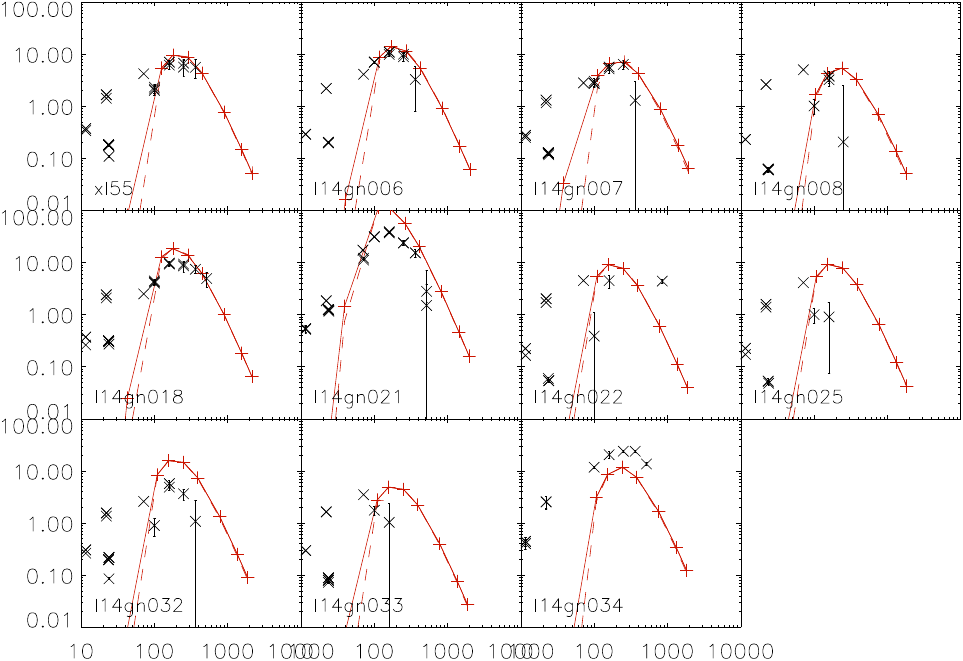}}
  \caption{IR SEDs of $z \sim 0.5$ LIRGs. Fig.~\ref{fig:IRspectra_phibss2_1} continued.
  \label{fig:IRspectra_phibss2_4}}
\end{figure*}

\begin{figure}
  \centering
  \resizebox{\hsize}{!}{\includegraphics{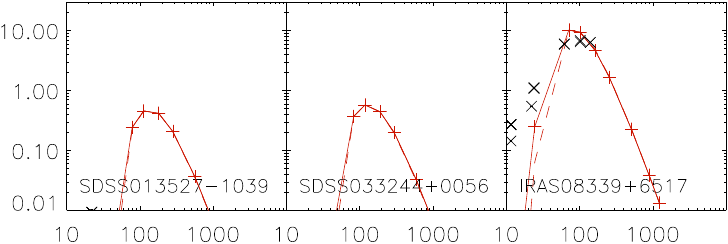}}
  \caption{IR SEDs of local LIRGs. Fig.~\ref{fig:IRspectra_phibss2_1} continued.
  \label{fig:IRspectra_dynamoe_2}}
\end{figure}

\end{appendix}
  
\end{document}